\documentclass[useAMS,usenatbib,onecolumn,usegraphicx]{mn2e}
\usepackage{times}
\usepackage{url}
\usepackage{color}

\newcommand{\bm}[1]{\mbox{\boldmath$#1$}}
\newcommand{\ave}[1]{\left\langle{#1}\right\rangle}
\newcommand{\skaco}[1]{\langle{#1}\rangle}
\newcommand{\tildel}{\tilde{\kappa}}
\newcommand{\colskip}{@{\hspace{0.3in}}}
\newcommand{\veclset}{\mbox{$\bm{l}_1,\bm{l}_2,\bm{l}_3$}}
\newcommand{\vecldset}{\mbox{$\bm{l}_1^\prime,\bm{l}^\prime_2,\bm{l}^\prime_3$}}

\newcommand{\tkappa}{\mbox{$\tilde{\kappa}$}}

\begin{document}
\title[Information content of WL bispectrum]
{
Information content of weak lensing power spectrum and bispectrum: including the 
non-Gaussian error covariance matrix
}

\author[I. Kayo, M. Takada and B. Jain]
{ 
Issha Kayo$^{1}$\thanks{E-mail: kayo@ph.sci.toho-u.ac.jp},
Masahiro Takada$^{2}$\thanks{E-mail: masahiro.takada@ipmu.jp} 
and Bhuvnesh Jain$^3$\thanks{E-mail: bjain@physics.upenn.edu}  \\
$^1$ Department of Physics, Toho University, 2-2-1 Miyama, Funabashi, 
Chiba 274-8510, Japan
\\
$^2$ Kavli Institute for the Physics and Mathematics of the Universe
 (Kavli IPMU, WPI),  The University of Tokyo, Chiba 277-8582, Japan
\\
$^3$ Department of Physics
and Astronomy, University of Pennsylvania, 
Philadelphia, PA 19104, USA
} 

\maketitle

\begin{abstract}
We address the amount of information in the non-Gaussian regime of weak
lensing surveys by modelling all relevant covariances of the power
spectra and bispectra, using 1000 ray-tracing simulation realizations
for a $\Lambda$ cold dark matter ($\Lambda$CDM) model and an analytical halo model.  
We develop a formalism to describe the covariance matrices of power
spectra and bispectra of all triangle configurations.
In addition to
the known contributions which extend up to six-point correlation
functions, we propose a new contribution `the halo sample variance
(HSV)' arising from the coupling of the lensing Fourier modes with
large-scale mass fluctuations on scales comparable with the survey
region via halo bias theory. We show that the model predictions are in
good agreement with the simulation once we take the HSV into
account.  The HSV gives a dominant contribution to the covariance
matrices at multipoles $l \ga 10^3$, which arises from massive haloes
with a mass of $\ga 10^{14}M_\odot$ and at relatively low redshifts
$z\la 0.4$.  Since such haloes are easily identified from a multi-colour
imaging survey, the effect can be estimated from the data.  By adding
the bispectrum to the power spectrum, the total information content or
the cumulative signal-to-noise ratio up to a certain maximum multipole
$l_{\rm max}$ of a few $10^3$, (S/N)$_{l_{\rm max}}$, is improved by 20--50 per cent, 
which is equivalent to a factor of 1.4--2.3 larger survey area
for the power spectrum measurement alone.  However, it is still smaller
than the case of a Gaussian field by a factor of 3 mostly due to the
HSV. 
Thus bispectrum measurements are useful for cosmology, but using
information from upcoming surveys requires that non-Gaussian
covariances are carefully estimated.
\end{abstract}

\section{Introduction}

The accelerated expansion of the Universe is the most tantalizing
problem in modern cosmology. Within Einstein's gravity theory, general
relativity, the cosmic acceleration can be explained by introducing dark
energy, which acts as a repulsive, rather than attractive, force to
expand the Universe.  Alternatively, the cosmic acceleration may be a
signature of the breakdown of general relativity on cosmological scales
\citep[see][for a review]{JainKhoury:10}. Many on-going and upcoming
wide-area galaxy surveys aim at testing dark energy and modified gravity
scenarios as the origin of cosmic acceleration; for example, the
Canada--France--Hawaii Telescope (CFHT) Weak Lensing Survey
\footnote{\url{http://www.cfhtlens.org/astronomers/content-suitable-astronomers}},
the Panoramic Survey Telescope \& Rapid Response System
(Pan-STARRS\footnote{\url{http://pan-starrs.ifa.hawaii.edu}}), the VLT
Survey Telescope (VST)
Kilo-Degree
Survey\footnote{\url{http://www.astro-wise.org/projects/KIDS/}}, the
Subaru Hyper Suprime-Cam Survey
\citep{Miyazakietal:06}\footnote{\url{http://www.naoj.org/Projects/HSC/index.html}},
the Dark Energy Survey
(DES\footnote{\url{http://www.darkenergysurvey.org}}), and in the next
decade, the Large Synoptic Sky Survey
(LSST\footnote{\url{http://www.lsst.org}}), the European Space Agency (ESA) Euclid satellite
mission\footnote{\url{http://sci.esa.int/science-e/www/area/index.cfm?fareaid=102}},
and the NASA Wide-Field Infrared Survey Telescope (WFIRST) satellite mission.
\footnote{\url{http://wfirst.gsfc.nasa.gov/}}

Among different cosmological probes, weak gravitational lensing or
cosmic shear is recognized as one of the most promising methods for
constraining the nature of dark energy, provided systematic errors are well
under control \citep[see][for
reviews]{BartelmannSchneider:01,SchneiderBook:06,HoekstraJain:08}. The
bending of light rays emitted from a distant galaxy due to the foreground
mass distribution causes the image to be distorted. The distortion
signal is too weak for us to measure in single galaxies, but we can use
a sufficiently large number of galaxy images, available from wide-field
survey, to detect the correlated shear signals existing in-between
different galaxy images. Weak lensing is a unique method of measuring
the total matter distribution including dark matter, free of galaxy bias
uncertainty, and allows a direct comparison of the measurement with
theory that is in most case about the statistical properties of the dark matter
distribution.  The theoretical predictions are
 obtained using N-body simulations
\citep[e.g.][]{SpringeletalNature:06} and/or analytical approaches
\cite[e.g.][]{Bernardeauetal:02,CooraySheth:02}. The cosmological weak
lensing signal has been measured by several groups \citep[e.g.][and
also see references therein]{Hamanaetal:03,HoekstraJain:08,Schrabbacketal:10}, and we are
waiting for measurements with much higher statistical
precision from upcoming surveys.

Most previous works on weak lensing, in theoretical and observational
studies, have focused on the shear two-point correlation function or
equivalently its Fourier transform, the power spectrum, as the
statistics to quantify the lensing field. Although these statistics
contain the full information when the field is a Gaussian random field
as in the cosmic microwave background (CMB) field
\citep{Komatsuetal:11}, it is not the case for the lensing field because
non-linear clustering in structure formation causes a coupling between
different Fourier modes, and the mass density field at redshifts
relevant for lensing surveys is not Gaussian. In fact, various studies
have shown that the information content carried by the power spectrum
might be saturated at multipole scales of a few $10^3$ [see 
\cite{Hamiltonetal:06,Takahashietal:09,Neyrincketal:09} for the 3D mass
density field, \cite{Satoetal:09} and \cite{Seoetal:11} for the lensing field, and
\cite{LeePen:08} for the result from the actual data]. In particular,
\cite{Satoetal:09} used 1000 ray-tracing simulation realizations for a
$\Lambda$ cold dark matter ($\Lambda$CDM) model to study the power spectrum covariance and the
information content of the power spectrum. They found that the
information content is reduced by a factor of 2 at multipoles $l\simeq
10^3 $ compared to the Gaussian case for a survey with typical source
redshift of $z_s\simeq 1$. Further, they showed that large-scale mass
density fluctuations of scales outside the simulation area contribute
significantly to non-Gaussian terms of the covariance. They developed a
formalism to describe the new non-Gaussian contribution by the number
fluctuations of massive haloes based on halo bias theory, which we
hereafter call the halo sample variance \citep[HSV; also
see][]{HuKravtsov03,TakadaBridle:07}.

Some fundamental questions remain unresolved: how important and useful
are the non-Gaussian signals in the lensing field for cosmology? Which
statistical method to extract the non-Gaussian signals is most useful?
Can we recover the Gaussian information content, which should have
existed in the linear field or the primordial field, by combining the
power spectrum and the non-Gaussian signals? For weak lensing, there is
additional expectation that the non-Gaussian signals will be useful for
cosmology, because the skewness, for example, has been shown complementary
to the power spectrum in its dependence on cosmological parameters
\citep{Bernardeauetal:97,JainSeljak:97,Hui:99,Jainetal:00,WhiteHu:00,
HamanaMellier:01,VanWaerbekeetal:01,CoorayHu:01b,TakadaJain:02,
TakadaJain:04, DodelsonZhang:05,KilbingerSchneider:05,
Sembolonietal:08,Bergetal:10,Munshietal:11,Piresetal:12}. The attempt to
measure the non-Gaussian signals from actual data was also made by
several groups \citep{Bernardeauetal:02,Zhangetal:03,Jarvisetal:04},
and the first significant detection was recently reported by 
\cite{Sembolonietal:11}, showing an improvement in cosmological
parameters compared to the two-point statistics alone.

In this paper, we study the
lensing bispectrum,
which contains the lowest-order
non-Gaussianity of the weak lensing field and is a natural extension
of the power spectrum. We consider all
triangle configurations available from a given range of multipoles
and their full covariance matrix including the non-Gaussian
contributions up to six-point correlation functions as well as the HSV
term, while only the
Gaussian errors have been assumed in most previous work
\citep{TakadaJain:04,Martinetal:12}. We use the 1000 simulation
realizations to study the usefulness and complementarity of the lensing
bispectrum compared to the power spectrum, and also develop an
analytical formula to describe the bispectrum covariance for a given
cosmology. In particular, we will
show that the HSV gives a significant contribution to
the bispectrum covariance at $l\ga $ a few $10^2$, and that the
bispectrum does carry additional information to the power spectrum even
in the presence of these significant correlations. 
Thus we will give a
quantitative answer to the fundamental questions above.

This paper is organized as follows. After briefly reviewing the lensing
power spectrum and bispectrum in Section~\ref{sec:preliminary}, we develop a
formulation to describe the bispectrum covariance in Section~\ref{sec:cov}.
In Section~\ref{sec:results}, we show the main results: we study the
bispectrum covariance using both simulations and
analytical model predictions. We quantify the information content
of the lensing bispectrum by including contributions from all
triangle configurations. Section~\ref{sec:conc} is devoted to discussion
and conclusion.

\section{Preliminaries: Lensing power spectrum and bispectrum}
\label{sec:preliminary}

In the context of cosmological gravitational lensing, the convergence
field is expressed as the weighted projection of the three-dimensional
density fluctuation field between source and observer \citep[see][for a
thorough review]{BartelmannSchneider:01,SchneiderBook:06}
\begin{equation}
\kappa(\bm{\theta})=\int_0^{\chi_H}\!\!d\chi W_{\rm GL}(\chi) 
\delta[\chi, \chi\bm{\theta}],
\label{eq:kappa}
\end{equation}
where $\bm{\theta}$ is the angular position on the sky, $\chi$ is the
comoving distance, and $\chi_H$ is the distance to the Hubble horizon.
We assume a flat geometry throughout this paper, and the radial
distance $\chi$ is equivalent to the comoving angular diameter distance.  
The comoving
distance $\chi(a)$ from an observer at $a=1$ to a source at $a$ is
expressed in terms of the Hubble expansion rate $H(a)$ as
$\chi(a)=\int^1_a\!\!da'/[H(a')a^{\prime 2}]$.  For source galaxies at a single redshift, the lensing efficiency function $W_{\rm GL}(\chi)$ is defined
as
\begin{equation}
W_{\rm GL}(\chi)\equiv \frac{3}{2}H_0^2\Omega_{\rm m0}a^{-1}\chi \left(
1-\frac{\chi}{\chi_s}
\right),
\end{equation}
where $\chi_s$ is the distance to the source galaxies.
See equation~(4) in
\cite{TakadaJain:09} for the lensing efficiency functions for tomographic
redshift bins. 
Under the flat-sky approximation, the Fourier transform
of the lensing field is defined as
\begin{equation}
\kappa(\bm{\theta})=\int\!\frac{d^2\bm{l}}{(2\upi)^2}\tilde{\kappa}_{\bm{l}}
{\rm e}^{{\rm i}\bm{l}\cdot\bm{\theta}}. 
\end{equation}
When the sky coverage of a survey is finite, we need to use the discrete
Fourier decomposition, rather than the infinite-range Fourier
decomposition \citep[also see appendix in][for details]{TakadaBridle:07}.

In this paper, we study the bispectrum of lensing field and the
covariance of the bispectrum.  The $n$-point power spectra relevant for
the bispectrum covariance are defined in terms of the ensemble averages
of the convergence fields in Fourier space as
\begin{eqnarray}
\skaco{\tilde{\kappa}_{\bm{l}_1}\tilde{\kappa}_{\bm{l}_2}
}&\equiv &(2\upi)^2 P(l_1)\delta_D^2(\bm{l}_1+\bm{l}_2),\\
\skaco{\tilde{\kappa}_{\bm{l}_1}\tilde{\kappa}_{\bm{l}_2}\tilde{\kappa}_{\bm{l}_3}}
&\equiv &(2\upi)^2 B(\bm{l}_1,\bm{l}_2,\bm{l}_3)
\delta_D^2(\bm{l}_1+\bm{l}_2+\bm{l}_3),\\
\skaco{\tilde{\kappa}_{\bm{l}_1}\tilde{\kappa}_{\bm{l}_2}
\tilde{\kappa}_{\bm{l}_3}\tilde{\kappa}_{\bm{l}_4}
}_c
&\equiv &(2\upi)^2 T(\bm{l}_1,\bm{l}_2,\bm{l}_3,\bm{l}_4)
\delta_D^2(\bm{l}_1+\bm{l}_2+\bm{l}_3),\\
\skaco{\tilde{\kappa}_{\bm{l}_1}\tilde{\kappa}_{\bm{l}_2}
\dots
\tilde{\kappa}_{\bm{l}_n}}_c
&\equiv &(2\upi)^2 P_{n }(\bm{l}_1,\bm{l}_2,\dots,\bm{l}_n)
\delta_D^2(\bm{l}_1+\bm{l}_2+\cdots+\bm{l}_n)\hspace{2em} \mbox{if $n\ge 5$}, 
\label{eq:ps_def}
\end{eqnarray}
where $\delta_D^2(\bm{l})$ is the Dirac delta function; $P$, $B$ and $T$
are the lensing power spectrum, bispectrum and trispectrum,
respectively; $P_n$ is the $n$-point power spectrum.  For the bispectrum
covariance, we need to include up to the six-point power spectra
$P_6$. The higher-order correlation function than the
bispectrum is the connected part of the $n$-point function, which
characterizes the non-Gaussianity of the lensing field and cannot be
expressed in terms of products of the power spectrum (or
any other lower-order correlation functions).
Exchange symmetry of wavevectors, $\bm{l}_i\leftrightarrow \bm{l}_j$,
reflects that the $n$-point correlation functions are
invariant under permutations of the arguments.
The delta functions come from the parallel translation invariance for a
statistically homogeneous field as is the convergence field.

The lensing power spectrum and bispectrum can be given as the weighted
line-of-sight projection of the three-dimensional power spectrum and
bispectrum of the underlying mass distribution. Employing Limber's
approximation \citep{Limber:54} and the flat-sky approximation, 
the lensing power spectrum and bispectrum are expressed as
\begin{eqnarray}
P(l)&=& \int_0^{\chi_H}\!d\chi W^2_{\rm GL}(\chi)\chi^{-2}
 P_\delta\left(k=\frac{l}{\chi};\chi \right),\\
B(\bm{l}_1,\bm{l}_2,\bm{l}_3)&=&\int^{\chi_H}_0\!d\chi 
W_{\rm GL}^3(\chi)\chi^{-4}\left. 
B_\delta(\bm{k}_1,\bm{k}_2,\bm{k}_3;\chi)\right|_{\bm{k}_i=\bm{l}_i/\chi},
\end{eqnarray}
where $P_\delta$ and $B_\delta$ are the power spectrum and bispectrum of
the mass distribution at each redshift $\chi(=\chi(z))$. Thus once the
$n$-point spectra of the mass density field
 are given for a given cosmological model, we
can compute the $n$-point spectra of the lensing field. The above
equations also mean that statistical properties of the lensing field
arise from those of the mass density field, since the prefactors
such as the $W_{\rm GL}(\chi)$ are pure geometrical quantities, not
statistical variables.

The power spectrum measurement for an actual survey is affected by the
intrinsic shape noise due to a finite sampling of source galaxy shapes:
\begin{equation}
P^{\rm obs}(l)=P(l)+\frac{\sigma_\epsilon^2}{\bar{n}_g},
\label{eq:ps_wshotn}
\end{equation}
where $\sigma_{\epsilon}$ is the rms of intrinsic ellipticities per
component, and $\bar{n}_g$ is the mean number density of source galaxies
per unit steradian. In the following, we will often omit the notation
${}^{\rm obs}$ to refer $P^{\rm obs}(l)$ for notational simplicity.
Throughout this paper we assume that  the orientation of intrinsic 
galaxy shape is
random and the shapes of different galaxies are uncorrelated; the
shape noise is  a Gaussian random field. 
 The
bispectrum is a measure of the non-Gaussianity, so is
not affected by the shape noise.

\section{Lensing Covariance}
\label{sec:cov}

The covariances of the lensing spectrum and bispectrum describe 
a measurement
accuracy of the spectra for a given survey. There are several sources
of the measurement errors: the shot noise arising due to a finite
sampling of galaxy shapes and the sample variance arising due to a
finite survey area. If the lensing field is Gaussian, the different
Fourier modes with $\bm{l}\ne\bm{l}'$
are independent, and therefore the sample variance is
determined by the number of independent Fourier modes for a given
multipole bin $l$ that are
available from the 
survey, yielding a simple formula of the sample variance contribution 
\citep[e.g.][]{Knox:95}. However, this is not the case for the lensing
field, because the lensing field is highly non-Gaussian at scales of interest
\citep{TakadaJain:04,TakadaJain:09,Satoetal:09}, 
and the different Fourier modes
correlate with each other.
In the following, we discuss theory for the lensing covariance matrices.

\subsection{Power spectrum covariance} 
\label{sec:pcov}

The power spectrum covariance has been well studied by previous works
\citep{Scoccimarroetal:99,CoorayHu:01,TakadaBridle:07,TakadaJain:09,Satoetal:09}. In
particular, \cite{Satoetal:09} derived an expression of the power
spectrum covariance including a new contribution from the mass
density fluctuations at larger scales than the survey area, and showed that the
formula well reproduces their ray-tracing simulation results.
According to this work, the power spectrum covariance is given as
\begin{eqnarray}
{\rm Cov}[P(l_i),P(l_j)]&=&\frac{2\delta^K_{l_i l_j}}{N_{\rm pairs}(l)}
\left[
P(l_i)+\frac{\sigma_\epsilon^2}{\bar{n}_g}
\right]^2+
\frac{1}{\Omega_{\rm
s}}\int_{\left|\bm{l}\right|\in l_i}\!\frac{d^2\bm{l}}{A(l_i)}
\int_{\left|\bm{l}'\right|\in
l_j}\!\frac{d^2\bm{l}'}{A(l_j)}T(\bm{l},-\bm{l},\bm{l}',-\bm{l}')
+{\rm Cov}^{PP}_{\rm HSV}(l_i,l_j; \Omega_{\rm s}),
\label{eq:pscov}
\end{eqnarray}
where $\delta^K_{l_il_j}$ is the Kronecker delta,
$\delta^K_{l_il_j}=1$ if $l_i=l_j$ within the bin width and otherwise
$\delta^K_{l_il_j}=0$; $\Omega_{\rm s}$ is the survey area in units of
steradian; $A(l_i)$ is the area of the above integration in Fourier
space, given as $A_s(l_i)\equiv \int_{\left| \bm{l}\right|\in
l_i}d^2\bm{l}$, where the integration range is confined to the
wavevectors satisfying the condition $l_i-\Delta l/2\le |\bm{l}|\le
l_i+\Delta l/2$ ($\Delta l$ is the bin width around the $i$th bin,
$l_i$); the third term ${\rm Cov}_{\rm HSV}$
is the new contribution which we call the halo sample variance (HSV)
contribution (see below). The quantity $N_{\rm pairs}(l_i)$ is the
number of independent pairs of two vectors $\bm{l}$ and $-\bm{l}$ in
Fourier space,
where the vector $\bm{l}$ has the length $l_i$ within the bin width
and ``independent'' here means different pairs discriminated by the
fundamental Fourier mode of a given survey, $l_f\simeq 2\upi/\Theta_{\rm
s}$ ($\Theta_{\rm s}$ is the angular scale of the survey area).
At the
limit $l_i\gg l_f$, $A(l_i)\simeq 2\upi l_i \Delta l$ and the number of
independent Fourier modes is given as
\begin{equation}
N_{\rm pairs}(l_i)\simeq \frac{2\upi l_i \Delta l}{(2\upi/\Theta_{\rm s})^2}
=\frac{\Omega_{\rm s}l_i\Delta l
}{2\upi}=2f_{\rm sky}l_i \Delta l, 
\label{eq:n_mode}
\end{equation}
where $f_{\rm sky}$ is the sky fraction defined as $f_{\rm sky}\equiv
\Omega_{\rm s}/4\upi$. See \cite{TakadaBridle:07} for a pedagogical
derivation of the power spectrum covariance based on the discrete
Fourier decomposition formulation (except for the third term ${\rm
Cov}^{PP}_{\rm HSV}$). In equation~(\ref{eq:pscov}), we ignored effects of
non-trivial survey geometry for simplicity.

The first and second terms on the r.h.s. of equation~(\ref{eq:pscov}) are the
standard covariance terms studied in most previous works.  The first
term describes the Gaussian covariance term that vanishes when $l_i\ne
l_j$, i.e. no correlation between different multipole bins. The second
term gives a non-Gaussian term arising from the lensing trispectrum
(four-point correlation function), which describes the mode coupling
between different multipole bins.  Both terms scale with survey area as
$1/\Omega_{\rm s}$; the amplitudes decrease with increasing the survey
area. It should also be noted that the Gaussian term depends on the
multipole bin width, while the non-Gaussian terms do not; a larger bin
width relatively reduces the Gaussian term contribution at the multipole
bin.

The third term of equation~(\ref{eq:pscov}) arises from the mode
coupling of the Fourier mode of our interest, $l_i$, with large-scale
modes of scales comparable with or even outside the survey region. Such
large-scale modes cannot be seen by an observer, but affect the power
spectrum estimation. If the entire survey region happens to be in an
overdense/underdense region, which is caused by the large-scale mass
density fluctuations, the number of massive haloes becomes
larger/smaller than the ensemble average according to the halo bias
theory \citep{MoWhite:96,Shethetal:01}.  Thus the number of massive
haloes found in a finite survey volume correlates with the mass density
fluctuations of scales comparable with or larger than the survey field
\citep{HuKravtsov03}.  To be more explicit, the number fluctuations of
haloes in a mass $M$ and in the redshift slice centred at $z$ are given
as
\begin{equation}
\delta N(M)=b(M)\frac{d^2V}{dz
 d\Omega}\Omega_{\rm s}\Delta z\frac{dn}{dM}\bar{\delta}_m(\Theta_{\rm s};z),
\label{eq:dN}
\end{equation}
where $d^2V/dz d\Omega$ is the comoving volume per unit redshift
interval and per unit solid angle, $d^2V/dz d\Omega=\chi^2$ for a flat
universe; $dn/dM$ is the ensemble-averaged mass function of haloes in the
mass range $[M,M+dM]$; $b(M)$ is the halo bias parameter;
$\delta_m(\Theta_s)$ is the mass density fluctuation  averaged
within the survey volume in the redshift slice which has area
$\Omega_{\rm s}$ and the redshift width $\Delta z$.
The lensing power spectrum amplitudes at small
angular scales are sensitive to the number of massive haloes in the
survey region and then correlates with the number fluctuations, which
results in the HSV. Note that the ensemble average of
the power spectrum is not affected by the
large-scale mode due to the fact $\ave{\bar{\delta}_m(\Theta_{\rm
s})}=0$. At the limit of $l_i,l_j\gg 1$,
the HSV contribution is given as
\begin{equation}
{\rm Cov}^{PP}_{\rm HSV}(l_i,l_j;\Omega_{\rm s})=
\int^{\chi_s}_0\!\!d\chi\left(\frac{d^2V}{d\Omega d\chi}\right)^2
\left[
\int\!\!dM\frac{dn}{dM}b(M)\left|\tilde{\kappa}_{l_i}\right|^2\right]
\left[
\int\!\!dM'\frac{dn}{dM'}b(M')\left|\tilde{\kappa}_{l_j}\right|^2
\right]
\left[
\int_0^{\infty}\!\!\frac{kdk}{2\upi}P_m^L(k)\left|
\tilde{W}(k\chi\Theta_{\rm s})
\right|^2
\right],
\label{eq:pscov_hsv}
\end{equation}
where $\tilde{W}(l)$ is the Fourier transform of the
survey window function; $P_m^L$ is the linear mass power spectrum;
$\tilde{\kappa}_{l}(\chi)$ is the Fourier transform of the convergence
field for which we assume a Navarro-Frenk-White (NFW) halo
\citep{Navarroetal:97} (see equation~28 in \citealt{OguriTakada:11} or section~3.2
in \citealt{TakadaJain:03} for the expression of $\tilde{\kappa}_l$).
For notational simplicity, we
omit to denote the redshift dependence of $dn/dM$, $b(M)$ and
$\tilde{\kappa}_l$.
In this paper, we simply consider the
window function given by $\tilde{W}(x)=2J_1(x)/x$, which corresponds to
a circle-shaped survey geometry with a radius of $\Theta_{\rm s}$.
For the halo model ingredients [$dn/dM, b(M)$ and the NFW profile], we
will throughout this paper employ the same models as used in
\cite{TakadaJain:09}.
Roughly speaking, the HSV term can be expressed as ${\rm Cov}_{\rm
HSV}^{PP}\sim P_{1h}(l_i)P_{1h}(l_j)\sigma^2_{m}(\Omega_{\rm s})$, where
$P_{1h}$ is the one-halo term of the lensing power spectrum and
$\sigma_{m}(\Omega_{\rm s})$ is the rms of the projected linear mass
density fluctuations smoothed with the angular scale of the survey area.
As implied from the above equation, the HSV affects the band
powers of different multipoles in the same way, and does not change the
shape of the power spectrum. This HSV term cannot be realized as long as
the discrete Fourier decomposition is used for deriving the power
spectrum covariance, because the large-scale modes outside the survey
region cannot be described by Fourier modes confined inside the survey
region.

Another important feature is that the HSV contribution depends on the survey
area via the integration of the linear mass power spectrum, $\int\!kdk
P_m^L(k) \left|\tilde{W}(k\chi\Theta_{\rm s})\right|^2$ unlike the other
terms which scale as $1/\Omega_{\rm s}$. For a power-law linear power
spectrum, $P^L_m(k)\propto k^n$, the HSV term is found to scale as ${\rm
Cov}^{PP}_{\rm HSV}\propto 1/(\Omega_{\rm s})^{1+n/2}$. Hence, for $n<0$,
which is indeed the case at $k\ge k_{\rm eq}$ ($k_{\rm eq}$ is the
wavenumber of the matter-radiation equality), the HSV amplitude decreases more
slowly with increasing area coverage than 
other terms.
On the other hand, when $n>0$ or $k\le k_{\rm eq}$, the HSV term
decreases more quickly. 
Thus the HSV term depends on the survey area in a non-trivial way, and
we will study the relative importance of the HSV term for different
survey areas.

There are even other sources of non-Gaussian errors arising from a
correlation of the lensing field in the weakly non-linear regime with the
mass density fluctuations of scales comparable with or larger than the
survey area.  This contribution can be formulated based on the
perturbation theory of structure formation, which is valid for the mass
density field in the weakly non-linear regime.  \cite{RimesHamilton:05}
first studied this effect for the 3D mass power spectrum, and named this
new contribution the beat-coupling mode \citep[also
see][]{Hamiltonetal:06,Sefusattietal:06,Takahashietal:09}. Furthermore,
\cite{dePutteretal:12} recently pointed out that the large-scale density
fluctuations cause an apparent modulation in the mean density estimated
from a finite survey region, and add an additional negative contribution to
the covariance. This term was shown to have a similar amplitude to the
beat-coupling mode. These large-scale mode contribution, which is
relevant for the weakly non-linear regime, differs from the HSV effect,
and 
\cite{TakadaJain:09} showed that the contribution to the lensing power
spectrum covariance is negligible compared to the non-Gaussian errors
arising from the four-point function (the second term in
equation~\ref{eq:pscov}) at multipoles of $l\ga $ a few $10^2$. Hence, in
this paper, we ignore the non-Gaussian errors arising from the
mode-coupling in the weakly non-linear regime for simplicity.

\subsection{Bispectrum covariance}
\label{sec:bcov}

As we discussed above, the lensing bispectrum is given as a function of
triangle configurations. An estimator of the lensing bispectrum
from the finite-area lensing survey can be found, by extending the
method developed in \cite{TakadaBridle:07} (see Appendix~\ref{app:bcov} for the
details), as
\begin{equation}
\hat{B}({l}_1,{l}_2,{l}_3)=
\frac{1}{\Omega_{\rm s}N_{\rm
trip}(l_1,l_2,l_3)}\sum_{\bm{q}_i}
\tilde{\kappa}_{\bm{q}_1}
\tilde{\kappa}_{\bm{q}_2}
\tilde{\kappa}_{\bm{q}_3}\Delta_{\bm{q}_{123}}(l_1,l_2,l_3),
\label{eq:bispest}
\end{equation}
where $\bm{q}_{123}\equiv \bm{q}_1+\bm{q}_2+\bm{q}_3$ and the summation
runs over all the pixels of $\bm{q}_1$, $\bm{q}_2$ and $\bm{q}_3$.
The function $\Delta_{\bm{q}_{123}}$ denotes the selection function
which is unity if each vector has a target length of
$l_i-\Delta l_i/2\le q_i\le l_i+\Delta l_i/2$ ($i=1,2,3$) and the three
vectors form the triangle configuration in Fourier space,
$\bm{q}_{123}=\bm{0}$; otherwise
$\Delta_{\bm{q}_{123}}(l_1,l_2,l_3)=0$. 
The prefactor
$1/\Omega_{\rm s}$ is from our definition of the discrete Fourier
decomposition \citep[see][]{TakadaBridle:07}.
The quantity $N_{\rm trip}$ is the number of
the triplets in Fourier space that form a given triangle configuration
specified by three side lengths $l_1,l_2,l_3$ with the bin widths.
This is calculated from the selection function as $N_{\rm
trip}=\sum_{\bm{q}_i}\Delta_{\bm{q}_{123}}(l_1,l_2,l_3)$. 
As described in \cite{Joachimietal:09} and Appendix~\ref{app:bcov}, 
for the limit of large multipole bins,
$l_1,l_2,l_3\gg l_f$, we can analytically estimate $N_{\rm trip}$ as
\begin{eqnarray}
N_{\rm
trip}(l_1,l_2,l_3)&\equiv &\sum_{\bm{q}_i; q_i\in l_i} \Delta_{\bm{q}_{123}}
\simeq \frac{\Omega^2_{\rm s}l_1l_2l_3\Delta l_1 \Delta l_2 \Delta
l_3}{2\upi^3 \sqrt{2l_1^2l_2^2+2l_1^2l_3^2+2l_2^2l_3^2-l_1^4-l_2^4-l_3^4}},
\label{eq:trip}
\end{eqnarray}
where $\Delta l_i$ is the bin width of the $i$th side length.
For the small-angle scales $l_i\gg 1$ (flat-sky
approximation limit) we are interested in, this equation gives a
good approximation to the Wigner- 3$j$ symbols which appears in the
bispectrum covariance derived under the full-sky approach \cite[see
equation 16 in][]{TakadaJain:04}.

The bispectrum covariance can be similarly defined as
\begin{eqnarray}
{\rm Cov}[B(\veclset),B(\vecldset)]&\equiv &
 \ave{\hat{B}(\veclset)\hat{B}(\vecldset)
}-B(l_1,l_2,l_3)B(l_1^\prime,l_2^\prime,l_3^\prime)\nonumber\\
&&\hspace{-13em}=\frac{1}{\Omega_{\rm s}^2N_{\rm trip}(l_1,l_2,l_3)
N_{\rm trip}(l_1^\prime,l_2^\prime,l_3^\prime)
}\sum_{\bm{q}_i; q_i\in l_i}
\sum_{\bm{q}_i^\prime; q_i\in l_i^\prime}
\left[
\ave{\tkappa_{\bm{q}_1}
\tkappa_{\bm{q}_2}
\tkappa_{\bm{q}_3}
\tkappa_{\bm{q}_1^\prime}
\tkappa_{\bm{q}_2^\prime}
\tkappa_{\bm{q}_3^\prime}
}\Delta_{\bm{q}_{123}}\Delta_{\bm{q}_{123}^\prime}\right]
-B(l_1,l_2,l_3)B(l_1^\prime,l_2^\prime,l_3^\prime).
\end{eqnarray}
The bispectrum covariance arises from the six-point correlation
function of the lensing field.

\begin{figure}
\centering
\includegraphics[width=0.5\textwidth]{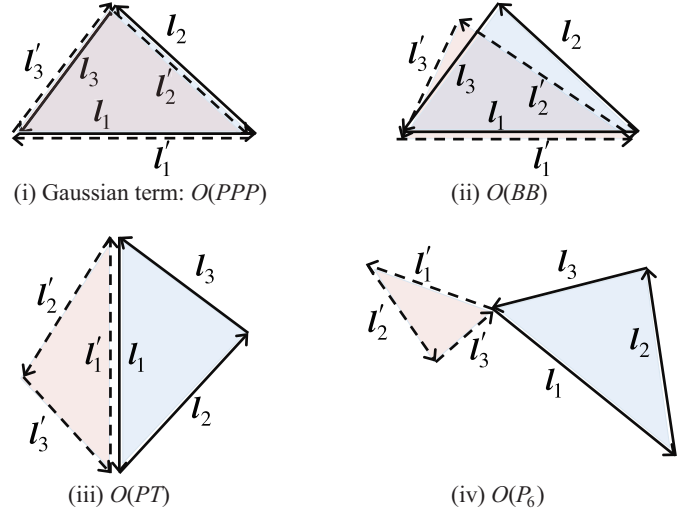}
\caption{
Illustration of the different terms of the bispectrum
 covariance.  The triangle configurations for two bispectra
in the covariance matrix,
$B(\veclset)$ and $B(\vecldset)$, 
are
 specified by sets of the three vectors $(\veclset)$ or $(\vecldset)$,
 denoted by the solid and dashed lines, respectively. The three vectors
 satisfy the triangle conditions $\bm{l}_1+\bm{l}_2+\bm{l}_3=\bm{0}$ and
 $\bm{l}_1'+\bm{l}_2'+\bm{l}_3'=\bm{0}$. 
(i) The Gaussian part of the bispectrum
 covariance, which arises only if the two triangle configurations have
 the same shape (within the coarseness of the bin widths).
 The two vectors with the same
 length but in the opposite direction such as $\bm{l}_1$ 
 and $\bm{l}_1^\prime$ yield the power spectrum after the ensemble
 average.
 Hence the Gaussian term amplitude is
 proportional to $P(l_1)P(l_2)P(l_3)$ and contributes to the diagonal
 terms of the covariance matrix.
(ii) Non-Gaussian part of the bispectrum covariance which arises if
 one side of the two triangles is in the same length and direction;
 here $\bm{l}_1=\bm{l}_1^\prime$ is shown as an example. The amplitude of this
 covariance term is proportional to $B(l_1,l_2,l_3)B(l'_1,l_2',l_3')$.
(iii) Non-Gaussian part that arises if one side length of
the two triangles is the same but in the opposite direction,
 $\bm{l}_1=-\bm{l}_1^\prime$. The amplitude is
 $O(PT)$; the figure shows a case of $P(l_1)
T(\bm{l}_1,\bm{l}_3,\bm{l}_2^\prime,\bm{l}_3^\prime)$.
The surrounding four vectors
($\bm{l}_2,\bm{l}_3,\bm{l}_2',\bm{l}_3'$ ) form a
quadrangular configuration satisfying the condition
$\bm{l}_2+\bm{l}_3+\bm{l}_2'+\bm{l}_3'=\bm{0}$, which gives the
 trispectrum contribution. 
(iv) Non-Gaussian part which arises for generic triangle configurations
 and therefore contributes to all diagonal and off-diagonal terms of
 the covariance matrix. The amplitude is proportional to the connected
 part of the six-point correlation function, $P_6$.
 As indicated, the 6 vectors of ($\bm{l}_1,\bm{l}_2,\bm{l}_3,\vecldset$)
 arise from the two triangles that form the six-point configuration in
 Fourier space, although 
the two vertices of the 6 points are collapsed to one
 point due to the triangle conditions. 
} \label{fig:bcov}
\end{figure}

We present the detailed derivation of the bispectrum covariance in
Appendix~\ref{app:bcov} based on the discrete Fourier decomposition
formulation.  Here we just give the expression of the bispectrum
covariance, which has three contributions, the Gaussian and
non-Gaussian errors and the HSV contribution:
\begin{eqnarray}
{\rm Cov}\left[B(l_1,l_2,l_3),B(l_1',l_2',l_3')\right]&=& 
{\rm Cov}_{\rm Gauss}
+
{\rm Cov}_{\rm NG}
+
{\rm Cov}^{BB}_{\rm HSV}\nonumber\\
&=&
\frac{\Omega_{\rm s}}{N_{\rm trip}(l_1,l_2,l_3)}
P(l_1)P(l_2)P(l_3)\left[
\delta^K_{l_1l_1^\prime}
\delta^K_{l_2l_2^\prime}
\delta^K_{l_3l_3^\prime}
+
\delta^K_{l_1l_1^\prime}
\delta^K_{l_2l_3^\prime}
\delta^K_{l_3l_2^\prime}
+
\delta^K_{l_1l_2^\prime}
\delta^K_{l_2l_1^\prime}
\delta^K_{l_3l_3^\prime}
+
\mbox{3 perms.}
\right]
\nonumber\\
&&+\frac{2\upi}{\Omega_{\rm s}}B(l_1,l_2,l_3)B(l_1',l_2',l_3')
\left[\frac{\delta^K_{l_1l_1'}}{l_1\Delta l_1}
+\frac{\delta^K_{l_1l_2'}}{{l_1\Delta l_1}}+\mbox{7 perms.}
\right]
\nonumber\\
&&+\delta^K_{l_1l_1'}\frac{2\upi}{\Omega_{\rm s}l_1\Delta l_1}
P(l_1)T(\bm{l}_2,\bm{l}_3,\bm{l}_2',\bm{l}_3')
+\delta^K_{l_1l_2'}\frac{2\upi}{\Omega_{\rm s}l_1\Delta l_1}
P(l_1)T(\bm{l}_2,\bm{l}_3,\bm{l}_1',\bm{l}_3')+\mbox{7 perms.}\nonumber\\
&&+ \frac{1}{\Omega_{\rm s}}
\int\!\!\frac{d\psi}{2\upi} ~
P_6(\bm{l}_1,\bm{l}_2,\bm{l}_3,\bm{l}_1',\bm{l}_2',\bm{l}_3'; \psi)
\nonumber \\
&&+ {\rm Cov}_{\rm HSV}^{BB},
\label{eq:bcov}
\end{eqnarray}
where the notation `NG' stands for the `non-Gaussian' error
contribution, and $P_6$ denotes the connected part of the six-point
correlation function.
Fig.~\ref{fig:bcov} shows a diagram picture of
these covariance terms from the first to the fourth lines on the
r.h.s. of equation~(\ref{eq:bcov}). When further including the intrinsic
shape noise contribution, we just replace the power spectra in the terms
of 
the above equation ($O(P^3)$ and $O(PT)$ terms)
with the power spectrum including the shot noise
contribution (equation~\ref{eq:ps_wshotn}).

The terms of the first line on the r.h.s.  are the Gaussian covariance
terms, which contribute only to the diagonal terms of the bispectrum
covariances.  The combination of the Kronecker deltas
$\delta^K_{l_1l_1'}\delta^K_{l_2l_2'}\delta^K_{l_3l_3'} $ is
non-vanishing only if the two triangle configurations are in the same
``shape'' within the bin widths. In particular, if triangle
configurations have symmetry such as isosceles or equilateral triangles,
the combination of Kronecker deltas (the terms in the square
bracket on the first line) yield a factor of 2 or 6 for isosceles and
equilateral triangles, respectively. 
The factors account for the
fact that different triangles transformed by parity and permutation
transformations ($l_i \leftrightarrow l_j^\prime$) are not independent
for a statistically homogeneous and isotropic field.
For a
general triangle configuration $l_i\ne l_j$, the factor becomes unity.
The prefactor $N_{\rm trip}(l_1,l_2,l_3)$ is given by
equation~(\ref{eq:trip}).

The terms from the second to the fourth lines
are the non-Gaussian error contributions, which arise
from the higher-order correlation functions of the lensing field.  The
coefficient of each term such as $2\upi/(\Omega_{\rm s}l_1\Delta l_1)$ is
from the number of independent configurations in Fourier space that
form a given configuration of six wavevectors ($\veclset,\vecldset$) in
Fig.~\ref{fig:bcov} (also see Appendix~\ref{app:bcov} for the mathematical
derivation). The angular integration in the fourth term including $P_6$ is over
the angle $\psi$ between the vectors $\bm{l}_1$ and $\bm{l}_1^\prime$
in order to include contributions over all the possible six-point
configurations in Fourier space. 
Note that the terms in the first, second and third lines
on the r.h.s. depend on the multipole bin widths such as $\Delta l_1$,
while the term including $P_6$ and the HSV do not depend on the bin
widths.

As in the power spectrum covariance (equation~\ref{eq:pscov}), the
number fluctuations of massive haloes due to the large-scale mass
fluctuations affect the bispectrum estimated from a finite area
survey. The HSV contribution to the
bispectrum covariance is given as
\begin{eqnarray}
{\rm Cov}[B_\kappa(\bm{l}_1,\bm{l}_2,\bm{l}_3),
B_\kappa(\bm{l}^\prime_1,\bm{l}^\prime_2,\bm{l}^\prime_3)]_{\rm HSV}
&=&\int\!d\chi\left(\frac{d^2V}{d\chi
    d\Omega} \right)^2
\left[
\int\!dM\frac{dn}{dM}b(M)
\tilde{\kappa}_M(l_1)
\tilde{\kappa}_M(l_2)
\tilde{\kappa}_M(l_3)
\right]
\nonumber\\
&&\times
\left[
\int\!dM'\frac{dn}{dM'}b(M')
\tilde{\kappa}_{M'}(l^\prime_1)
\tilde{\kappa}_{M'}(l^\prime_2)
\tilde{\kappa}_{M'}(l^\prime_3)
\right]
\int\frac{kdk}{2\upi}P_m^L\left(k;\chi\right)
\left|\tilde{W}(k\chi\Theta_s)\right|^2.
\label{eq:bcov_hsv}
\end{eqnarray}
This contribution has not been realized in previous works.  By
comparing with ray-tracing simulations, we will show below that the HSV
contribution is dominant over other covariance terms at $l\ga 1000$,
in the non-linear regime, and adding the HSV term to the model
predictions significantly improves agreement with the simulation
results. Again note that the HSV term scales with the survey area in a
non-trivial way via the linear mass power spectrum (see the discussion below
equation~\ref{eq:pscov_hsv}), while the other terms scale with survey area as
$1/\Omega_{\rm s}$.

\begin{figure}
\centering
\includegraphics[width=0.33\textwidth]{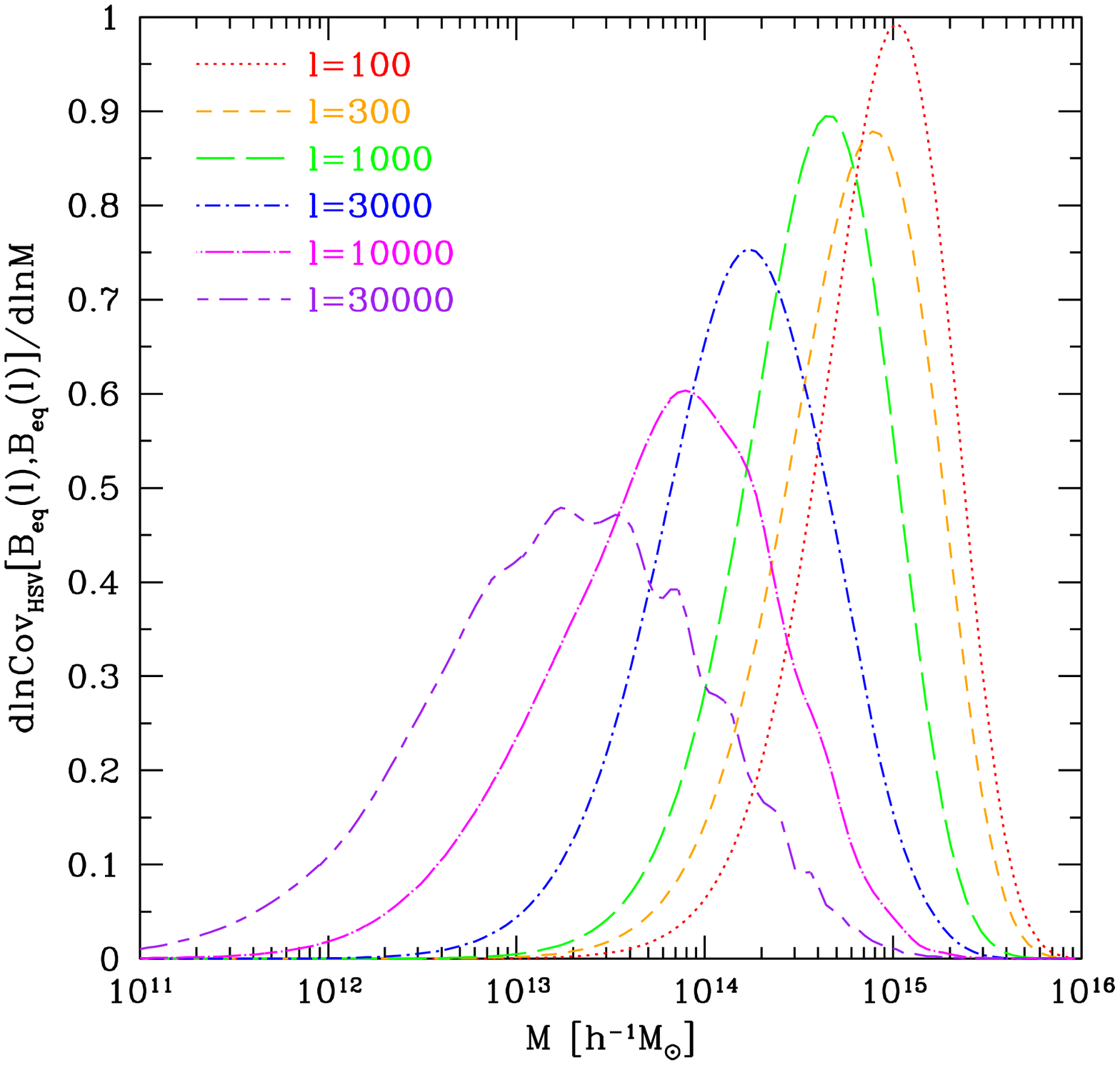}
\includegraphics[width=0.33\textwidth]{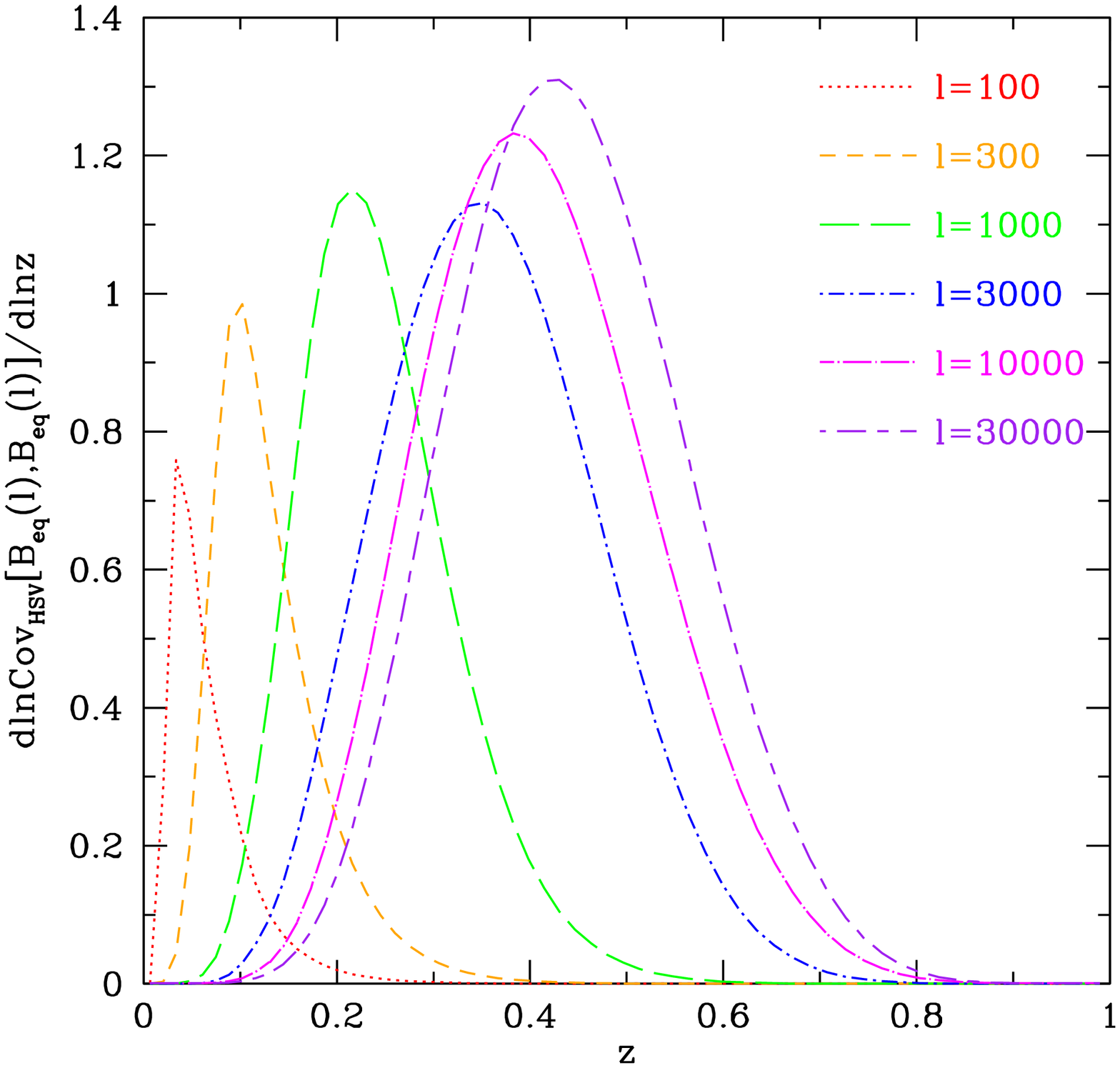}
\includegraphics[width=0.33\textwidth]{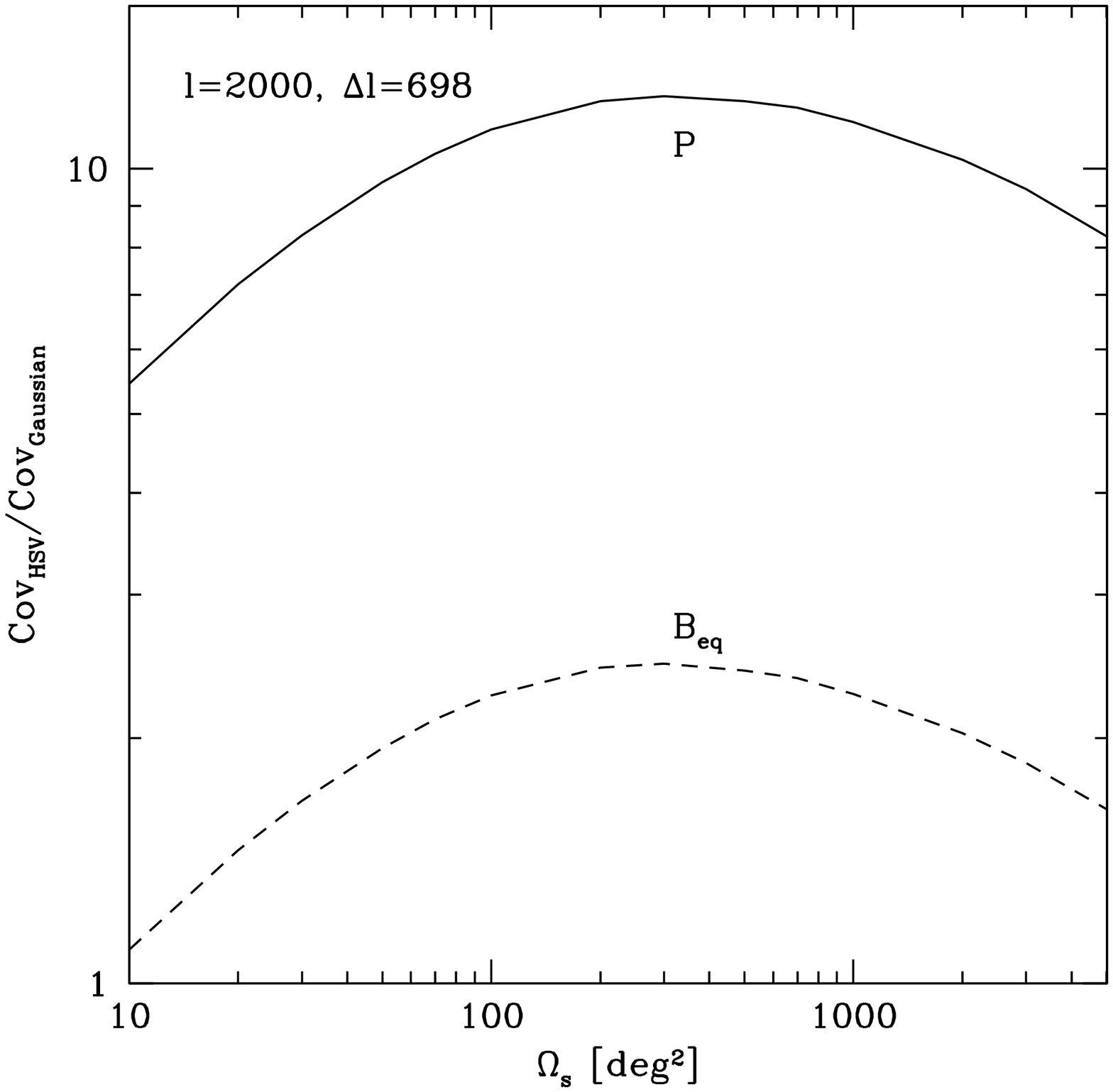}
\caption{Left-hand panel: the plot shows how haloes of different masses
 contribute to the halo sample variance (HSV) terms of the
 bispectrum covariance for a given equilateral triangle
 configuration. We used the halo model expression (equation~\ref{eq:bcov_hsv}) to
 compute the fractional contributions.  For the triangle with side
 length $l = 1000$, haloes with masses $\ga 10^{14}M_\odot$ 
 dominate the contribution to the HSV. Middle: Similar plot,
 but for the redshift distribution of the bispectrum covariance. For the
 triangle with $l=1000$, haloes at $z\la 0.4$ give the dominant
 contribution.  Right: The survey area dependence of the HSV
 contribution for the power spectrum and the bispectrum, relative to that
 of the Gaussian error, which scales as ${\rm Cov}_{\rm Gauss}\propto
 1/\Omega_{\rm s}$. Here we consider the scale of $l=2000$ for $P(l)$
 and $B_{\rm eq}(l)$, employ the flat sky approximation, and assume the
 survey geometry given by $\Omega_{\rm s}=\upi \Theta_{\rm s}^2$.
 }\label{fig:bcov_hsv_dlnM}
\end{figure}

The left-hand and middle panels of Fig.~\ref{fig:bcov_hsv_dlnM} show which
haloes of mass and redshift range contribute to the HSV term of the
bispectrum covariance for an equilateral triangle configuration of a
given side length. For the triangle with $l=1000$, haloes with masses
$\ga 10^{14}M_\odot$ and at redshift $z\la 0.4$ give a dominant
contribution to the HSV effect. These haloes are relatively easy to be
identified from a concentration of galaxies on the sky, X-ray or the
Sunyaev--Zel'dovich effect. In other words, identifying such massive haloes
in a survey region and comparing the number with the expected number for
a fiducial cosmological model will help understand the HSV effect for
the weak lensing observables in the survey region \citep[also see][for
the similar discussion]{TakadaBridle:07}. The right-hand panel shows how the
HSV terms scale with the survey area, in comparison with the
Gaussian covariance term for the power spectrum covariance and the
bispectrum covariance of equilateral triangles. Here we consider the
particular multipole bin $l=2000$, and
assume the flat-sky approximation and a circular survey geometry, $\Omega_{\rm
s}=\upi \Theta_{\rm s}^2$, for simplicity.
The plot shows that the HSV term decreases more slowly than other
terms as  $\Omega_{\rm s}$ increases up to a few 100 deg$^2$.

\subsection{Cross-covariance between power spectrum and bispectrum}
\label{sec:pbcov}

Since the power spectrum and bispectrum are not independent for the
non-Gaussian field, we need to account for the cross-covariance between
the two observables in order not to double-count the information
content. 
We can derive the cross-covariance (see Appendix~\ref{app:pbcov})
similarly to the power spectrum and the bispectrum covariance as
\begin{eqnarray}
{\rm Cov}\left[P^{\rm est}(l),B^{\rm est}(l_1,l_2,l_3)\right]
&=&{\rm Cov}^{PB}_{NG} + {\rm Cov}^{PB}_{\rm HSV}\nonumber\\
&=&\delta^K_{ll_1}\frac{4\upi}{\Omega_{\rm
 s}l_1\Delta l_1}P(l)B(l,l_2,l_3) + \mbox{2 perms.}
+\frac{1}{\Omega_{\rm s}}\int\!\frac{d\psi}{2\upi} ~
P_5(\bm{l},-\bm{l},\bm{l}_1,\bm{l}_2,\bm{l}_3; \psi)
+ {\rm Cov}^{PB}_{\rm HSV},
\label{eq:pbcov}
\end{eqnarray}
where $P_5$ is the five-point correlation function, and $\psi$ is the
angle between the vectors $\bm{l}$ and $\bm{l}_1$ as in
equation~(\ref{eq:bcov}). The HSV term is given as
\begin{eqnarray}
{\rm Cov}[P_\kappa(l),
B_\kappa(\bm{l}_1,\bm{l}_2,\bm{l}_3),
]_{\rm HSV}
&=&\int\!d\chi\left(\frac{d^2V}{d\chi
    d\Omega} \right)^2
\left[
\int\!dm\frac{dn}{dM}b(M)
\left|\tilde{\kappa}_M(l)\right|^2
\right]
\nonumber\\
&&\times
\left[
\int\!dM\frac{dn}{dM'}b(M')
\tilde{\kappa}_{M'}(l_1)
\tilde{\kappa}_{M'}(l_2)
\tilde{\kappa}_{M'}(l_3)
\right]
\int\frac{kdk}{2\upi}P_m^L\left(k;\chi\right)
\left|\tilde{W}(k\chi\Theta_s)\right|^2.
\label{eq:pbcov_hsv}
\end{eqnarray}
We will use these equations when computing the total information content
for a combined measurement of the lensing power spectrum and bispectrum
for a given survey.  

\subsection{Halo sample variance contribution to the $n$-point
  correlation function measurement} 
\label{sec:hsv}

As we have seen for the HSV contributions to the lensing spectrum
covariances (equations~\ref{eq:pscov_hsv}, \ref{eq:bcov_hsv} and
\ref{eq:pbcov_hsv}), the equations have similar forms at the
small-angle limit, where the one-halo term is dominated in the halo model
picture. Extending these findings, we can find the HSV contribution to
the covariance matrix between any $n$- and $n'$-point correlation
functions in Fourier space:
\begin{eqnarray}
{\rm Cov}[P_n(\bm{l}_1,\bm{l}_2,\dots,\bm{l}_n),
P_{n'}(\bm{l}_1^\prime,\bm{l}_2^\prime,\dots,\bm{l}_{n'}^\prime)]_{\rm HSV}&=&\int\!d\chi\left(\frac{d^2V}{d\chi
   d\Omega} \right)^2
\left[
\int\!dM\frac{dn}{dM}b(M)
\tilde{\kappa}_M(l_1)
\tilde{\kappa}_M(l_2)
\cdots
\tilde{\kappa}_M(l_n)
\right]\nonumber\\
&&\hspace{-4em}\times \left[
\int\!dM\frac{dn}{dM'}b(M')
\tilde{\kappa}_{M'}(l_1^\prime)
\tilde{\kappa}_{M'}(l_2^\prime)
\cdots
\tilde{\kappa}_{M'}(l_{n'}^\prime)
\right]\int\frac{kdk}{2\upi}P_m^L\left(k;\chi\right)
\left|\tilde{W}(k\chi\Theta_s)\right|^2.
\label{eq:pmpn_hsv}
\end{eqnarray}
Roughly speaking, the amplitude of the HSV term simply scales as ${\rm
Cov}_{\rm HSV}\sim P^{\rm 1h}_n(l_i)P^{\rm
1h}_{n'}(l_j^\prime)\sigma_m^2(\Theta_s)$. Thus any $n$-point correlation
functions at small angle scales can be affected by the large-scale mass
fluctuations of scales 
comparable with or outside the survey area.

We should also emphasize that the HSV contribution affects
measurements of any two- or three-dimensional correlation functions
from a finite area survey, and can be very important if one is
interested in the small-scale signals which are sensitive to the
abundance of haloes in the finite survey region \citep[e.g. see][for a
similar discussion on the SZ power spectrum
measurement]{Shawetal:09,ZhangSheth:07}. 

\subsection{Halo model predictions for the lensing covariances}
\label{sec:halomodel}

As we have described up to the preceding section, the power spectrum and
bispectrum covariance calculations require to compute the four-, five- and
six-point correlation functions in addition to the power spectrum and
bispectrum.
For the power spectrum and bispectrum, some theoretical models are proposed
by comparing with simulations,
e.g. \cite{Smithetal:03} and \cite{ValageasNishimichi:11a} for the power
spectrum and \cite{ScoccimarroFrieman:99} and \cite{ValageasNishimichi:11b} for
the bispectrum. The higher-order functions are, however, fairly uncertain
because there are fewer studies to compare the model predictions with
simulations \citep[see e.g.][for an attempt to compute the kurtosis which is
the collapsed four-point function]{TakadaJain:02}, partly because the
higher-order correlations require a substantial amount of computational
costs.  Instead of pursuing a reliable model for the higher-order
correlation functions, in this paper we employ the halo model approach
to compute the higher-order functions
\citep{PeacockSmith:00,Seljak:00,MaFry:00,Scoccimarroetal:01,CooraySheth:02}
in which the correlations of the mass distribution are expressed as two
separate contributions: correlations of dark matter particles within the
same halo and correlations between particles in different haloes.  We
have found that, up to the four-point correlation functions, the halo model
predictions are accurate at 10--30 per cent level in the amplitude compared to N-body
simulations \citep[in particular for lensing fields; ]
[]{TakadaJain:02,TakadaJain:03a,TakadaJain:03}.  Since our purpose of
this paper is to assess the importance of  non-Gaussian error
contributions to the bispectrum covariance matrices, we consider that
the halo model approach is adequate enough. 

We know that most of the lensing information comes from small angle scales
in the non-linear clustering regime 
to which the one-halo term, the correlation arising from the same  halo,
provides a dominant contribution.  In addition, the non-Gaussian errors
are important only at the small scales, as can be explicitly found from
fig.~5 in \cite{Satoetal:09}.  For these reasons, we include only the
one-halo terms to compute the non-Gaussian error contributions to the
lensing covariances, which significantly simplifies the computation.
Although the $n$-point correlation function depends on $n$ wavevectors
such as $P_n(\bm{l}_1,\bm{l}_2,\dots,\bm{l}_n)$, the one-halo term does
not depend on any angle between the vectors, but rather depends only on
the length of each vector; $P_n^{\rm 1h}(\bm{l}_1,\bm{l}_2,
\dots,\bm{l}_n)=P_n^{\rm 1h}(l_1,l_2,\dots,l_n)$, reflecting spherical
mass distribution around halo in a statistical average sense.  To be
more explicit, assuming the Limber's approximation, the one-halo term of
the $n$-point correlation function can be computed as follows \citep[see
around equation~30 in][]{TakadaJain:03a}:
\begin{equation}
P^{\rm 1h}_n(l_1,l_2,\dots,l_n)=
\int_0^{\chi_s}\!\!d\chi \frac{d^2V}{d\chi d\Omega}
\int\!dM\frac{dn}{dM}\tkappa_M(l_1)\tkappa_M(l_2)\cdots\tkappa_M(l_n).
\end{equation}
The one-halo term is computed by a two-dimensional
integration, as the Fourier transform of the lensing field due to an NFW
halo, $\tkappa_M(l)$, can be computed analytically for a given halo with
mass $M$ and at redshift $z$.

As can be found from equations~(\ref{eq:pscov}), (\ref{eq:bcov}) and
(\ref{eq:pbcov}), some of the non-Gaussian terms can be further
simplified; e.g. one of the non-Gaussian terms in the power spectrum
covariance (equation~\ref{eq:pscov}) can be simplified as
\begin{equation}
\frac{1}{\Omega_{\rm
s}}\int_{\left|\bm{l}\right|\in l_i}\!\frac{d^2\bm{l}}{A(l_i)}
\int_{\left|\bm{l}'\right|\in
l_j}\!\frac{d^2\bm{l}'}{A(l_j)}T(\bm{l},-\bm{l},\bm{l}',-\bm{l}')
\simeq \frac{1}{\Omega_{s}}T^{\rm 1h}(l_i,l_i,l_j,l_j),
\end{equation}
where we have assumed that the lensing trispectrum does not largely
change within the multipole bin $\Delta l$ around the bins $l_i$ and
$l_j$. Thus the above approach allows a faster computation of 
the power spectrum and bispectrum covariances.

\section{Results: Comparison with ray-tracing simulations}
\label{sec:results}

\subsection{Ray-tracing simulations}

To study the lensing covariance matrices, we use 1000 realizations of
ray-tracing simulations  for a $\Lambda$CDM
model in \cite{Satoetal:09}. Although the simulations were done for
various source redshifts ranging from $z_s=0.6$ to 3, we use the outputs
of $z_s=1$ in this paper.  In brief, each realization has an area of
$5\times 5$ deg$^2$ ($\Omega_{\rm s}=0.0076$ sr) in 
square shaped geometry. 
The $\Lambda$CDM model adopted is characterized by 
cosmological parameters: the matter density parameter $\Omega_{\rm m}=0.238$,
the baryon density parameter $\Omega_{\rm b}=0.042$, the initial spectral
index $n_s=0.958$, the amplitude of the density fluctuations
$\sigma_8=0.76$ and the Hubble constant of $H_0=100h~$km
s$^{-1}$Mpc$^{-1}$ with $h=0.732$. The linear matter power spectrum used
to set the initial conditions of N-body simulations is computed by the
public code
CAMB \citep{Lewisetal:00}.
It was shown that the ray-tracing simulations
are reliable to within a 5 per cent accuracy up to multipole $l \simeq
6000$ for the power spectrum and up to $l \simeq 4000$ for the
bispectrum. See
\cite{Satoetal:09} and \cite{ValageasSatoNishimichi:12} for more details.\footnote{The simulation data is available at
\url{http://www.a.phys.nagoya-u.ac.jp/~masanori/HSC/}}

As can be found from fig.~1 in \cite{Satoetal:09}, the ray-tracing
simulations were done in a light cone of area $5\times 5$ deg$^2$, viewed from an observer position ($z=0$). The
projected mass density fields in intermediate-redshift slices were
generated from N-body simulations which have a larger simulation box
than the volume covered by the light cone. Hence the lensing fields have
contributions from the mass density field of scales outside the
ray-tracing simulation area, although, exactly speaking, the modes
outside the N-body simulation box were not included (see
Appendix~\ref{app:volume} for the effect). Thus the
ray-tracing simulations allow us to study the HSV effect on the
covariance matrices.

\subsection{Measuring power spectrum, 
bispectrum and  the covariance matrices from simulations}
\label{app:blsim}

In each simulation realization, the lensing convergence field,
$\kappa(\bm{\theta})$, is given on $2048\times 2048$ grids.  We used the
FFT method to compute the Fourier transformed field, $\tkappa(\bm{l})$.
The
fundamental mode of the discrete Fourier decomposition is $l_f =72
(=2\upi/5^\circ)$ and the Nyquist frequency is $\sim70000$, which is
large enough compared to the resolution limit of the N-body simulations.
We use multipole bins that are logarithmically spaced by $\Delta \ln
l=0.3~ (\Delta \log l\simeq 0.13)$, which significantly reduces the
number of triangle configurations (the number of different bispectra) we
need to consider, compared to $\Delta l=1$ as in the CMB case.
We consider 16 multipole bins in total; the first bin is in the range
$l=[72,97.2]$, and the 16th bin is in the range $l=[6481.2, 8748.8]$, so
$l_{\rm min}=72$ and $l_{\rm max}=8748.8$.

The power spectrum of the $i$th multipole bin $l_i$, $P(l_i)$, is
estimated from each realization by computing an azimuthal average of the
estimator, $\tkappa_{\bm{l}}\tkappa_{-\bm{l}}$, where $|\bm{l}|$ resides
in the target bin. We then averaged the estimated power spectra in 1000
realizations to estimate the ensemble-averaged power spectrum
(corresponding to the power spectrum for the area of $1000\times
25=25000$ deg$^2$). Then we computed the scatters among the power
spectra of 1000 realizations in order to estimate the covariance matrix
of the power spectra.  Hence the covariance matrix is for an area of 25
deg$^2$.

The bispectrum
is given as a function of triangle configuration. We use three side
lengths $(l_1,l_2,l_3)$ to parametrize triangle configuration, where the
triangle conditions are given as $|l_j-l_k|\le l_i \le l_j+l_k$.
Although the multipole bin used has a logarithmically spaced bin width,
we impose the triangle conditions on the central values of the multipole
bins. In addition we impose the condition $l_1\le l_2\le l_3$ so that
every triangle configuration is counted once.  For the 16 multipole bins
above, we have 204 triangle configurations in total.

\begin{figure}
\centering
\includegraphics[width=0.5\textwidth]{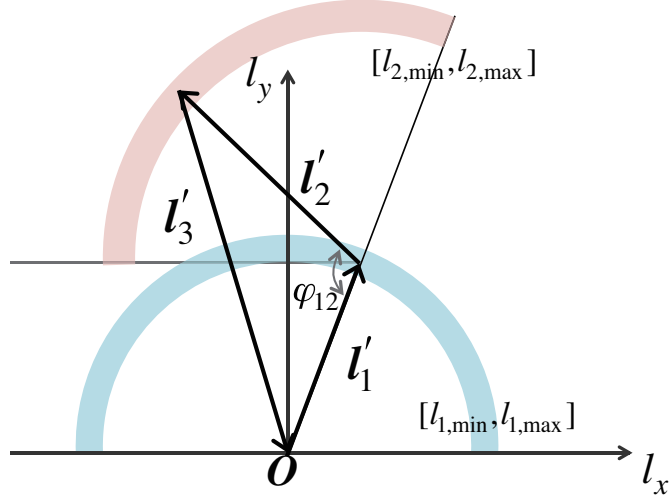}
\vspace{0em}
\caption{Illustration describing how to choose three vectors in
 Fourier space that satisfy triangle configurations within multipole
 bin widths. The triangle configuration is specified by three side
 lengths which have central values ($l_1,l_2,l_3$) and the widths
 $\Delta\ln l=0.3$.  The three vertices chosen are used in estimating the
 bispectrum from ray-tracing simulations by averaging the estimator
 ${\rm
 Re}[\tkappa_{\bm{l}_1^\prime}\tkappa_{\bm{l}_2^\prime}\tkappa_{\bm{l}_3^\prime}]
 $ over the triangles (see text for details).  
 First, the first vector $\bm{l}_1^\prime$ is chosen from the
 annulus (shaded region), which is in the upper half plane and has the
 radius in the range $[l_{1,\rm min},l_{1,\rm max}]$ (a bin of
 $l_1$). Then, the second vector $\bm{l}_2^\prime$ is chosen from the
 annulus which has position angle in the range $\arg(\bm{l}_1^\prime)\le
 \arg(\bm{l}_2^\prime) \le \upi$ and has radius in the range $[l_{2,\rm
 min},l_{2,\rm max}]$.  For the given pair of $\bm{l}_1^\prime$ 
and $\bm{l}_2^\prime$, the third vector $\bm{l}_3^\prime$ is determined
 by the triangle condition
 $\bm{l}_3^\prime=-\bm{l}_1^\prime-\bm{l}_2^\prime$; if $l_3^\prime$ is
 in the range $l_{\rm min}\le l_3^\prime\le l_{\rm max}$, the triplet
 of ($\vecldset$) is accepted and otherwise discarded.
The angle
 $\varphi_{12}$ is used for discussion in Appendix~\ref{app:bcov}. 
} \label{fig:triang}
\end{figure}

For a triangle configuration that is specified by the side lengths
$(l_1,l_2,l_3)$ (with the bin widths), we can estimate the bispectrum
from each ray-tracing simulation by averaging the estimator ${\rm
Re}[\tkappa_{\bm{l}_1^\prime} \tkappa_{\bm{l}_2^\prime}
\tkappa_{\bm{l}_3^\prime}]$ over all the triplets ($\vecldset$) which
satisfy the triangle conditions; the length of each vector is in the
triangle bin such as $l_{1,\rm min}\le l_1^\prime \le l_{1,\rm
max}$. Note that, as long as the triplets of $(\vecldset)$ in Fourier
space have the same side lengths $l_1,l_2,l_3$ within the bin widths,
all the triangles transformed by the rotation and
parity transform are equivalent to yield the same bispectrum due to
the rotation and parity invariance for a statistically
homogeneous and isotropic field.  In our simulations, the
Fourier transformed convergence field, $\tkappa(\bm{l})$, is given on
$2048\times 2048$ grids in Fourier space, where the grids are linearly
spaced by the fundamental mode, $l_f=2\upi/\Theta_{\rm s}=72$.
To have an efficient computation over 1000 realizations, we first built
the table of three vectors (grids), $(\vecldset)$, where each triplet
satisfies the triangle conditions
($|\bm{l}^\prime_j-\bm{l}^\prime_k|\le l_i^\prime \le
|\bm{l}^\prime_j+\bm{l}^\prime_k|$) and is assigned to one of the
triangle configurations binned by three side lengths
($l_1,l_2,l_3$). Then we used the {\em same} table of triplets for
the 1000 realizations to compute the average and scatters of the
estimated bispectra as a function of triangle configurations.

To be more precise, we built the table of three vectors (grids) in the
way illustrated in Fig.~\ref{fig:triang}.  First, we choose the first
vector $\bm{l}_1^\prime$ from one of the grids in the upper half of
Fourier space, i.e. $0\le \arg(\bm{l}_1^\prime)<\upi$, by imposing the
condition that the length $l_1^\prime $ is in the range of the multipole
bin, $l_{\rm min}\le l_1^\prime < l_{\rm max}$.  Then we survey for the
second vector $\bm{l}_{2}^\prime$ from the region where the length is
$l_{\rm min}\le l_2^\prime < l_{\rm max}$ and the position angle
satisfies $\arg(\bm{l}_1^\prime)\le \arg(\bm{l}_2^\prime)< \upi$ ({more
precisely, $l_1'\le l_2'$ if and only if
$\arg(\bm{l}_1')=\arg(\bm{l}_2')$}). For a given pairs of
$\bm{l}_1^\prime$ and $\bm{l}_2^\prime$, we choose the third vector
$\bm{l}_3^\prime $ from the condition $\bm{l}_3^\prime
=-\bm{l}_1^\prime-\bm{l}_2^\prime $ and then accept the triplet of
$(\vecldset)$ (otherwise discard it) if the length $l_3^\prime$ is in
the range satisfies the condition $l_{\rm min}\le l_3^\prime < l_{\rm
max}$. 
Then we assign each set of three vectors,
$(\vecldset)$, to one of the triangle configuration bins labelled by
$(l_1,l_2,l_3)$ {by sorting $l_1'$, $l_2'$ and $l_3'$ in
the ascending order.} For any of the sets of three vectors chosen in
this way, two of the three vectors are in the upper half plane of
Fourier space, while the remaining one is in the lower half plane.
Hence, we miss triangles with configurations for which
two vectors are in the lower plane and the other in the upper.  
However, we can recover
these triangles by just flipping the signs of all three
vectors and obtain the same value of bispectrum, due to the symmetry
$\tkappa_{\bm{l}}=\tkappa_{-\bm{l}}^*$, which comes from the real
condition of the lensing field. Thus, we do not need to count the latter cases,
but just twice the number of actually counted triangles to obtain
$N_{\rm trip}$, the number of independent triplets.
For some of the following results, we will use the  measured $N_{\rm
trip}$ when computing the Gaussian error contributions to the
bispectrum covariance (the first term in equation~\ref{eq:bcov}). Although
equation~(\ref{eq:trip}) gives a good approximation to $N_{\rm trip}$ for the
limit of $l_i\gg 1$, the $N_{\rm trip}$ directly estimated from the
simulation properly takes into account the effect of discrete grids in
the lensing map.

Using the table of three vectors ($\vecldset$) obtained in the process
above, we estimate the bispectrum by averaging
Re$[\tildel_{\bm{l}_1'}\tildel_{\bm{l}_2'}\tildel_{\bm{l}_3'}]$ from
each realization.  Note that the value of $\tkappa_{\bm{l}^\prime_i}$ is
taken from the field at the grid that has two coordinate components
($l^\prime_{ix},l^\prime_{iy}$), not from the field at the grid denoted
by the arrow of the vector $\bm{l}^\prime_i$ in Fig.~\ref{fig:triang}.
Although we take the real part of
$\tildel_{\bm{l}_1'}\tildel_{\bm{l}_2'}\tildel_{\bm{l}_3'}$ for the
average, this is not essential, because the estimated bispectrum
satisfies the real condition to a very good approximation after the
average over many triangles. 

The dimensions of the resulting covariance matrices
are: $16\times 16$, $204\times 204$ and $16\times 204$ (or $204\times
16$) 
for the power
spectrum covariance, the bispectrum covariance and the cross-covariance,
respectively. The 1000 realizations are enough to compute the
204$\times$204 elements of the covariance matrix \citep[see appendix
of][]{Takahashietal:11}, although a larger number of the realizations
are ideally needed for a more accurate estimate of the covariance
matrix. 
The situation will be worse in a case where more triangle
configurations are considered, e.g. as in the case of lensing tomography
where different redshift slices are further needed to include
\citep{TakadaJain:04}. Hence an analytical approach to compute the
covariance matrices is useful.

In fact, \cite{Hartlapetal:07} pointed out that, by
assuming a multivariate Gaussian distribution for a statistical
variable, the number of realizations used to estimate the covariance is
important.  They showed that the inverse covariance matrix can be biased
if the covariance matrix is estimated from a finite number of the
realizations.  In our case, we use 1000 realizations to estimate the
bispectrum covariance matrix for 204 triangle configurations (the
dimension of the bispectrum covariance is $204\times 204$), which may
result in an overestimate of 10 per cent for the signal-to-noise ratio (S/N) for the
bispectrum. However, we will show below the
simulation results without any correction because 
the bias is not large comparing to the
accuracy of our concern or we do not know whether the bispectrum estimators
obey the multi-variate Gaussian distribution.

\subsection{Comparison of the simulation results and the halo model predictions}

\begin{figure}
\centering
\includegraphics[width=0.5\textwidth]{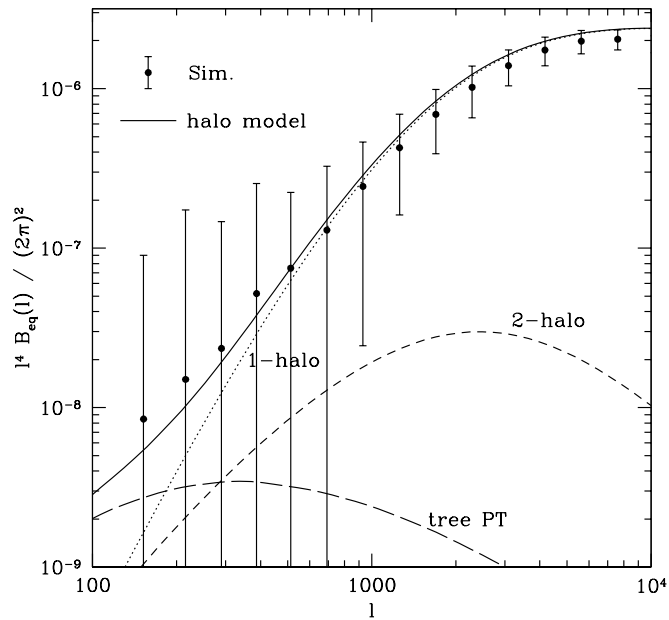}
\caption{The
lensing bispectrum 
for equilateral triangle configuration as a function of the side length
 $l$, where the multipole bins are
 logarithmically spaced by $\Delta \ln l=0.3$.
The data points with error bars are the bispectrum measured from the
 1000 ray-tracing simulations with the source
 redshift $z_s=1$. Each simulation covers an area of $25$ deg$^2$.
 The error bars show the scatters of the 1000 realizations, which
 therefore correspond to the measurement errors expected for the area of
 25 deg$^2$.  The solid curve shows the halo model prediction, while the
 dotted, short-dashed and long-dashed curves show the one-, two- and three-halo
 term contributions to the bispectrum. For the three-halo term, we used the
 tree-level perturbation theory prediction. 
Note that the bispectra for $l \ga 4000$ may be affected by 
the resolution limit of the ray-tracing simulations.
}
 \label{fig:bleq}
\end{figure}
Fig.~\ref{fig:bleq} plots an example of the measured bispectrum ({\it
points} with error bars) and the halo model prediction ({\it lines}) for
equilateral triangle configurations. The halo model agrees fairly well
with the simulation results, although it underestimates the bispectrum
amplitude around $l\sim$ a few 100 and overestimates at $l\ga 1000$.
An improvement of the halo model accuracy may be available by refining
the halo model calculation, as performed in
\cite{ValageasSatoNishimichi:12}. However, we do not
pursue this possibility in this paper.

\begin{figure}
\centering
\includegraphics[width=0.5\textwidth]{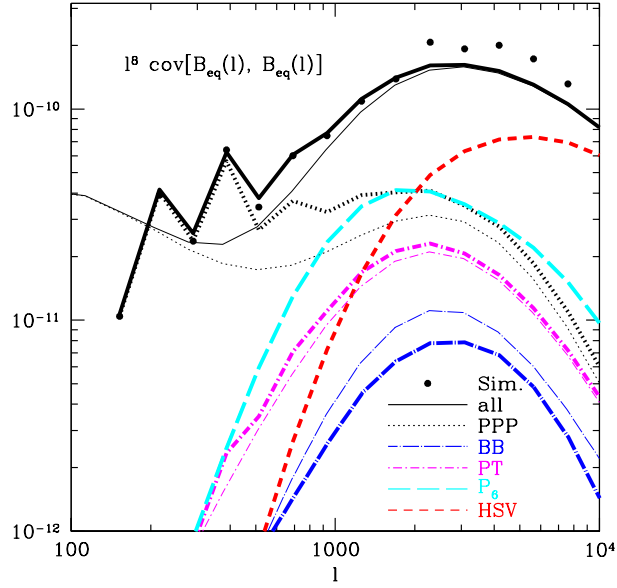}
\caption{Diagonal terms of the bispectrum covariance matrix for
 equilateral triangle configurations as a function of the side length
 $l$, as in the previous figure.  The covariance amplitude is shown in
 the unit of $l^6{\rm Cov}[B_{\rm eq}(l), B_{\rm eq}(l)]$, since $l^3
 B_{\rm eq}(l)$ gives the contribution to the skewness $\ave{\kappa^3}$.
 The points are the simulation results estimated from the scatters of
 1000 simulations. The other curves are the halo model predictions,
 which are computed based on the method described in
 Section~\ref{sec:bcov}. The dotted, long dot--dashed, short dot--dashed, and
 long dashed curves are the contributions that are proportional to $P^3,
 B^2, PT$ and $P_6$, respectively (also see Fig.~\ref{fig:bcov}).
For thick curves, we used the power spectra $P(l)$, the
 bispectra $B(l)$ and the number of triangles directly estimated from
 the simulations.  For comparison, the thin curves show the results if
 we use the halo model for $P(l)$ and $B(l)$ as well as
 equation~(\ref{eq:trip}) for the number of triangles. 
The short dashed
 curve is the HSV contribution, which dominates over other terms at
 multipole bins, $l \ga 2000$. The solid curves are the total
 contribution, the sum of all the terms. } \label{fig:bcov_eq}
\end{figure}

\begin{figure}
\centering
\includegraphics[width=0.46\textwidth]{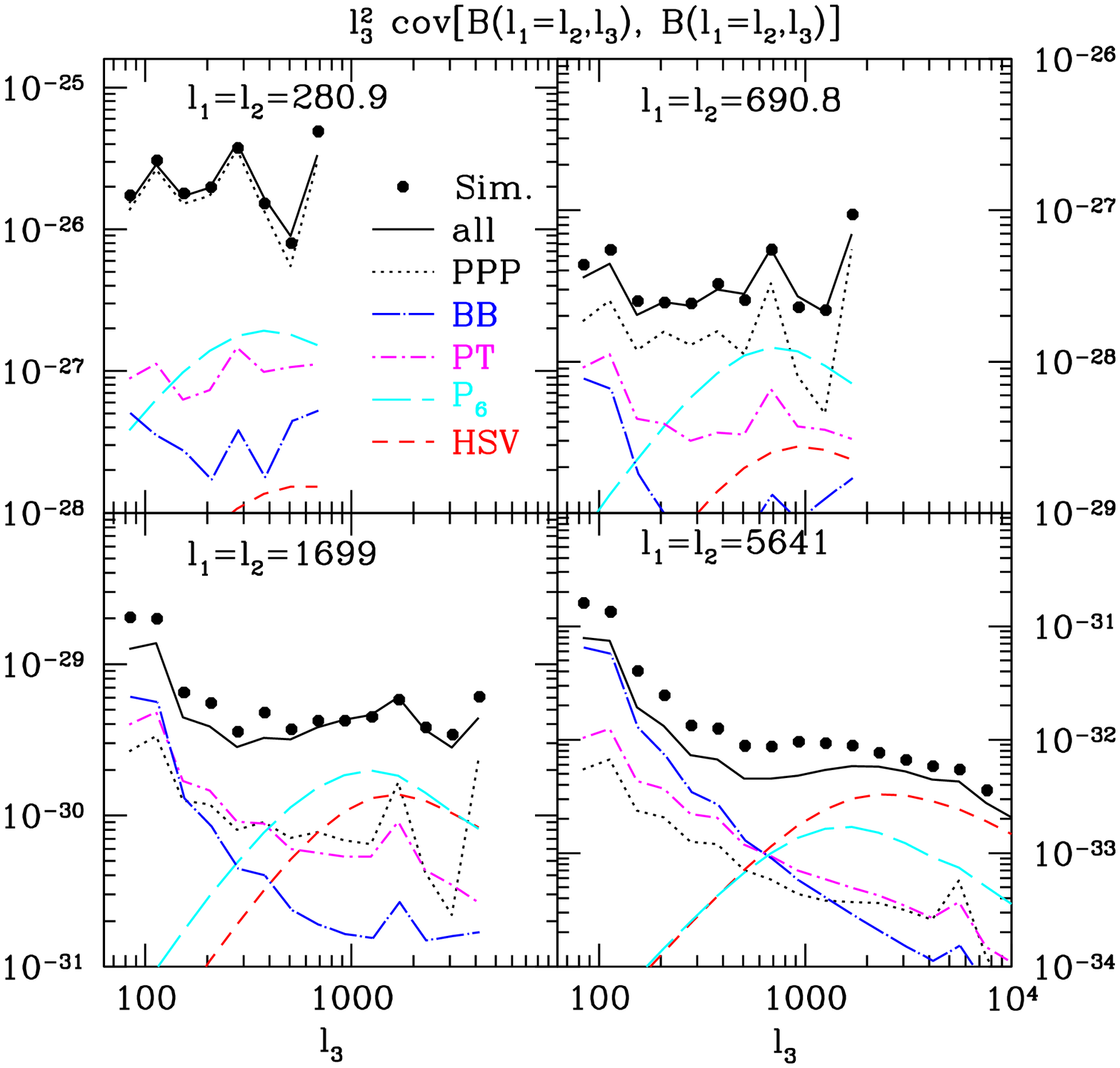}
\includegraphics[width=0.46\textwidth]{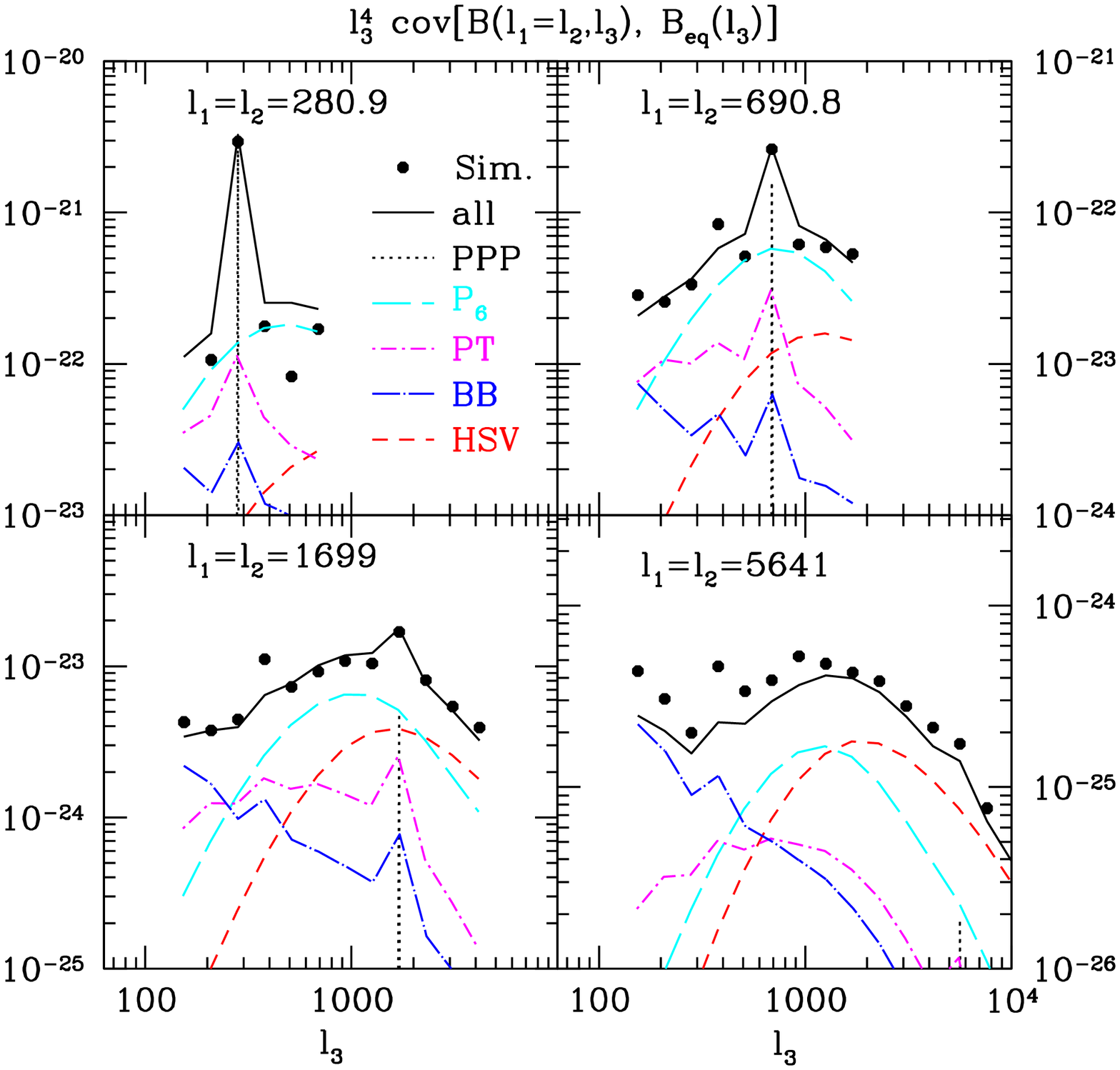}
\caption{Similar to the previous plot, but for different triangle
 configurations. Left-hand panels: the covariance matrix for isosceles
 triangle configurations, $B(l_1, l_2, l_3)$, with $l_1=l_2$. The
 different panels are for different side lengths $l_1$ and $l_2$; within
 each panel, the covariance matrix is shown as a function of $l_3$. 
 Note that
 the covariance matrix is shown in the multipole range where the
 triangle conditions $|l_j-l_k|\le l_i\le l_j+l_k$ are satisfied, but
 the condition $l_1\le l_2\le l_3$ is not imposed. 
 Right-hand panels: the covariance matrix elements between the bispectra of
 isosceles and equilateral triangles. 
The Gaussian terms of $P^3$ denoted by the vertical dotted lines appear at a
 particular value of $l_3$, where the two triangles become the same,
 i.e. equilateral triangles with $l_1=l_2=l_3$. 
}  \label{fig:bcov_iso}
\end{figure}

\begin{figure}
\centering
\includegraphics[width=0.5\textwidth]{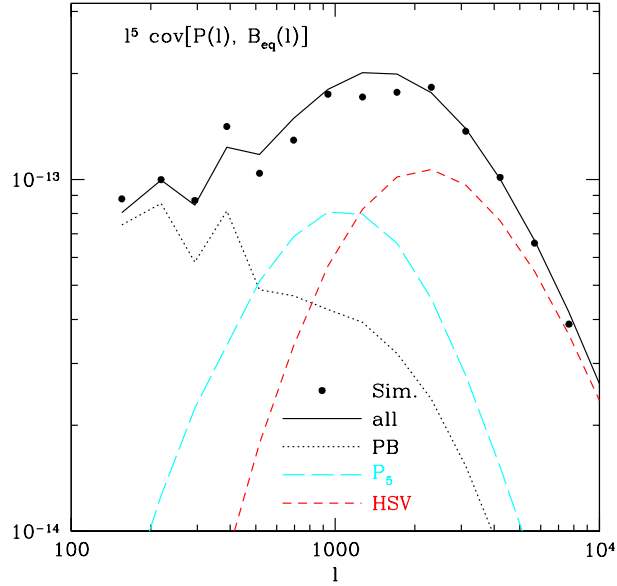}
\caption{
Cross-covariance between the power spectrum and the
 bispectrum of equilateral triangles, ${\rm Cov}[P(l),B_{\rm eq}(l)]$,
 as a function of $l$. The covariance amplitude is plotted in the
 units of $l^5{\rm Cov}[P,B_{\rm eq}]$, because $l^2P$ and $l^3B$
 contribute to 
$\ave{\kappa^2}$ and $\ave{\kappa^3}$, respectively. Note that there
 is no Gaussian error contribution to the cross-covariance, because it
 arises from the five-point correlation functions. The different curves are
 the model predictions that are computed based on the method  in
 Section~\ref{sec:pbcov}. } \label{fig:pbcov}
\end{figure}

\begin{figure}
\centering
\includegraphics[width=0.9\textwidth]{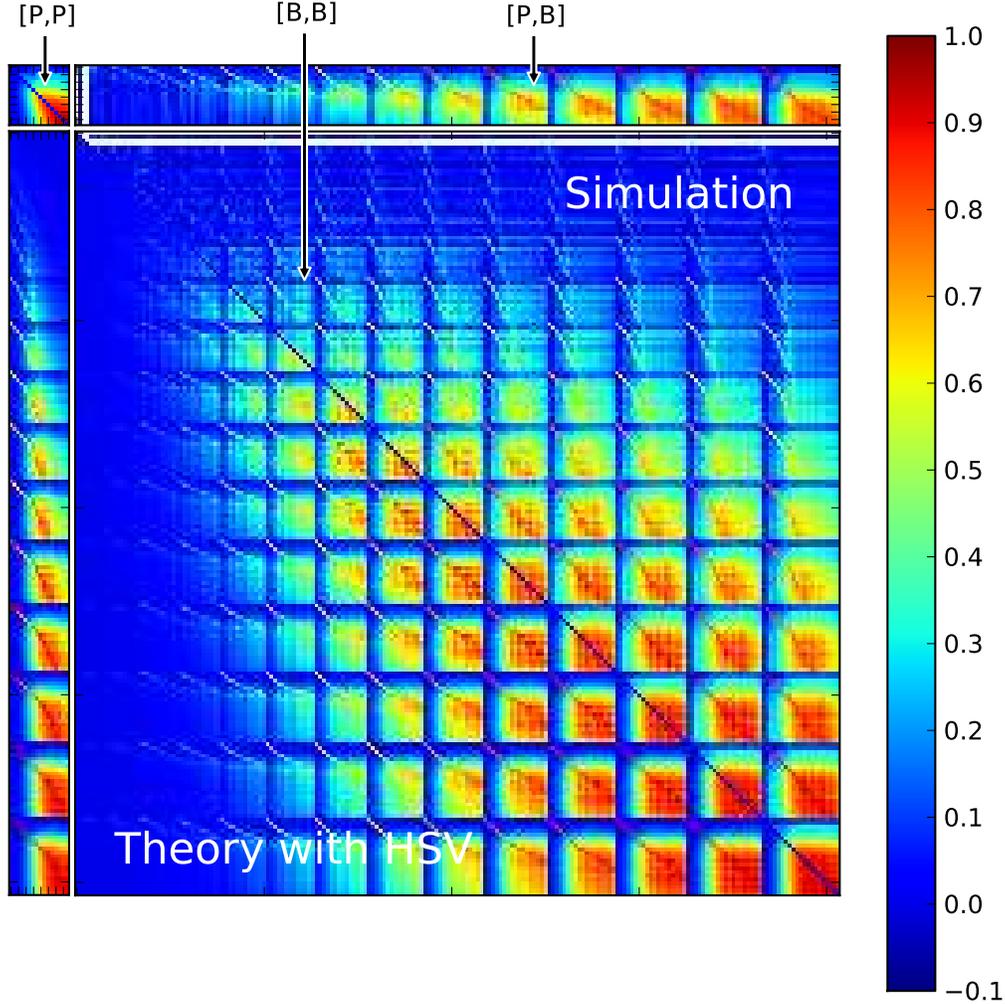}
\caption{Correlation coefficients $r^{XY}_{ij}$ (defined by
equation~\ref{eq:r_XY}) for the power spectrum and bispectrum covariance and
the cross-covariance between the power spectrum and the bispectrum,
where $X,Y=P$ or $B$ and the indices $i$ or $j$ denote the multipole bin
or the triangle configuration, e.g. $X_i=B(\bm{l}_i)$.  Note that the
diagonal elements $r^{XX}_{ii}=1$ by definition.  The upper-right matrix
elements show the simulation results, while the lower-left elements show
the halo model predictions with the HSV effect.  The upper-left
square-shaped panel ($16\times 16$ elements) shows $r^{PP}_ij$ for the
power spectrum covariance. The lower-right panel ($204\times 204$) shows
the bispectrum covariance matrix $r^{BB}_{ij}$.  The upper-right or
lower-left rectangular-shape panels ($16\times 204$ or $204\times 16$)
show the cross-covariance matrix $r^{PB}_{ij}$ or $r^{BP}_{ij}$.  As
the multipole becomes larger, the correlation coefficients have larger
values and approach $r_{ij}^{XY}\simeq 1$.  }
 \label{fig:cov2D}
\end{figure}

\begin{figure}
\centering
\includegraphics[width=0.5\textwidth]{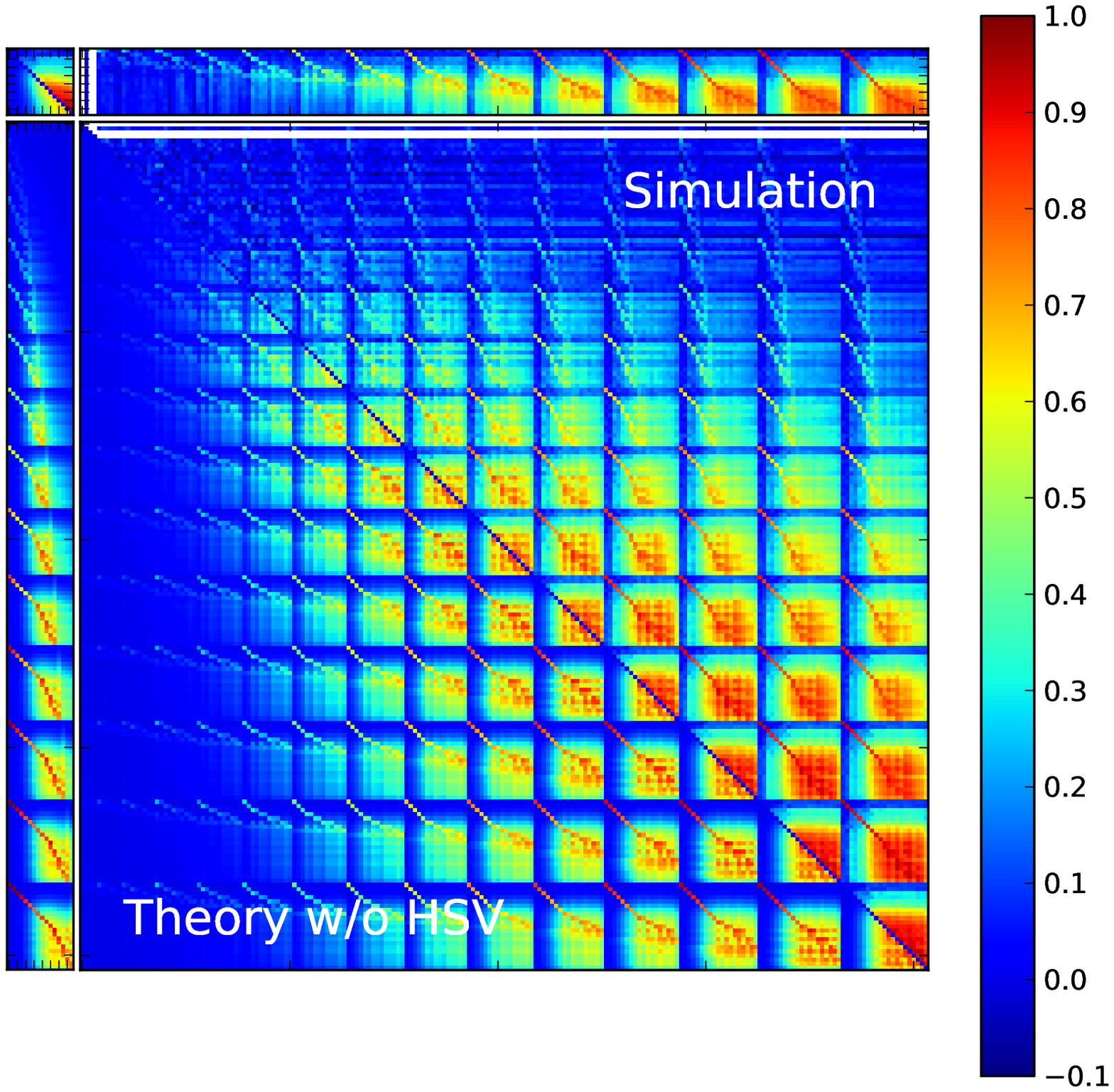}
\caption{Similar plot to the previous figure, but the halo model
 predictions without the HSV effect are shown (the lower-left matrix
 elements). Note that the simulation results (the upper-right elements)
 are the same as in the previous figure. 
It is evident that the halo model
 predictions are well below the simulation measurements for large multipoles owing 
 to the missing HSV terms.  
} \label{fig:cov2D_woHSV}
\end{figure}

In Figs~\ref{fig:bcov_eq}--\ref{fig:pbcov}, we study the covariance
matrices of the lensing bispectra for some representative triangle
configurations of the 204 triangles. Note that, in these results, the
two triangle configurations of the covariance have the same length(s)
for at least one side length (e.g. $l_1=l_1^\prime$). Hence the
covariance terms $O(BB)$ and $O(PT)$ in Fig.~\ref{fig:bcov} are not
vanishing for the  covariances shown in these figures. 
First, in Fig.~\ref{fig:bcov_eq}, we study the diagonal terms of the
bispectrum covariance matrix for equilateral triangles, ${\rm
Cov}[B_{\rm eq}(l),B_{\rm eq}(l)]$, as a function of the side length
$l$.  The points are the simulation results.  The jagged feature
at small $l$ bins is due to the effect of discrete pixels in the lensing
maps. The different curves are the halo model predictions for the
covariance matrix (equations~\ref{eq:bcov} and \ref{eq:bcov_hsv}). 
The dotted curve shows the Gaussian error
contribution that scales with $P(l)^3$, where we use the number of
triangles directly computed from the simulated lensing map.
The long dot--dashed, short
dot--dashed, long dashed and short dashed curves are the different non-Gaussian
terms. For these calculations, we use the $P(l)$
and $B_{\rm eq}(l)$ directly estimated from the 1000 ray-tracing
simulations, while we use the halo model in Section~\ref{sec:halomodel} to
compute the the higher-order functions.
The solid curve is the total power of the
covariance matrix, the sum of the terms above. The figure clearly shows that the
non-Gaussian errors become significant at $l\ga \mbox{a few } 100$,
and the model predictions are in good agreement with the simulation
results if we include the HSV term. The HSV term dominates the
other terms at $l\ga 1000$. These findings are similar to the results
in the power spectrum covariance \citep[see figs 5--7 in][]
{Satoetal:09}.

Fig.~\ref{fig:bcov_iso} shows the covariance matrices
for isosceles triangle configurations (left panel) and the off-diagonal
components between different triangle configurations, equilateral and
isosceles triangles (right panel). For this plot, we
do not use the condition $l_1\le l_2 \le l_3$ for presentation
purposes, but all the triangles shown here are indeed included in the
covariance matrix elements we will use in the following analysis. Even for
these more general triangle configurations, the model predictions
including the HSV effect are in good agreement with the simulations.
One may find that the covariance amplitudes or some non-Gaussian terms
peak at some particular value
of $l_3$ in the $x$-axis. This happens when the isosceles triangles
have higher symmetry, equilateral shape; the number of
independent triangles is smaller for higher-symmetry triangles,
leading to the greater covariance amplitudes. 

In Fig.~\ref{fig:pbcov}, we study the cross-covariance between the power
spectrum and the bispectrum of equilateral triangles, ${\rm Cov}[P(l),
B_{\rm eq}(l)]$, as a function of the multipole bin $l$. There is no
Gaussian error contribution because the cross-covariance arises from
the five-point functions. The figure again shows that
the model predictions including the HSV term well reproduces the
simulation results. 

In Fig.~\ref{fig:cov2D}, we compare the halo model predictions with the
simulation results for all the matrix elements of the power spectrum
covariance, the bispectrum covariance and the cross-covariance, in one
figure. 
To do this, we use the correlation coefficients of the covariance
 matrices defined as
\begin{equation}
 r^{XY}_{ij}\equiv\frac{{\rm Cov}[X_i, Y_j]}{\sqrt{{\rm Cov}[X_i, X_i]{\rm
  Cov}[Y_j, Y_j]}},
\label{eq:r_XY}
\end{equation}
where $X$ and $Y$ are the power spectrum or the bispectrum, and the
subscript $i$ or $j$ in $X$ or $Y$ 
denote the $i$th multipole bin or the $i$th triangle configuration; 
$X_i=P(l_i),
X_i=B(\bm{l}_i)$, and so on. 
The diagonal components $r_{ii}=1$ by definition. If $r_{ij}=0$,
it means no correlation between the spectra $X_i$ and $Y_j$, while the 
the higher values of $r_{ij}$  mean 
stronger correlations. 
For illustration
purposes, we use the following indices of the 204 triangles so that the
different triangles are indexed 
in increasing order of $l_3$:
\begin{eqnarray}
\Delta(i_{l_1},i_{l_2},i_{l_3})&=&
(1,1,1),\nonumber\\
&&(1,1,2), (1,2,2), (2,2,2),\nonumber\\
&&(1,1,3), (1,2,3), (1,3,3), (2,2,3), (2,3,3), (3,3,3),\nonumber\\
&&(1,3,4), \cdots,\nonumber\\
&&\vdots\nonumber\\
&&(1,16,16), \cdots, 
(14,14,16), (14,15,16), (14,16,16), (15,15,16),
(15,16,16), (16,16,16),
\label{eq:triang_order}
\end{eqnarray}
where we have used the 16 logarithmically spaced multipole
bins of $l$. Note that, for a given $l_3$-bin, the other multipole bins
$(l_1,l_2 )$ are listed in increasing order of $l_1$ ($l_1\le l_2$ for
each triangle index). 
With this triangle index, the higher-index triangle
configuration corresponds to the triangles having higher multipoles or
larger side lengths. The figure shows
that the halo model well reproduces the two-dimensional features of
the covariance matrices seen from the simulations.\footnote{
In Fig.~\ref{fig:cov2D} we fully used the halo model to compute the
covariance matrix 
elements including the power spectrum and bispectrum, unlike in
Figs~\ref{fig:bcov_eq}--\ref{fig:pbcov}.}
Most of the off-diagonal terms of the bispectrum covariance 
consist of only the term $O(P_6)$ and the HSV term, because in
general the shapes of two triangles are different from each other
in contrary to the cases of Figs~\ref{fig:bcov_eq} and \ref{fig:bcov_iso}.
The correlation coefficients become greater at higher
multipoles, almost $r_{ij}\simeq 1$.
For comparison, Fig.~\ref{fig:cov2D_woHSV} shows the results without
the HSV term in the halo model predictions (the simulation results are
the same to the previous figure), where the discrepancy is clear.

\subsection{Information content of the lensing bispectrum}

As we have studied, the non-linear structure formation induces
non-Gaussian errors in the weak lensing field, provoking significant
correlations between the power spectra of different multipoles and the
bispectra of different triangle configurations as well as significant
cross-correlations between the power spectra and the bispectra.
Then a more fundamental, important question arises. How much additional
information do the lensing bispectra carry to the lensing power spectrum?
Can joint measurements of the power spectra and bispectra 
recover the information content of the Gaussian field, which the
primordial density field of structure formation should have had as in
the CMB case? In this section, we address these questions.

A useful quantity to quantify the impact of the non-Gaussian errors is
the expected signal-to-noise ratio (S/N) for a measurement of the
lensing power spectra and bispectra in a given survey that is
characterized by its area
and shot noise parameters. The S/N is sometimes called the {\it
information content} \citep{Tegmarketal:97} \citep[also see][and
references therein]{TakadaJain:09}. For the power spectrum measurement,
the S/N is defined as
\begin{equation}
\left(\frac{S}{N}\right)_P^2\equiv 
\sum_{l_i,l_j<l_{\rm max}} P(l_i) \left[\bm{C}^P\right]^{-1}_{ij}P(l_j),
\label{eq:snps1}
\end{equation}
where the summation indices $i,j$ run over multipole bin indices up to a
given maximum multipole $l_{\rm max}$, and $[\bm{C}^P]^{-1}$ is the
inverse of the power spectrum covariance matrix. 
The inverse of S/N  is equivalent to a precision of measuring 
the logarithmic
amplitude of the power spectrum 
 up to a given maximum multipole $l_{\rm max}$, 
assuming that the shape of the power spectrum
is perfectly known. The S/N is independent of the multipole bin width
as long as the bin width is fine enough to capture the shape of the
lensing power spectrum (on the other hand, the relative
strength of the non-Gaussian errors to the Gaussian errors 
depends on the width).
Similarly, the S/N for the bispectrum measurement or the information
content about the bispectrum amplitude is defined as
\begin{equation}
\left(\frac{S}{N}\right)_B^{2}=\sum_{\{l_i\},\{l_j\}\le l_{\rm max}}B_i\left[
\bm{C}^B
\right]^{-1}_{ij}B_j,
\label{eq:snbs1}
\end{equation}
where the summation indices $i,j$ run over triangle configurations, and
we include all the triangle configurations whose side lengths are smaller
than a given maximum multipole $l_{\rm max}$. 

We also consider the S/N for a combined measurement of the lensing
power spectra and bispectra up to a given $l_{\rm max}$. In the presence
of the non-Gaussian errors, the total S/N is not simply a sum of
the two estimates of S/N of the power spectra and the bispectra due to the
cross-covariance.  To
study this, we first define the data vector for the joint measurement as
\begin{equation}
\bm{D}=\left\{P_1,P_2,\cdots,P_{n_b}, B_1,B_2,\cdots,B_{i_{\rm triang, max}}\right\}.
\end{equation}
The covariance matrix for the data vector
$\bm{D}$ is given as 
\begin{equation}
{\bm C}^{P+B}=
\left(
\begin{array}{cc}
\bm{C}^P&\bm{C}^{PB}\\
\bm{C}^{PB}&\bm{C}^B
\end{array}
\right),
\label{eq:cov-p+b}
\end{equation}
where the $\bm{C}^{PB}$ is the cross-covariance between the power
spectrum and the bispectrum.
Then, the total (S/N)$_{P+B}$ is similarly defined as
\begin{equation}
\left(\frac{S}{N}\right)^2_{P+B}=\sum_{i,j\le l_{\rm max}}
D_i \left[\bm{C}^{P+B}\right]^{-1}_{ij}D_j.
\label{eq:snp+b}
\end{equation}
%

\begin{figure*}
 \includegraphics[width=0.48\textwidth]{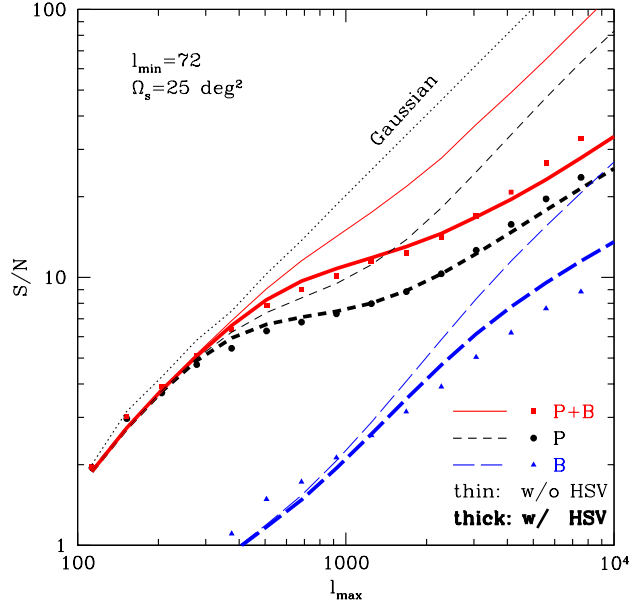} 
\caption{
Cumulative S/N for the power spectrum ($P$),
the bispectrum ($B$) and the joint measurement ($P+B$) for a survey area
of 25 deg$^2$ and source redshift $z_s=1$.  They are shown as functions
of the maximum multipole $l_{\rm max}$, where the power spectrum and/or
bispectrum information are included over $l_{\rm min}\le l \le l_{\rm
max}$ (see equations~\ref{eq:snps1}, \ref{eq:snbs1} and \ref{eq:snp+b}).  The
minimum multipole is set to $l_{\rm min}=72$. We do not include the
shape noise contamination here -- it is shown in the next figure.  The
circle, triangle and square symbols are the simulation results for $P$,
$B$ and $P+B$ measurements, respectively, computed from the 1000
realizations.  The thick short-dashed, long-dashed and solid curves
are the corresponding halo model predictions.
The corresponding thin curves are the results without the HSV
 contributions. 
For comparison, the dotted curve shows the S/N for the power spectrum
for the Gaussian field, which the primordial density field should have
contained.  Note that the simulation results for $B$ and
$P+B$ could be overestimated by about 10 per cent due to a finite number of
 the simulation realizations used to estimate the covariance matrices
\citep{Hartlapetal:07}.}  \label{fig:sn}
\end{figure*}

Fig.~\ref{fig:sn} shows the expected S/N for measurements of the
power spectra and the bispectra for a survey area of 25 square degrees
(i.e. the area of the ray-tracing simulation), as a function of the
maximum multipole $l_{\rm max}$ up to which the power spectrum and/or
bispectrum information are included. The minimum multipole is
fixed to $l_{\rm min}=72$. We do not include the shot noise
contamination to the error covariance matrices, so the results solely
correspond to the cosmological information contents.  The circle,
triangle and square symbols are the simulation results for the S/N
of the power spectra, the bispectra and the joint measurements,
respectively, which are computed using the 1000 realizations. The thick/thin
short-dashed, long-dashed and solid curves are the halo model
predictions with/without the HSV terms.
First of all, the lensing bispectra add new information content to the
power spectrum measurement.
To be more quantitative, adding the bispectrum measurement
increases the S/N by about 50 per cent for $l_{\rm max}\simeq 10^3 $
compared to the power spectrum measurement alone. Note that the $l_{\rm
max}$ of a few thousands is the typical maximum multipole for
upcoming weak lens surveys. This improvement is equivalent to about
2.3 larger survey area for the power spectrum measurement alone; that
is, the same data sets can be used to obtain the additional
information, if the bispectrum measurement is
combined with the power spectrum measurement.  Secondly, the halo model
predictions are in nice agreement with the simulation results.  Note
that the total S/N for the joint measurement ($P+B$) is close to the
linear sum of the S/N values ((S/N)$_P$ and (S/N)$_B$), not the sum of
their squared values (S/N)$^2$, due to the significant cross-covariance
between $P$ and $B$ \citep[see Appendix C in][for the similar discussion]{TakadaBridle:07}. If ignoring the cross-covariance, adding the
bispectrum measurement does not much improve the S/N (only by
5 per cent or so). Hence it is important to take into account the correlation
between the two measurements.

Next, let us compare the result above with the case of a Gaussian
random field, which is shown in the dashed curve in Fig.~\ref{fig:sn}.
The S/N for a Gaussian field is equivalent to the number of
independent Fourier modes up to a $l_{\rm max}$ for a given survey
area. 
The figure clearly shows that the joint measurement of the power
spectrum and the bispectrum does not recover the full information
content of the Gaussian field. This implies that the higher-order
statistics beyond the bispectrum are also important to recover the full
information content. One may argue that the initial memory of the field
cannot be recovered due to the non-linear structure formation. However,
we would like to note that, if ignoring the HSV contribution to the
covariance, adding the bispectrum can recover about 75 per cent of the Gaussian
information, as shown by the thin curves. 
Hence the loss of the information contents is mostly due to the 
the HSV contribution. 
As discussed in Section~\ref{sec:hsv},
the HSV alters the overall amplitude but preserves the shape of the
lensing spectra.  Hence the HSV may give the worst case degradation of
the amplitude parameter, but may not cause any serious degradation of
parameters that are sensitive to the shapes of the lensing spectra. A
genuine impact of the HSV on cosmological parameters needs to be
further studied and this is our future work.

\begin{figure*}
 \includegraphics[width=0.45\textwidth]{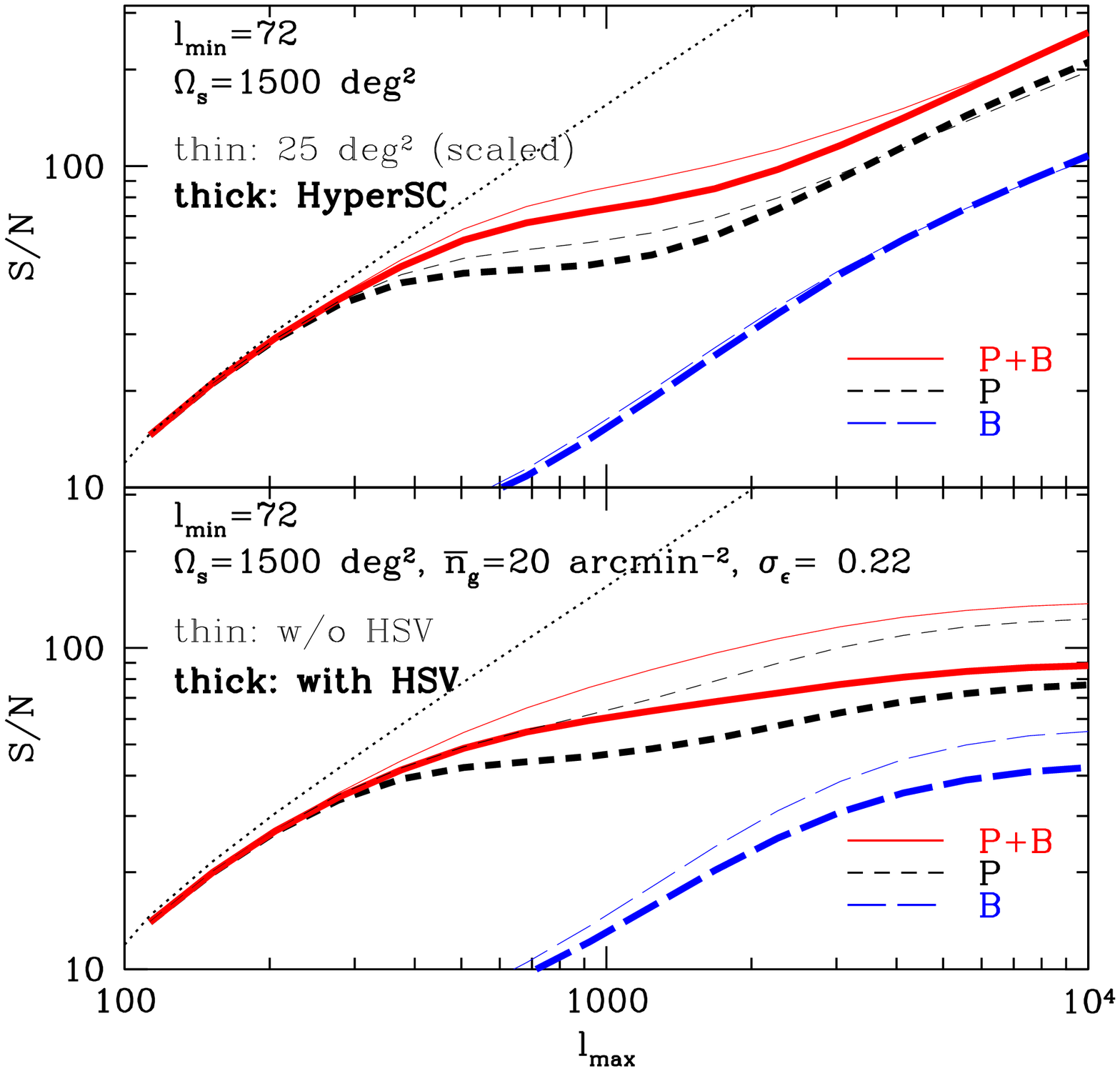}
 \includegraphics[width=0.45\textwidth]{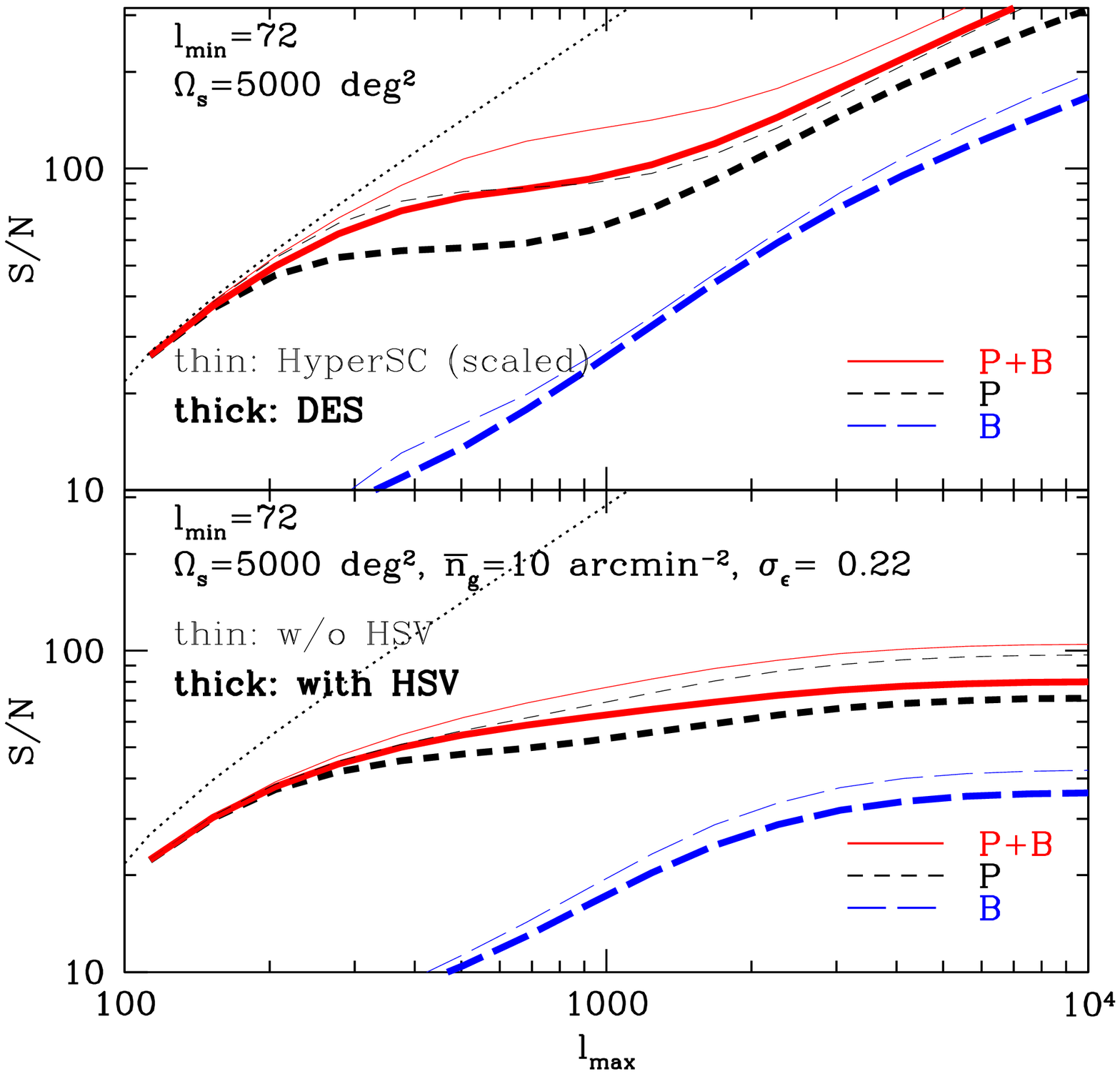}
 \caption{Cumulative S/N, as in the previous
 figure, but for upcoming weak lensing surveys,
 the Subaru Hyper Suprime-Cam (HyperSC) survey and the 
DES in the left- and right-hand panels, respectively. These surveys
 are characterized by survey parameters: survey area ($\Omega_{\rm
 s}$), mean source redshift ($\bar{z}_s$), and mean (effective)
 number density of source galaxies ($\bar{n}_g$). We assumed
 $\Omega_{\rm s}=1500$ deg$^2$, $\bar{z}_s=1$, and $\bar{n}_g=20$
 arcmin$^{-2}$ for the HyperSC survey and $\Omega_{\rm s}=5000$
 deg$^2$, $\bar{z}_s=0.7$, and $\bar{n}_g=10$ arcmin$^{-2}$ for the
 DES.  We set $\sigma_{\epsilon}=0.22$ for the rms intrinsic
 ellipticity per component for both the surveys.  The upper and lower
 plots in each panel show the results without and with the shot noise
 contamination.  The thin curves in the upper plot of the left-hand panel
 shows the S/N values obtained by scaling the results for $25$
 deg$^2$ in Fig.~\ref{fig:sn} assuming S/N $\propto \Omega_{\rm
 s}^{1/2}$.  The thin curves in the upper plot of the right-hand panel are
 similar, but obtained by scaling the HyperSC results in the left panel
 to 5000 deg$^2$. The lower plots in
 each panel show the results including the shot noise contribution to the
 covariance, but the thin/thick lines are without/with the
 HSV terms.
} \label{fig:sn_hsc}
\end{figure*}

\begin{table}
\begin{center}
Expected cumulative S/N for the upcoming weak lensing surveys\\
\begin{tabular}
{l|\colskip llll\colskip llll}\hline\hline
Survey &\multicolumn{4}{|c|}{Subaru HyperSC Survey }
&\multicolumn{4}{|c|}{DES}\\
$l_{\rm max}(\simeq)$& $1000$ 
& $1000$ 
& $2000$ 
& $2000$ 
& $1000$ 
& $1000$ 
& $2000$ 
& $2000$ \\
shot noise ($\sigma_\epsilon$)& w/o
& with
& w/o
& with
& w/o
& with
& w/o
& with
\\
\hline
(S/N)$_P$ & 53 & 48 & 74 & 57 
& 75 & 56 & 116 & 63\\
(S/N)$_B$ & 19 & 16 & 35 & 26 
& 33 & 20 & 59 & 29\\
(S/N)$_{P+B}$ & 78 (48\%) &64 (33\%) & 98 (32\%)& 72 (26\%)
& 103 (37\%)& 66 (18\%)& 145 (25\%)& 73 (16\%)\\
\hline
\end{tabular}
\caption{Cumulative S/N of the power spectrum ($P$), the
 bispectrum $(B)$ and the joint measurement ($P+B$) expected for the
 upcoming weak lensing surveys, the Subaru Hyper Suprime-Cam survey
 (HyperSC) and
 the DES, as in Fig.~\ref{fig:sn_hsc}. Here we consider
 $l_{\rm max}\simeq 1000$ and 2000 for the maximum multipole (more
 exactly, the bins of $l_{\rm max}=1245$ and $2268$). The column denoted
 by `w/o' or `with' in the row `shot noise' gives the S/N values
 with and without the intrinsic shape noise contribution to the
 covariances.  
The percentage in the parenthesis shows the improvement of
 S/N for the joint measurement ($P+B$) compared to the power
 spectrum alone ($P$).
 \label{tab:sn}}
\end{center}
\end{table}

In Fig.~\ref{fig:sn_hsc} and Table~\ref{tab:sn}, we show the S/N
expected for the upcoming wide-field weak lensing surveys, the Subaru
Hyper Suprime-Cam (HyperSC) survey and the DES,
which are characterized by the survey area, the mean source redshift and the
mean number density of source galaxies of $\Omega_{\rm s}=1500$
sq. degrees, $\bar{z}_s=1$ and $\bar{n}_g=20$ arcmin$^{-2}$ for the
HyperSC survey, while $\Omega_{\rm s}=5000$ deg$^2$, $\bar{z}_s=0.7$
and $\bar{n}_g=10$ arcmin$^{-2}$ for the DES, respectively. 
Here we employ the halo model to compute the S/N and assume a circular
survey geometry for simplicity. The figure and table show that
these surveys promise a significant detection of the
lensing bispectrum; (S/N) $\simeq 26$ or $29$ for the HyperSC or the DES,
respectively, when assuming $l_{\rm max} \simeq 2000$ and
including the shot noise effect. It also means that the theoretical
prediction of the lensing bispectrum needs to be as accurate as a few
per cent for the upcoming surveys. We find that the bispectrum
adds new information, increasing the total S/N by about 20 -- 30 per cent
compared to the power spectrum alone, which is equivalent to a factor of
1.4 -- 1.7 larger survey area. 

Fig.~\ref{fig:sn_hsc} also shows that the HSV is significant for
these surveys.  The thin curves in the upper plots of the
left-panel are the S/N computed by scaling the values
for 25 deg$^2$ in Fig.~\ref{fig:sn} assuming S/N $\propto \Omega_{\rm s}^{1/2}$. 
Since the covariance terms except for the HSV term scale as $1/\Omega_{\rm
s}$, the differences between the thick and thin curves 
are due to the HSV term which depends on the survey area via
the shape of the matter power spectrum convolved with the survey window
function (see the discussion below equation~\ref{eq:pscov_hsv}).
The S/N values are smaller than the naively-scaled
results, which means that the HSV decreases more slowly than
$1/\Omega_{\rm s}$.  The upper plot of the right panel
shows the similar plot, but for the DES results with scaled values
for the HyperSC in the left panel. The S/N are smaller than the scaled
HyperSC results, because
the typical source redshift of $\bar{z}_s=0.7$ for the DES is lower than
that of the HyperSC of $\bar{z}_s=1$ and the DES is more sensitive to
the non-linear density fluctuations.
The HSV has significant influence on the S/N even in the presence 
of the shot noise as shown in the lower panels of the figure.
Although the shot noise leads to a saturation of the S/N at large
multipoles, note that the systematic
effects such as the highly non-linear clustering effect and/or the
baryonic effect become more significant at these high multipoles
\citep{White:04,ZhanKnox:04,HutererTakada:05,Hutereretal:06}.

\subsection{Principal component analysis of the lensing covariance matrices}

A principal component analysis (PCA) of the power spectrum and
bispectrum covariance matrices is useful to quantify how the different
power spectra and/or bispectra are correlated with each other and how
many independent modes or triangles contribute to most of the
information contents \citep{TakadaJain:09} \citep[also see][for the
the 3D bispectrum case]{Scoccimarro:00}. Since
the covariance matrix is symmetric by definition, it can always be
decomposed as
\begin{equation}
 C^X_{ij}=\sum_{a} S^X_{ai}(\lambda^X_a)^2S^X_{aj},
\end{equation}
where $X$ is the lensing power spectrum or bispectrum, $\lambda_a^X$ is
the $a$th eigenvalue or principal component,
$(\bm{S}^X)^{-1}=(\bm{S}^X)^T$,
$\sum_k S^X_{ik}S^X_{jk}=\delta^K_{ij}$ and $\bm{S}^X$ is normalized so
as to satisfy $\sum_{j}(S^X_{ij})^2=1$.  The matrix $S^X_{ai}$ is
considered as the projection matrix as it describes how the power in the
$i$th multipole bin or triangle configuration is projected onto the
$a$th eigenmode.  Using this representation, the inverse of the
covariance matrix is given by $[\bm{C}^X]^{-1}_{ij}=S_{ia}^X
(1/\lambda^X_a)^2S_{ja}^X$. Hence, the S/N values (equations~\ref{eq:snps1}
or \ref{eq:snbs1}) can be rewritten as
\begin{equation}
\left(\frac{S}{N}\right)^2_X=\sum_{a}\left\{
\frac{1}{\lambda_a^X}\sum_{i}S^X_{ai}X_i
\right\}^2,
\label{eq:snpca}
\end{equation}
where $X_i$ is either $P(l_i)$, $B(\bm{l}_i)$ or the joint
($P+B$). Thus, since $(1/\lambda_a^X)\sum_{i}S_{ai}^XX_i$ can be
considered as the S/N for the $a$th eigenmode, the above equation
expresses the total S/N (for a given $l_{\rm max}$) as a sum of
contributions from the independent eigenmodes.
We can then re-order the eigenmodes such that the lower eigenmode has
the higher contribution to the S/N.

\begin{figure}
  \begin{center}
   \includegraphics[width=0.49\textwidth]{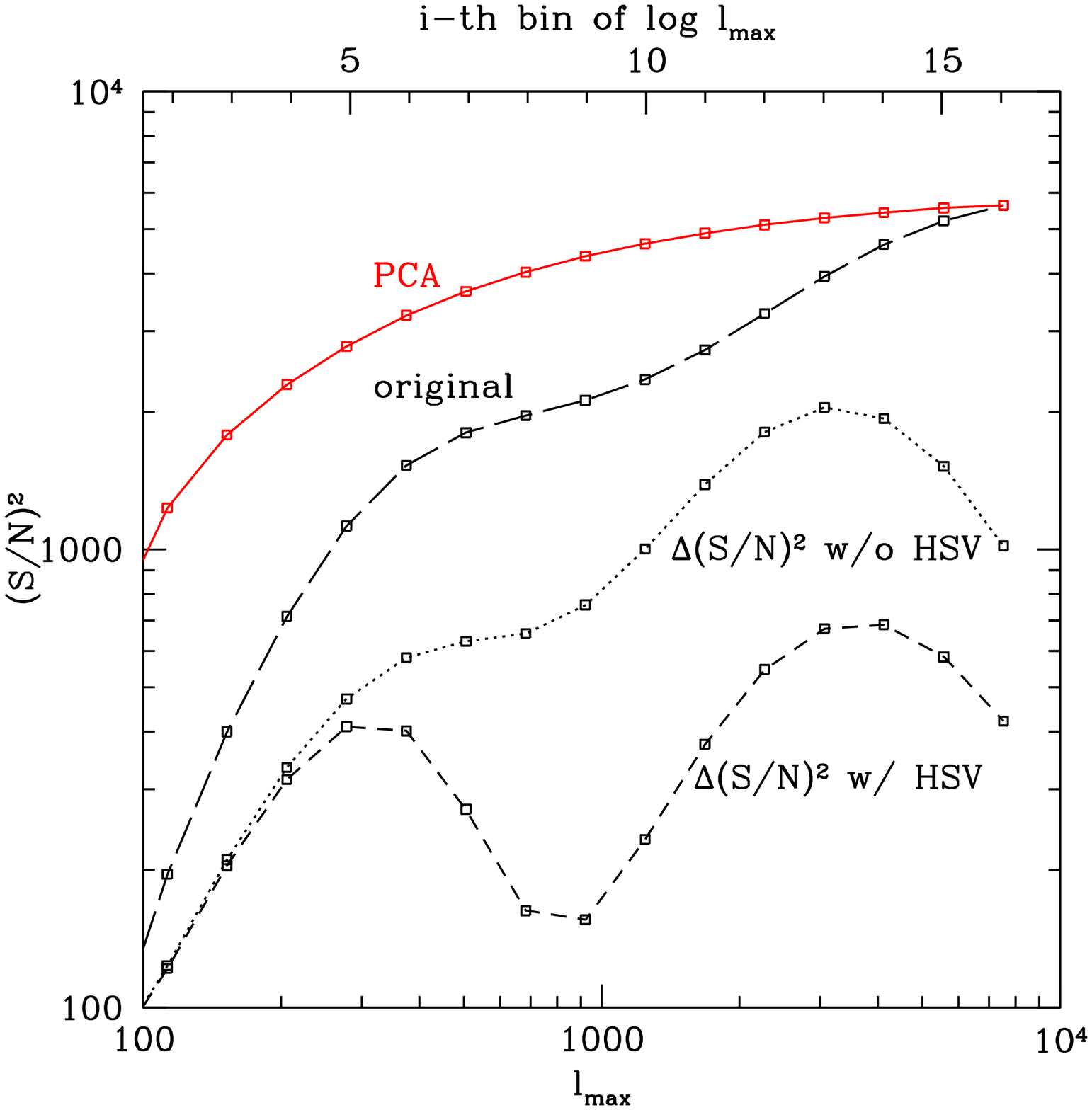}
   \includegraphics[width=0.49\textwidth]{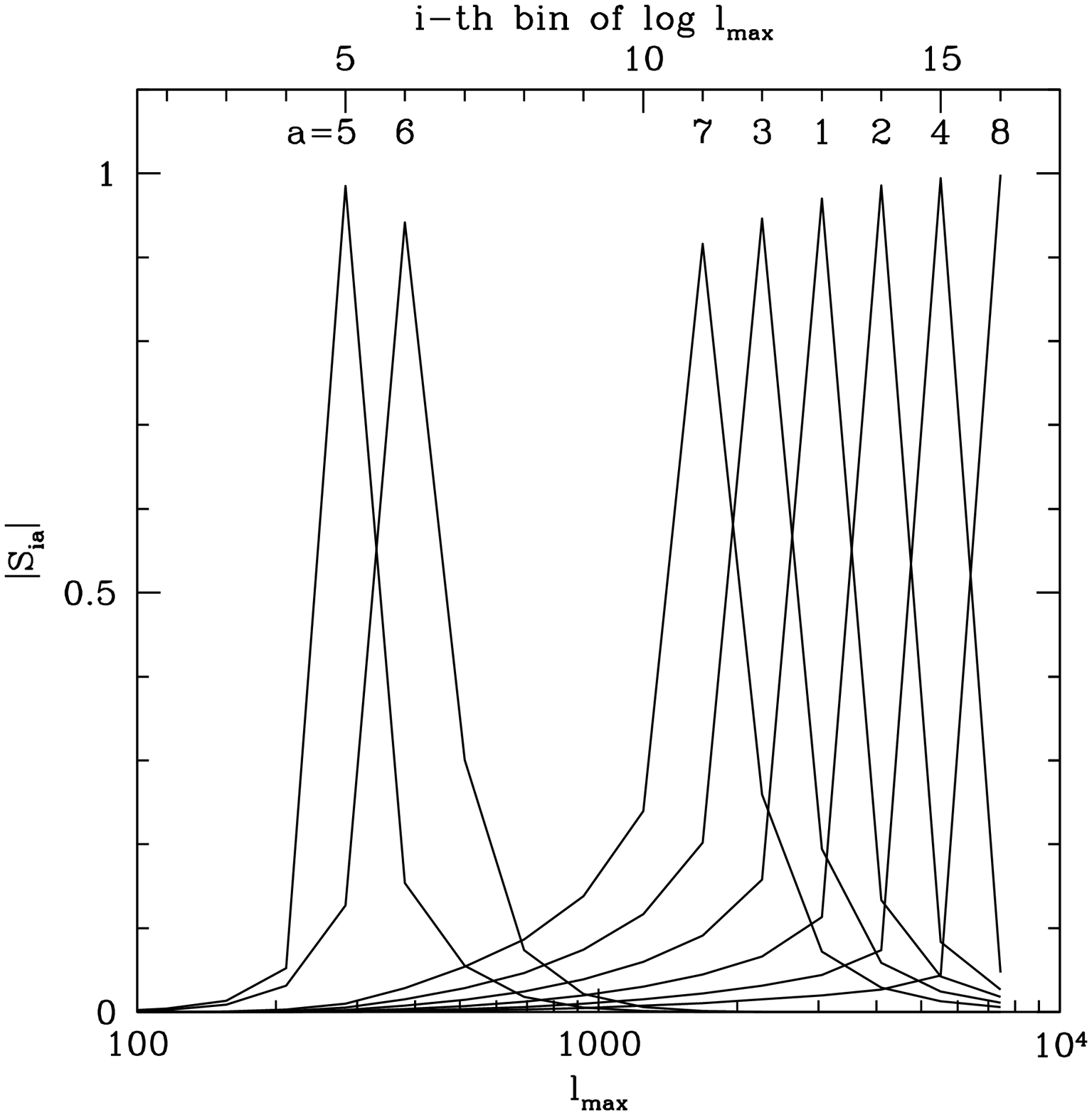}
  \end{center}
\caption{PCA of the power spectrum
 covariance for the 16 logarithmically spaced multipole bins over $72\le
 l\le 8748.8$, for the Subaru HyperSC-type survey with 1500 deg$^2$ and
 with the shot noise.  Left-hand panel: the long-dashed curve shows the
 cumulative (S/N)$^2$ for the power spectrum as in the left-lower panel
 of Fig.~\ref{fig:sn_hsc}, but for (S/N)$^2$ instead of S/N (the
 points denote the central value of each multipole bin).  The
 short-dashed and dotted curves show the differential contribution to
 the (S/N)$^2$ at each multipole bin, with and without the HSV effect,
 respectively. The solid curve shows how the (S/N)$^2$ value is
 recovered by adding the PCA eigenmodes.  Note that the
 PCA results are shown as a function of the order of the PCA eigenmodes,
 where the PCA eigenmodes are ranked in decreasing order of the
 differential contribution to the (S/N)$^2$ (see equation~\ref{eq:snpca}).
 Right-hand panel: the projection matrix $|S_{ai}|$ for the first eight
 eigenmodes, where the index $a$ denotes the $a$th eigenmode.  }
 \label{fig:pspca}
\end{figure}

Fig.~\ref{fig:pspca} shows the PCA results for the power spectrum
measurement for the Subaru HyperSC-type survey including the shot noise
effect as in Fig.~\ref{fig:sn_hsc}.  This figure can be compared with
Fig. 5 in \cite{TakadaJain:09}, where the HSV effect was not
included. The short-dashed and dotted curves show each contribution of
the $i$th multipole bin to the total (S/N)$^2$ with and without the
HSV, respectively.  It should be stressed that the power spectrum of $l
\sim 10^3$ has a local minimum, due to the significant HSV contribution.

The long-dashed curve shows the total (S/N)$^2$ as a function of the
maximum multipole $l_{\rm max}$, up to $l_{\rm max}=8748.8$ (the 16-th
multipole bin). On the other hand, the solid curve shows how the
cumulative (S/N)$^2$ value increases by adding the $a$th eigenmode
(equation~\ref{eq:snpca})\footnote{First, we made the principal component
analysis of the power spectrum covariance matrix including up to the
multipole bin $l=8748.8$ (i.e. $16\times 16$ matrix). Then, we
re-ordered the eigenmodes in increasing order of the differential
(S/N)$^2$ value. The solid curve in Fig.~\ref{fig:pspca} shows how the
cumulative S/N value increases by adding a new $a$th eigenmode.}. The
figure shows that, among the 16 multipole bins, including about seven
eigenmodes (about half of the multipole bins) can recover about 90 per cent of
the total (S/N)$^2$. The other eigenmodes are relatively less important
due to strong correlations between the different multipole bins. The
right panel shows the projection matrix $|S_{ia}|$ for each multipole
bin, showing how the neighboring multipole bins are correlated with each
other.  The projection matrix for each eigenmode peaks at some multipole
bin, but has tails to different multipole bins. In particular, the
eigenmodes around a few $10^3$ have longer tails, reflecting significant
correlations between different multipole bins due to the non-Gaussian
errors.

\begin{figure}
  \begin{center}
   \includegraphics[width=0.6\textwidth]{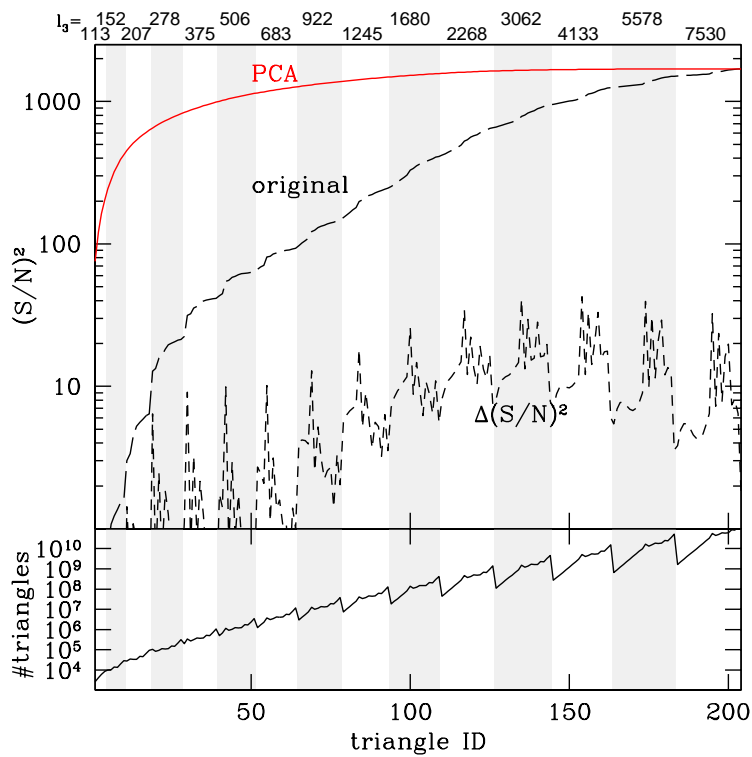}
  \end{center}
\caption{PCA analysis of the bispectrum covariance, for the 204
 triangle configurations over $72\le l\le 8748.8$, for the HyperSC-type
 survey as in the previous figure. The $x$-axis denotes either the
 triangle index or the PCA eigenmodes. The triangle indices are in the
 order given by equation~(\ref{eq:triang_order}); the larger indices
 correspond to triangles with the larger side length of $l_3$.
 The vertical shaded regions denote the triangles with 
 the same $l_3$ bin, as indicated by the top label of `$l_3$'-bin
 values.  Upper panel: 
 the short-dashed curve shows the differential (S/N)$^2$ of
 each triangle configuration, while the long dashed curve shows the
 cumulative (S/N)$^2$ as in the left-lower panel of
 Fig.~\ref{fig:sn_hsc}, but for the finer multipole bins (stepped by
 each triangle configuration). The solid curve shows the cumulative
 (S/N)$^2$ as one adds the different PCA eigenmodes. About 70
 eigenmodes out of 204 triangles can recover 90 per cent of the
 total (S/N)$^2$ of the bispectrum measurement over $72\le l\le 8748.8$.
 Lower panel: the solid curve shows the number of independent
 triangles available from Fourier modes for the HyperSC-type survey
 (1500 deg$^2$), for each
 triangle configuration in the $x$-axis. 
}
 \label{fig:bspca}
\end{figure}

Fig.~\ref{fig:bspca} shows the PCA analysis for the bispectrum
covariance matrix, where we include the 204 triangles over $72\le l\le
8748.8$.  The different triangle configurations contribute to the total
(S/N)$^2$ in a complex way. We use the triangle index, given by
equation~(\ref{eq:triang_order}), and the higher-index triangle corresponds
to triangle with larger side length of `$l_3$', under the condition
$l_1\le l_2\le l_3$.  However, for a given $l_3$ bin, either of the two
side lengths $l_1$ or $l_2$ can be much smaller than $l_3$, and such
a triangle configuration generally has a
smaller contribution to (S/N)$^2$. As
a result, the differential (S/N)$^2$ curve has jagged features.  The
sold curve shows the cumulative (S/N)$^2$ by adding the new PCA
eigenmodes, and manifests that about 70 eigenmodes, one third of
the 204 triangles, carry 90 per cent of the total (S/N)$^2$.

\begin{figure}
\begin{center}
 \includegraphics[width=0.48\textwidth]{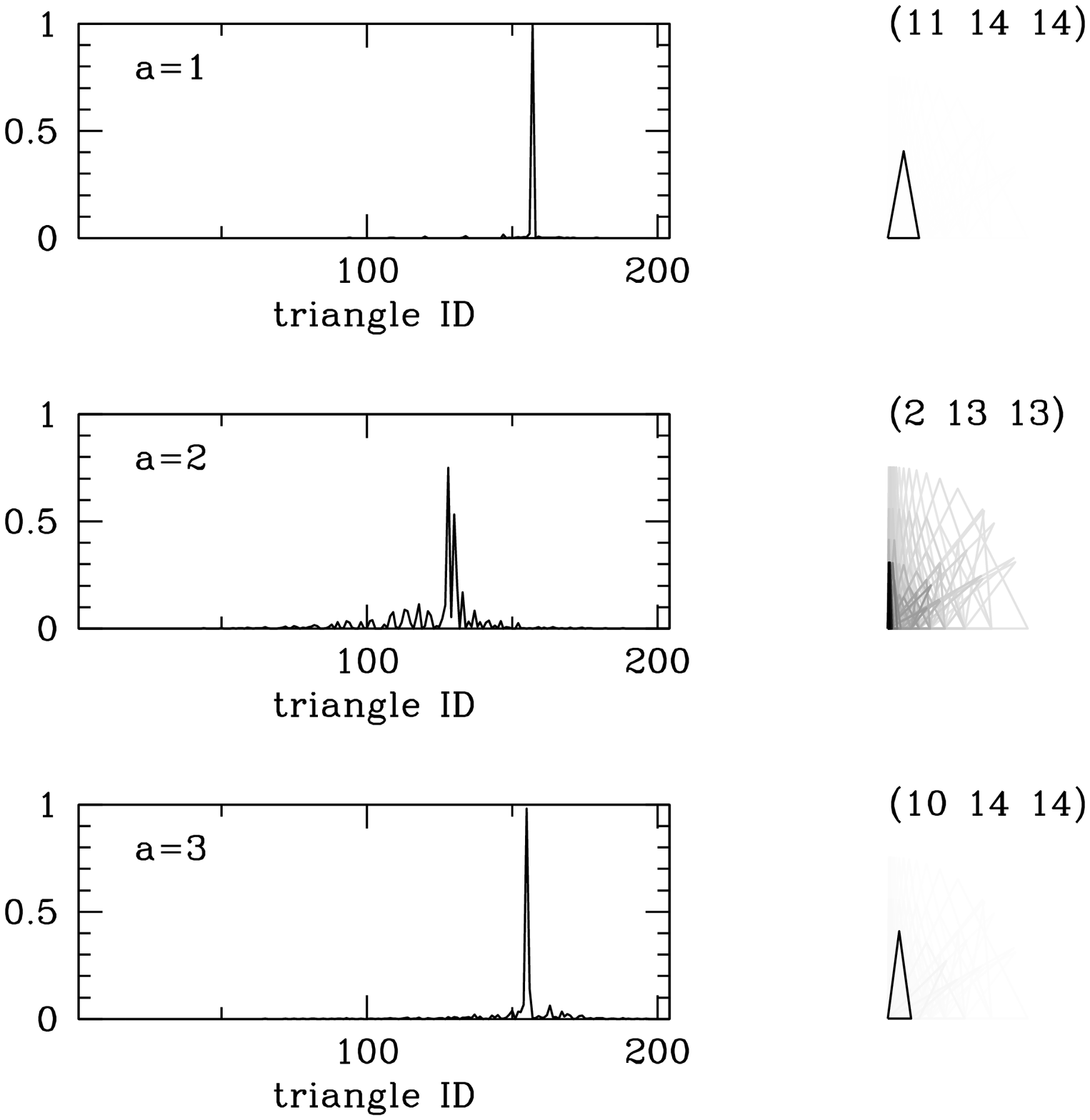}
 \includegraphics[width=0.48\textwidth]{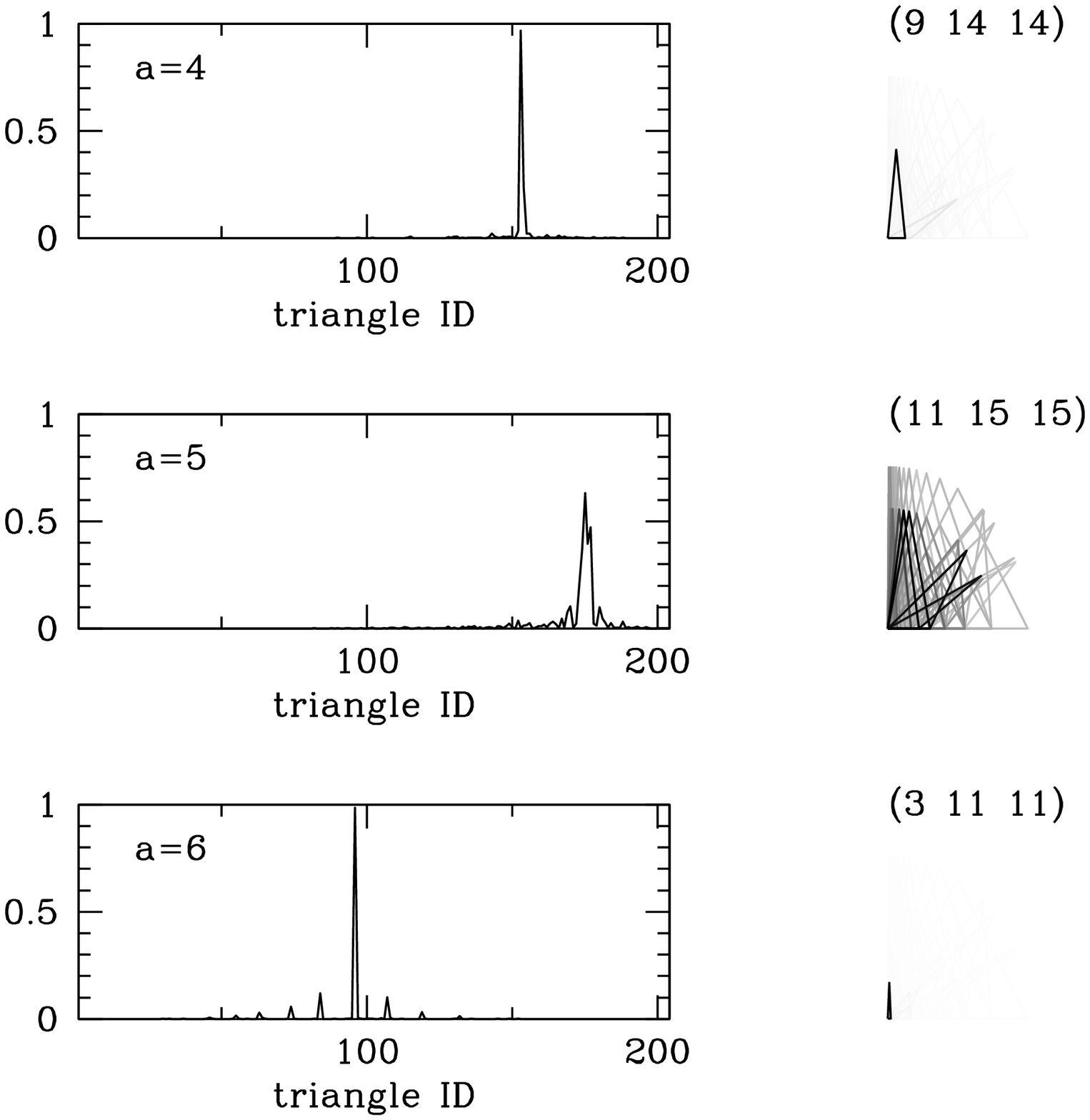}
\caption{The curve in each panel shows which triangle
configurations contribute to each
PCA eigenmode for the bispectrum measurement in
Fig.~\ref{fig:bspca}. We consider the first 6 PCA eigenmodes. They are
labeled by the index $a$ from the highest contribution to the (S/N)$^2$
and the absolute values of $|S_{ia}|$ are plotted, where
$S_{ia}$ is the projection matrix of the $a$th eigenfunction onto the
$i$th triangle configuration. The triangle configurations shown on the
right side of each panel are the triangle configurations with
$S_{ia}\neq 0$, where the triangle sizes are plotted on the linear scale
and the thickness of the lines corresponds to the
amplitude of $|S_{ia}|$. The configurations with
the largest contribution are labelled in the side length bins 
($l_1,l_2,l_3$) at top. Note that we are
using logarithmically-spacing multipole bins in this paper, so about
3/4 triangles available in the Fourier space are isosceles triangles.
\label{fig:bcov_6pca} }
\end{center}
\end{figure}

Fig.~\ref{fig:bcov_6pca} shows which triangle configurations contribute
to the first 6 PCA eigenfunctions, which therefore have the 6 highest
contributions to the (S/N)$^2$. More exactly, each panel shows the
projection matrix $|S_{ia}|$ for the $a$th eigenmode
($a=1,2,\dots,6$). In most panels, the projection matrix $|S_{ia}|$ has
a single or a few peaks, reflecting that the higher-S/N eigenmode
arises mostly from a single or a few different triangles. For example,
the highest (S/N)$^2$ eigenfunction arises from isosceles triangles,
which have side lengths of a few $10^3$ (11- and 14-th multipole bins).
However, note that we have used rather wide, logarithmically-spaced
multipole bins, and most of triangle configurations available from
Fourier space are close to isosceles triangle configurations. Hence, an
exact shape of triangles giving a large contribution to the total
(S/N)$^2$ slightly changes with the bin width. If we take a finer
multipole bin, the triangles with the highest (S/N)$^2$ may differ from
isosceles triangles.

\section{Conclusion and Discussion}
\label{sec:conc}

We have studied the covariance matrices of the lensing
power spectrum and the bispectrum by using both 1000 ray-tracing
simulations and the analytical halo model.
We have found that there are significant non-Gaussian error
contributions to the lensing covariance matrices; the power spectra or
the bispectra at higher multipoles than a few hundreds are highly
correlated with each other. In particular, we have shown that the mass
density fluctuation at scales comparable with or outside a survey
region causes significant non-Gaussian error contributions, which we call
the HSV and has not been fully
studied in the literature. With the HSV contributions, the halo
model predictions reproduce the covariance matrices measured
from the simulations (Figs~\ref{fig:bleq}--\ref{fig:cov2D}).

Then, we have addressed how much information the lensing bispectrum adds
to the power spectrum by including all the triangle
configurations available up to a maximum multipole
$l_{\rm max}$. 
Adding the bispectrum
measurement improves the S/N by about 20--50 per cent compared to the power
spectrum alone measurement, at $l_{\rm max}$ of a few $10^3$
(Fig.~\ref{fig:sn}).
We have also studied the prospects for upcoming weak lensing surveys,
including the shot noise contamination due to intrinsic galaxy shapes
(Fig.~\ref{fig:sn_hsc} and Table~\ref{tab:sn}). 
The improvement
in S/N is
equivalent to about 1.4--2.3 larger survey area for the power spectrum
measurement alone. Hence, our results show that the same imaging data
can be used to improve the constraining power of cosmological
parameters, if we combine the power spectrum and bispectrum
measurements. We have also found that the HSV effect
is significant, leading to a large degradation in the S/N.
By using a PCA of the covariance
matrices, we have shown that about 1/3 eigenmodes of all the triangle
configurations over
the range of $10^2\la l\la 10^4$ ($70$ eigenmodes compared to the
204 triangles for the multipole binning we assumed) carry most of
the total information for the bispectrum measurement
(Fig.~\ref{fig:bspca}).  Thus, our results give a quantitative answer
to the longstanding question of how the bispectrum can be
useful and complementary to the power spectrum by fully taking into
account non-Gaussian errors and all triangle configurations.

Future surveys such as the Subaru Hyper Suprime-Cam survey or the DES
allow for a significant
detection of the lensing bispectra
(Fig.~\ref{fig:sn_hsc} and Table~\ref{tab:sn}).
However, there are some practical issues for making the bispectrum measurement
feasible for future surveys. First, we must take into account
effects of the complex survey geometry and masked regions such as
saturated bright stars. This can be done
by extending the method in \cite{Hikageetal:11} for the power spectrum
measurement to the bispectrum, but has not yet been fully addressed
in the literature. It is also important to explore a method of cleanly
decomposing three-point correlations of $E/B$ modes in the presence of
the survey window function. 
An alternative approach is to measure the real-space
three-point correlation functions, rather than the bispectrum, which does
not suffer from the geometrical problem. However, there are
even more significant correlations between the three-point correlations
of different triangles, and the estimate of the covariance matrices
involves multi-dimensional integrations of the bispectrum covariance if
one uses the halo model approach. Although the alternative to the theoretical
approach may be to use a sufficient number of simulations, 
it will not be feasible to construct the covariance matrices for many
cosmological models. Hence, which of the Fourier- or real-space
is more useful/tractable for the three-point correlation
measurements is still an open issue. 

Although we have focused on the S/N as a measure of the information
content, this is just one measure to quantify the complementarity of the
lensing bispectrum.  Specifically, the S/N quantifies the expected
precision of the amplitude, assuming that the shapes of the power
spectrum of the bispectrum are known.  Hence, the S/N is not
necessarily the best measure to quantify the power of the bispectrum for
constraining cosmological parameters, especially parameters that are
sensitive to the shape of the bispectrum \citep[also see][for the
similar discussion]{TakadaJain:09}.  For instance, \cite{TakadaJain:04}
showed that the bispectrum can significantly improve the accuracy of
parameter estimate from the power spectrum measurement alone, by
efficiently breaking parameter degeneracies, while the S/N of the
bispectrum itself is smaller than that of the power spectrum by a factor
of few up to $l_{\rm max}$ of a few $10^3$ (see Fig.~5). Although they
ignored the non-Gaussian error contributions to the covariances, the
relative amplitude of the S/N between the power spectrum and the
bispectrum is similar to our case with the full non-Gaussian terms,
because the non-Gaussian terms degrade both the power spectrum and the
bispectrum.  Thus we can expect a similar improvement in cosmological
parameters when including the bispectrum information, even with the
inclusion of non-Gaussian errors.  Exploring the genuine usefulness of
the lensing bispectrum for cosmology is very interesting -- the full
forecasts for various cosmological parameters will be presented in
future work.

In particular, to do such parameter forecasts, it would be much more
interesting and useful if including  tomographic information of the
lensing field that is available from photometric redshift of
source galaxies. It has been found that the lensing tomography can
significantly improve the constraining power by recovering the
radial-direction information of the lensing field 
\citep[e.g.][]{TakadaJain:04}. However, for the lensing bispectrum
tomography, we need to further include all the triangle configurations
in different multipole and redshift bins in order to estimate the full
potential. For example, if we consider three redshift bin tomography,
the total number of triangles is $3^3\times 204=5508$ for the
same multipole binning as used in this paper. Thus, 
even 1000 realizations are not enough to reliably estimate the covariance
matrices of the lensing bispectra.
 Hence we believe that the analytical formula we developed in
this paper are essential to address these issues. 

Perhaps as important as the improvement in statistical precision, the
bispectrum might enable a self-calibration
of  systematic errors inherent in the lensing tomography measurements
such as photometric redshift errors and imperfect galaxy shape
measurement.
Lensing bispectra depend on systematic
errors in a different way from the power spectrum, but the two spectra
share the same large-scale structure, hence the same cosmological
parameters that describe the underlying true cosmology. Thus combining
the power spectrum and bispectrum measurements can be used to
self-calibrate  systematic errors and improve  cosmological
constraints \citep{Hutereretal:06}.
Again, it is important to fully take into account
the non-Gaussian errors in order to quantify how well
self-calibration works for upcoming lensing surveys.

Planned multi-colour imaging surveys can also be used to measure other
cosmological probes such as the abundance of galaxy clusters and the
correlation functions of galaxies. Recently, \cite{OguriTakada:11}
proposed a new method of using the halo--shear correlations or the
stacked lensing signals around massive haloes as a function of the
cluster redshifts, and showed that the stacked lensing tomography allows
us to constrain cosmological measurements to a high precision.  In halo
(cluster)--shear correlations, the signals arise from the large-scale
structure at the cluster redshift.  As we have shown, massive clusters
are key observables for understanding the HSV effect on the lensing power
spectra. It would be interesting to pursue how the lensing power
spectrum and bispectrum measurements can be combined with the cluster
observables to improve cosmology parameter estimate by calibrating the
HSV effect. The formulation we have developed in this paper can be
straightforwardly extended to the halo--shear bispectra such as the
halo--shear--shear and halo--halo--shear three-point correlation functions.

Finally, there are several effects that we have ignored in this paper
and need to be studied. One is that we only considered a simple
circular-shaped survey geometry to compute the HSV term.  For a general
geometry of the survey, the Fourier transformed survey window function
becomes anisotropic and causes additional apparent correlations between
the different spectra. Since the large-scale density fluctuations which
contribute to the HSV are in the linear regime, it may be
straightforward to take into account the effect by extending our
formulation. Secondly, we have used the flat-sky approximation,
which is not valid for very wide surveys such as the LSST. For
a full-sky survey as the ultimate case, the HSV effect is caused by the
horizon-scale perturbations with wavenumbers comparable to the
matter-radiation equality wavenumber $k_{\rm eq}$.
In this case, the HSV terms decrease more rapidly with increasing 
survey area than other covariance terms. Hence the HSV effect may not be that
significant for such an all-sky survey. This is worth studying and the
formulation and the method we developed in this paper can be used for
such studies.

\section*{Acknowledgments}
We would like to thank an anonymous referee for many important suggestions.
We also thank Gary Bernstein, Elisabeth Krause, Takahiro
Nishimichi and Masanori Sato for useful discussion.  This work is supported in part by JSPS
KAKENHI (Grant Number: 23340061 and 24740171), by JSPS Core-to-Core
Program `International Research Network for Dark Energy', by World
Premier International Research Center Initiative (WPI Initiative), MEXT,
Japan, by the FIRST program `Subaru Measurements of Images and
Redshifts (SuMIRe)', CSTP, Japan, and by NSF grant AST-0908027 and DOE
grant DE-FG02-95ER40893.

\bibliographystyle{mn2e}
\bibliography{ms_ref}

\appendix 

\section{Derivation of the bispectrum covariance}
\label{app:bcov}

We derive equations of the bispectrum covariance
matrix given in terms of the lensing spectra (power spectrum, bispectrum
and the higher-order functions), following the method developed in
\cite{TakadaBridle:07}.

\subsection{Discrete Fourier decomposition}

The lensing power spectrum is estimated from the
Fourier transform of the lensing convergence field, $\tkappa_{\bm{l}}$.
When the Fourier decomposition is done in a finite
survey region, the Fourier modes are by nature discrete
and the fundamental mode is limited
by the size of the survey area, $l_f=2\upi/\Theta_{\rm s}$, where the
survey area is given by $\Omega_{\rm s}=\Theta_{\rm s}^2$ (we assume a
square survey geometry). For this case, the convergence
field can be expanded using the discrete Fourier decomposition as
\begin{equation}
\kappa(\bm{\theta})=\frac{1}{\Omega_{\rm
 s}}\sum_{\bm{l}}\tkappa_{\bm{l}}e^{i\bm{l}\cdot\bm{\theta}},
\end{equation}
where the summation runs over the combination of integers $(n_x,n_y)$
for $\bm{l}= (2\upi/\Theta_{\rm s})(n_x,n_y)$. The prefactor
$1/\Omega_{\rm s}$ is our convention, motivated by the fact that the
discrete Fourier decomposition has the limit of the continuous Fourier
decomposition; $(1/\Omega_{\rm
s})\sum_{\bm{l}}\tkappa_{\bm{l}}e^{i\bm{l}\cdot\bm{\theta}} \rightarrow
\int\!d^2\bm{l}/(2\upi)^2\tkappa_{\bm{l}}e^{i\bm{l}\cdot\bm{\theta}}$ for
the limit of $\Theta_{\rm s}\rightarrow \infty$.
If the Fourier transform is confined in the survey region like
here; the modes of scales outside the survey region are out of
consideration and the HSV cannot be taken into account.

In the discrete Fourier
decomposition, the orthogonal relation for eigen function
$e^{i\bm{l}\cdot \bm{\theta}}$ is modified as
\begin{equation}
\int_{\Omega_{\rm
 s}}\!d^2\bm{\theta}e^{i(\bm{l}-\bm{l}')\cdot\bm{\theta}}=\Omega_{\rm s}
\delta^K_{\bm{l}-\bm{l}'},
\end{equation}
where the integration range is confined to the survey region, and 
$\delta^K_{\bm{l}}$ is the Kronecker-type delta;
$\delta^K_{\bm{l }}=1$ if $\bm{l}=\bm{0}$, and otherwise
$\delta^K_{\bm{l}}=0$. The orthogonal relation above suggests that the
Kronecker delta should be replaced with the Dirac delta function
for the limit of $\Theta_{\rm s}\rightarrow \infty$: $\Omega_{\rm
s}\delta^K_{\bm{l}-\bm{l}'}\rightarrow (2\upi)^2\delta^D(\bm{l}-\bm{l}')$. 
Hence the definitions of the lensing spectra are modified for a
finite-area survey from equation~(\ref{eq:ps_def}); e.g.
\begin{eqnarray}
&&\ave{\tkappa_{\bm{l}_1}\tkappa_{\bm{l}_2}}=\Omega_{\rm
 s}\delta^K_{\bm{l}_1+\bm{l}_2}P(l_1), \nonumber\\
&&\ave{\tkappa_{\bm{l}_1}\tkappa_{\bm{l}_2}\tkappa_{\bm{l}_3}
}=\Omega_{\rm s}\delta^K_{\bm{l}_1+\bm{l}_2+\bm{l}_3}
B(l_1,l_2,l_3), 
\end{eqnarray}
and similar expressions for the higher-order functions.

\subsection{Bispectrum estimator and the covariance matrix}
\label{app:bcov2}

The bispectrum is given as a function of triangle configurations. We use
a parametrization of three side lengths ($l_1,l_2,l_3$) to specify a
triangle configuration. The three parameters are enough, because all the
triangles transformed by rotation, permutation
and parity transformations in Fourier space are equivalent to yielding
the same bispectrum in an ensemble average sense for a homogeneous and
isotropic field as expected for the lensing field. Hence, within the
framework of the discrete Fourier decomposition we described above, we
can define an estimator of the bispectrum:
\begin{equation}
\hat{B}({l}_1,{l}_2,{l}_3)=
\frac{1}{\Omega_{\rm s}N_{\rm
trip}(l_1,l_2,l_3)}\sum_{\bm{q}_i}
\tilde{\kappa}_{\bm{q}_1}
\tilde{\kappa}_{\bm{q}_2}
\tilde{\kappa}_{\bm{q}_3}\Delta_{\bm{q}_{123}}(l_1,l_2,l_3),
\label{eq:bispest2}
\end{equation}
where we have introduced the abbreviated notation $\bm{q}_{123}\equiv
\bm{q}_1+\bm{q}_2+\bm{q}_3$ and the summation runs over all the grids 
of $\bm{q}_1,\bm{q}_2$ and $\bm{q}_3$ in Fourier space. 
The function $\Delta_{\bm{q}_{123}}(l_1,l_2,l_3)$ denotes the selection
function defined so that $\Delta_{\bm{q}_{123}}=1$ if each vector has
the target length, $l_i-\Delta l_i/2\le q_i\le l_i+\Delta l_i/2$
($i=1,2,3$) as well as the three vectors satisfy the triangle condition,
$\bm{q}_{123}=\bm{0}$; otherwise 
$\Delta_{\bm{q}_{123}}=0$.
The quantity $N_{\rm trip}$ is the number of triplets of grids (vectors)
which form the triangle configuration of ($l_1,l_2,l_3$).

We average the estimator of $\tkappa_{\bm{q}_1}
\tkappa_{\bm{q}_2}\tkappa_{\bm{q}_3}$ over all the triangles that have
the side lengths of $(l_1,l_2,l_3)$ within the bin widths.
For the limits of $l_i\gg l_f$, 
the number of independent
triplets of a target triangle configuration, $N_{\rm trip}$, can be
estimated as 
\begin{eqnarray}
N_{\rm
trip}(l_1,l_2,l_3)&\equiv &\sum_{\bm{q}_i} \Delta_{\bm{q}_{123}}(l_1,l_2,l_3),
\nonumber\\
&\approx& 2\times \frac{ (2\upi l_1\Delta l_1) \times 
(l_2 \Delta \varphi_{12} \times \Delta l_2)}{
(2\upi/\Theta_{\rm s})^4}=\frac{\Omega_{\rm s}^2}{4\upi^3}  l_1 l_2 \Delta
l_1\Delta l_2
 \Delta \varphi_{12},  
\label{eq:trip_app}
\end{eqnarray}
where ``independent triplets'' means different combinations of
triplets that form the target triangle configuration, and $\Delta l_i$
is the bin width around each multipole bin, $\varphi_{12}$ is the angle
between the two side lengths $q_1$ and $q_2$ in
Fig.~\ref{fig:triang}, and $\Delta\varphi_{12}$ is the angle extent
allowed by the multipole bins as we explain below. 
First, $2\upi l_1\Delta l_1/(2\upi/\Theta_{\rm s})^2$ is the
number of grids in the annulus, which has the radius of $l_1$ with width of
$\Delta l_1$, as in the coefficient in the power spectrum covariance
(see equation~\ref{eq:n_mode}).  Then, for a grid of the vector
$\bm{q}_1$, we can find other two vertices (two grids), denoted by the
vectors $\bm{q}_2$ and $\bm{q}_3$ in Fig.~\ref{fig:triang}, so that the
two vectors have the target lengths, $q_2=l_2$ and $q_3=l_3$ within the
bin widths. Given the bin widths, the number of grids, which are covered
by variations in the two vectors $\bm{q}_2$ and $\bm{q}_3$, can be
estimated as $(l_2\Delta\varphi_{12}\times \Delta l_2)/(2\upi/\Theta_{\rm
s})^2$, where $\Delta\varphi_{12}$ is the variation in the angle
$\varphi_{12}$ allowed by the variations in the two vectors $\bm{q}_1$
and $\bm{q}_2$.  Finally, the prefactor $2$ in equation~(\ref{eq:trip_app})
accounts for counter-part triplet of grids that are transformed by
parity transformation $\varphi_{12} \rightarrow -\varphi_{12}$.
As can be found from Fig.~\ref{fig:triang},
the angle $\varphi_{12}$ is given as
\begin{equation}
\cos\varphi_{12}=\frac{q_1^2+q_2^2-q_3^2}{2q_1q_2},
\end{equation}
where the two vectors have the target lengths within the bin widths:
$l_i-\Delta l_i/2\le q_i\le l_i+\Delta l_i/2$ ($i=1,2$). 
If varying
the side length $q_3$ by $\Delta l_3$, we can find
the angle variation as
\begin{eqnarray}
\Delta\varphi_{12}&\simeq &(\sin\varphi_{12})^{-1}
\frac{l_3\Delta l_3}{l_1l_2}
= \frac{2l_3\Delta l_3}
{\sqrt{
2l_1^2l_2^2+2l_1^2l_3^2+2l_2^2l_3^2-l_1^4-l_2^4-l_3^4}}. 
\label{eq:delphi}
\end{eqnarray}
Substituting this equation into equation~(\ref{eq:trip_app}),
the number of independent triplets that form the triangle
configuration of ($l_1,l_2,l_3$), equation~(\ref{eq:trip}), is obtained.
The triplet number has a symmetric property under permutation of
$l_1\leftrightarrow l_2$ and so on.  As shown in \cite{Joachimietal:09},
this expression gives a good approximation, better than 1 per cent accuracy, to
the full-sky expression that is given by the Wigner-3$j$ symbol, for the
limits of $l_i\gg 1$ (i.e. the flat-sky limit).

The ensemble average of the bispectrum estimator (\ref{eq:bispest2}) is
found to yield the lensing bispectrum:
\begin{eqnarray}
\ave{\hat{B}(\bm{l}_1,\bm{l}_2,\bm{l}_3)}&=&
\frac{1}{\Omega_{\rm s}N_{\rm
trip}(l_1,l_2,l_3)}\sum_{\bm{q}_i}
\ave{\tilde{\kappa}_{\bm{q}_1}
\tilde{\kappa}_{\bm{q}_2}
\tilde{\kappa}_{\bm{q}_3}}\Delta_{\bm{q}_{123}}(l_1,l_2,l_3)
\nonumber\\
&=&\frac{1}{\Omega_{\rm s}N_{\rm trip}(l_1,l_2,l_3)}\sum_{\bm{q}_i}
B(l_1,l_2,l_3)
\Omega_{\rm
s}\delta^K_{\bm{q}_{123}}\Delta_{\bm{q}_{123}}(l_1,l_2,l_3)\nonumber\\
&\simeq & \frac{1}{N_{\rm trip}}B({l}_1,{l}_2,{l}_3) 
\sum_{\bm{q}_i}\delta^K_{\bm{q}_{123}}\Delta_{\bm{q}_{123}}(l_1,l_2,l_3)\nonumber\\
&=& B({l}_1,{l}_2,{l}_3),
\end{eqnarray}
where we have assumed on the third line that the lensing bispectrum does
not largely change within the multipole bins $\Delta l_i$, and on the
forth line that the Kronecker delta $\delta^K_{\bm{q}_{123}}$
automatically holds together with the selection function
$\Delta_{\bm{q}_{123}}$; i.e., $\delta^K_{\bm{q}_{123}}=1$ when
$\Delta_{\bm{q}_{123}}=1$.

The bispectrum covariance is defined as
\begin{eqnarray}
{\rm Cov}\left[B(l_1,l_2,l_3),B(l_1',l_2',l_3')\right]\equiv
\frac{1}{\Omega_{\rm s}N_{\rm trip}}
\frac{1}{\Omega_{\rm s}N_{\rm trip}^\prime}
\sum_{\bm{q}_i}\sum_{\bm{q}'_i}
\skaco
{\tilde{\kappa}_{\bm{q}_1}
\tilde{\kappa}_{\bm{q}_2}
\tilde{\kappa}_{\bm{q}_3}
\tilde{\kappa}_{\bm{q}^{\prime}_1}
\tilde{\kappa}_{\bm{q}^{\prime}_2}
\tilde{\kappa}_{\bm{q}^{\prime}_3}}
\Delta_{\bm{q}_{123}}
\Delta_{\bm{q}_{123}^{\prime}}
-B(l_1,l_2,l_3)B(l_1',l_2',l_3'),
\label{eq:defbcov}
\end{eqnarray}
where we omitted the arguments in $N_{\rm trip}$ and
$\Delta_{\bm{q}_{123}}$ such as
$N_{\rm trip}^\prime \equiv N_{\rm trip}(l_1^\prime,l_2^\prime,l_3^\prime)$ for
notational simplicity. The bispectrum covariance arises from the
six-point correlation function. 

The six-point correlation function can be further computed as
\begin{eqnarray}
\skaco
{\tilde{\kappa}_{\bm{q}_1}
\tilde{\kappa}_{\bm{q}_2}
\tilde{\kappa}_{\bm{q}_3}
\tilde{\kappa}_{\bm{q}^{\prime}_1}
\tilde{\kappa}_{\bm{q}^{\prime}_2}
\tilde{\kappa}_{\bm{q}^{\prime}_3}}&=&
\Omega_{\rm s}^3 P(q_1)P(q_3)P({q}^{\prime}_2)
\delta^K_{\bm{q}_{12}}
\delta^K_{\bm{q}_{31'}}
\delta^K_{\bm{q}_{2'3'}}
+
\mbox{14 perms.}
\nonumber\\
&&
+ \Omega_{\rm s}^2
B(q_1,q_2,q_3)B(q_1',q_2',q_3')\delta^K_{\bm{q}_{123}}\delta^K_{\bm{q}_{1'2'3'}}
+
\mbox{9 perms.}
\nonumber \\
&&
+\Omega_{\rm s}^2P(q_1)T(\bm{q}_3,\bm{q}_1',\bm{q}_2',\bm{q}_3')\delta^K_{\bm{q}_{12}}
\delta^K_{\bm{q}_{31'2'3'}}
+
\mbox{14 perms.}
\nonumber \\
&&
+\Omega_{\rm s}P_6(\bm{q}_1,\bm{q}_2,\bm{q}_3,\bm{q}_1',\bm{q}_2',\bm{q}_3')
\delta^K_{\bm{q}_{1231'2'3'}},
\label{eq:6kappap}
\end{eqnarray}
where $\delta^K_{\bm{q}_{1231'}}\equiv
\delta^K_{\bm{q}_1+\bm{q}_2+\bm{q}_3+\bm{q}_1'}$ and so on.  Inserting the
six-point correlation function above into equation~(\ref{eq:defbcov}) yields
\begin{eqnarray}
{\rm Cov}\left[B(l_1,l_2,l_3),B(l_1',l_2',l_3')\right]&=&\frac{1}{N_{\rm trip}}
\frac{1}{N_{\rm trip}^\prime}
\sum_{\bm{q}_i}\sum_{\bm{q}'_i}\Delta_{\bm{q}_{123}}(l_1,l_2,l_3)
\Delta_{\bm{q}_{123}^\prime}(l_1^\prime,l_2^\prime,l_3^\prime)
\nonumber\\
&&\times
\left\{
\Omega_{\rm s}
P(q_1)P(q_2)P(q_3)
\left[
\delta^K_{\bm{q}_{11'}}
\delta^K_{\bm{q}_{22'}}
\delta^K_{\bm{q}_{33'}}
+\delta^K_{\bm{q}_{11'}}
\delta^K_{\bm{q}_{23'}}
\delta^K_{\bm{q}_{32'}}
+\delta^K_{\bm{q}_{12'}}
\delta^K_{\bm{q}_{21'}}
\delta^K_{\bm{q}_{33'}}
+\mbox{3 perms.}
\right]\right.
\nonumber\\
&&\hspace{1em}
+B(q_1,q_2,q_1')B(q_3,q_2',q_3')\delta^K_{\bm{q}_{121'}}\delta^K_{\bm{q}_{32'3'}}
+\mbox{8 perms.}
\nonumber \\
&&\hspace{1em}
+P(q_1)T(\bm{q}_2,\bm{q}_3,\bm{q}_2',\bm{q}_3')\delta^K_{\bm{q}_{11'}}
\delta^K_{\bm{q}_{232'3'}}
+\mbox{8 perms.}
\nonumber \\
&&\hspace{1em}
\left.
+\frac{1}{\Omega_{\rm s}}P_6(\bm{q}_1,\bm{q}_2,\bm{q}_3,\bm{q}_1',\bm{q}_2',\bm{q}_3')
\delta^K_{\bm{q}_{1231'2'3'}}\right\},
\label{eq:6kappa}
\end{eqnarray}
where we have dropped terms including the Kronecker delta
such as $\delta^K_{\bm{q}_{12}}$, because we are not interested in such
triangles with $\bm{q}_3=0$ under the triangle conditions $\bm{q}_{123}=\bm{0}$.
As we explained in Fig.~\ref{fig:bcov}, the bispectrum covariance has
different contributions that arise from the terms proportional to $P^3$,
$O(BB)$, $O(PT)$ and $P_6$, respectively. These terms are further
simplified as shown below.

\subsubsection{Gaussian error contribution to the bispectrum covariance}

First, we consider the term in equation~(\ref{eq:6kappa}) proportional to
the power spectra cubed $P^3$. We call this term the Gaussian
error contribution, although the bispectrum itself is a measure of
non-Gaussianity in the lensing field. Many previous works only
considered this term when studying the lensing bispectrum, e.g. for
parameter forecasts, mainly for simplicity. The Gaussian term is
simplified as   
\begin{eqnarray}
{\rm Cov}_{\rm Gauss}
&=&\frac{\Omega_{\rm s}}{N_{\rm trip}}\frac{1}{N_{\rm trip}^\prime}
\sum_{\bm{q}_i}\sum_{\bm{q}_i'}\Delta_{\bm{q}_{123}}(l_1,l_2,l_3)
\Delta_{\bm{q}_{123}^{\prime}}(l_1',l_2',l_3')
P(q_1)P(q_2)P(q_3)
\left[
\delta^K_{\bm{q}_{11'}}\delta^K_{\bm{q}_{22'}}\delta^K_{\bm{q}_{33'}}
+
\mbox{5 perms.}
\right]\nonumber\\
&&
\simeq \frac{
\Omega_{\rm s}}{N_{\rm trip}N_{\rm trip}^\prime}
\sum_{\bm{q}_i}\Delta_{\bm{q}_{123}}(l_1,l_2,l_3)
P(q_1)P(q_2)P(q_3)\left[
\delta^K_{l_1l_1'}
\delta^K_{l_2l_2'}
\delta^K_{l_3l_3'}
+\mbox{5 perms.}
\right]
\nonumber \\
&&
=\frac{
\Omega_{\rm s}}{N_{\rm trip}}
P(l_1)P(l_2)P(l_3)\left[
\delta^K_{l_1l_1'}
\delta^K_{l_2l_2'}
\delta^K_{l_3l_3'}
+\mbox{5 perms.}
\right], 
\label{eq:bcov_gauss}
\end{eqnarray}
where we use the facts that the power spectrum does not largely
change within the bin width and that the terms including the Kronecker
delta are non-vanishing only
if the two triangles of the two bispectra have the same shape. The
terms including the Kronecker delta give 1, 2 or 6 for general,
isosceles and equilateral triangles, respectively, which correspond to
the factor $\Delta$ in equation~(28) of \cite{TakadaJain:04}. The
Gaussian error covariance term contributes only to the diagonal terms of
the bispectrum covariance matrix. Recalling the fact $N_{\rm
trip}\propto 1/\Omega_{\rm s}^2$ (equation~\ref{eq:trip}), this contribution
scales with survey area as ${\rm Cov}_{\rm
Gauss}\propto 1/\Omega_{\rm s}$.

\subsubsection{Non-Gaussian error contributions to the bispectrum covariance}

Other terms in equation~(\ref{eq:6kappa}) are non-Gaussian error
contributions to the bispectrum covariance that arise from the
higher-order functions. In the following, we derive further simplified
expressions for each term.

\paragraph{Non-Gaussian terms of $O(B^2)$}
We consider the
terms that are proportional to the bispectra
squared, the terms of $O(B^2)$. 
Here we
consider the first term of $O(B^2)$ terms in equation~(\ref{eq:6kappa}) as an
example as
\begin{eqnarray}
\frac{1}{N_{\rm trip}}
\frac{1}{N_{\rm trip}^\prime}
\sum_{\bm{q}_i}\sum_{\bm{q}'_i}B(q_1,q_2,q_1')B(q_3,q_2',q_3')
\delta^K_{\bm{q}_{121'}}\delta^K_{\bm{q}_{32'3'}}\Delta_{\bm{q}_{123}}(l_i)
\Delta_{\bm{q}_{123}'}(l_i^\prime)
\nonumber \\  
&&\hspace{-26em}\simeq 
\frac{1}{N_{\rm trip}}
\frac{1}{N_{\rm trip}^\prime}
B(l_1,l_2,l_1')
B(l_3,l_2',l_3')\sum_{\bm{q}_i}\sum_{\bm{q}_i'}
\delta^K_{\bm{q}_1+\bm{q}_2+\bm{q}_1^\prime}
\delta^K_{\bm{q}_3+\bm{q}_2^\prime+\bm{q}_3^\prime}
\Delta_{\bm{q}_{123}}(l_i) \Delta_{\bm{q}_{123}'}(l_i')\nonumber\\
&&\hspace{-26em}
=\frac{1}{N_{\rm trip}}
\frac{1}{N_{\rm trip}^\prime}
B(l_1,l_2,l_1')
B(l_3,l_2',l_3')
\sum_{\bm{q}_1,\bm{q}_2,\bm{q}_3}\Delta_{\bm{q}_{123}}(l_i)
\delta^K_{l_3l_1^\prime}
\left\{
\sum_{\bm{q}_2^\prime,\bm{q}_3^\prime}
\delta^K_{\bm{q}_3+\bm{q}_2^\prime+\bm{q}_3^\prime}
\Delta_{\bm{q}_{3}+\bm{q}_2'+\bm{q}_3'}(l_i')
\right\}
\nonumber\\
&&\hspace{-26em}
=\frac{1}{N_{\rm trip}}
\frac{\delta^K_{l_3l_1^\prime}
}{N_{\rm trip}^\prime}
B(l_1,l_2,l_1')
B(l_3,l_2',l_3')
\sum_{\bm{q}_1,\bm{q}_2,\bm{q}_3}\Delta_{\bm{q}_{123}}(l_i)
\left\{
\sum_{\bm{q}_2^\prime,\bm{q}_3^\prime}
\Delta_{\bm{q}_{3}+\bm{q}_2'+\bm{q}_3'}(l_i')
\right\}\nonumber\\
&&\hspace{-26em}
\simeq \frac{1}{N_{\rm trip}}
\frac{\delta^K_{l_3l_1^\prime}}{N_{\rm trip}^\prime}
B(l_1,l_2,l_1')
B(l_3,l_2',l_3')
\sum_{\bm{q}_1,\bm{q}_2,\bm{q}_3}\Delta_{\bm{q}_{123}}(l_i)
\left[
\frac{2
\Delta l_2^\prime l_2^\prime\Delta\varphi_{32'}}
{(2\upi/\Theta_{\rm s})^2}
\right]\nonumber\\
&&\hspace{-26em}
=\delta^K_{l_3l_1^\prime}
B(l_1,l_2,l_3)
B(l_1',l_2',l_3')
\frac{\Omega_{\rm s}}{(2\upi)^2}
\frac{2\Delta l_2^\prime l_2^\prime\Delta\varphi_{32'}}{N_{\rm trip}^\prime(l_1',l_2',l_3')}
\nonumber\\
&&\hspace{-26em}
=
\delta^K_{l_3l_1^\prime}
\frac{2\upi}{\Omega_{\rm s}l_1'\Delta l_1'}B(l_1,l_2,l_3)
B(l_1',l_2',l_3').
\label{eq:bb1}
\end{eqnarray}
In the above calculation, we use several simplifications using the
triangle condition and the Kronecker deltas. In the second
line of the r.h.s., the product of Kronecker deltas,
$\delta^K_{\bm{q}_1+\bm{q}_2+\bm{q}_1'}\delta^K_{\bm{q}_3+\bm{q}_2^\prime
+\bm{q}_3'}$, is non-vanishing only if $l_3=l_1^\prime$, because of the
conditions imposed by the selection functions
$\Delta_{\bm{q}_{123}}(l_i)$ and $\Delta_{\bm{q}^\prime_{123}}(l_i')$;
i.e. $l_1-\Delta l_1/2\le q_i\le l_1+\Delta l_1/2\le$,
$\bm{q}_1+\bm{q}_2+\bm{q}_3=\bm{0}$, and so on. In other words, the term
above is non-vanishing only if the two triangle configurations have the
same length on their one side,
$l_3=l_1'$ in this case (see Fig.~\ref{fig:bcov}). Hence, in the second
line, we introduce the Kronecker delta $\delta^K_{l_3l_1'}$,
and also use the Kronecker delta
$\delta^K_{\bm{q}_1+\bm{q}_2+\bm{q}_1'}$ to drop the summation over
$\bm{q}_1'$ (then we also use the triangle conditions,
$\bm{q}_1'=-\bm{q}_1-\bm{q}_2=\bm{q}_3$ due to
$\bm{q}_{1}+\bm{q}_2+\bm{q}_3=\bm{0}$). In the third line, we drop
one Kronecker delta $\delta^K_{\bm{q}_3+\bm{q}_1'+\bm{q}_2'}$
because it is automatically satisfied by the selection function
$\Delta_{\bm{q}_3+\bm{q}_1'+\bm{q}_2'}$. The curly bracket is intended
to mean that the summation $\sum_{\bm{q}_2',\bm{q}_3'}$ is done before
another summation $\sum_{\bm{q}_i}$. In the forth line, we computed the
summation
$\sum_{\bm{q}_2',\bm{q}_3'}\Delta_{\bm{q}_3+\bm{q}_2'+\bm{q}_3'}(l_1,'l_2',l_3')
$, which gives the number of triplets of grids satisfying the selection
function. For the limits of $q_2',q_3'\gg l_f$, we can use the similar
calculations given by equations~(\ref{eq:trip_app}) and
(\ref{eq:trip}). With the selection function
$\Delta_{\bm{q}_3+\bm{q}_2'+\bm{q}_3'}(l_1,'l_2',l_3')$, we can find
that, for a given vector $\bm{q}_3$, the summation gives the number of
paired grids ($\bm{q}_2',\bm{q}_3'$) within the bin widths, $l_2'-\Delta
l_2'/2\le q_2'\le l_2'+\Delta l_2'/2$ and $l_3'-\Delta l_3'/2\le q_3'\le
l_3'+\Delta l_3'/2$. According to the similar calculation
(equation~\ref{eq:delphi}), the number of the paired grids is estimated as
$2\times (l_2'\Delta l_2' \Delta\varphi_{32'})/(2\upi/\Theta_{\rm s})^2$,
where $\Delta \varphi_{32'}$ is the variation in the angle between
the two side lengths $l_2'$ and $l_3$ in the triangle of
($l_3,l_2',l_3'$) due to the variations of $l_2'$ and $l_3$ within the
bin widths.  In the fifth line, we used
$\sum_{\bm{q}_i}\Delta_{\bm{q}_{123}}=N_{\rm trip}(l_1,l_2,l_3)$.
In this step, using the fact $l_3=l_1'$ via the
Kronecker delta $\delta^K_{l_3l_1'}$, we also replaced
$B(l_1,l_2,l_1')B(l_3,l_2',l_3')$ with
$B(l_1,l_2,l_3)B(l_1',l_2',l_3')$, the product of the original bispectra
taken in the covariance calculation.  In the sixth line, we further
simplified $2(l_2'\Delta l_2\Delta\varphi_{32'})/[N_{\rm
trip}(l_1',l_2',l_3')(2\upi/\Theta_{\rm s})^2]$ by using the similar
equations to equations~(\ref{eq:delphi}) and (\ref{eq:trip}) and the fact
$l_3=l_1'$ to obtain the coefficient $2\upi/[\Omega_{\rm s}l_1'\Delta
l_1']$.

Performing the similar calculations to other terms in equation~(\ref{eq:6kappa}),
the terms of $O(B^2)$ in the bispectrum covariance can be reduced to 
\begin{equation}
{\rm Cov}[B(l_i),B(l_i')]_{{\rm NG}, O(B^2)}=
\frac{2\upi}{\Omega_{\rm s}} B(l_1,l_2,l_3)B(l_1',l_2',l_3')
\left[\frac{\delta^K_{l_1l_1'}}{l_1\Delta l_1}+\frac{\delta^K_{l_1l_2'}}{l_1\Delta l_1}
+ \mbox{7 perms.} \right].
\label{eq:bcovbb}
\end{equation}
These terms scale with survey area as ${\rm Cov}_{{\rm NG},
O(B^2)}\propto 1/\Omega_{\rm s}$. The terms contribute to diagonal terms
of the bispectrum covariance as well as some off-diagonal terms when the
two triangle configurations have one same length side within the bin widths
(see Fig.~\ref{fig:bcov} for the diagrammatic illustration).

\paragraph{Non-Gaussian terms of $O(PT)$}
Next let's
consider the terms proportional to $O(PT)$. Similarly to the calculation in
equation~(\ref{eq:bb1}), the first term of equation~(\ref{eq:6kappa}) is
simplified as
\begin{eqnarray}
\frac{1}{N_{\rm trip}}
\frac{1}{N_{\rm trip}^\prime}\sum_{\bm{q}_i}\sum_{\bm{q}_i'} 
P(q_1)T(\bm{q}_2,\bm{q}_3,\bm{q}_2',\bm{q}_3')\delta^K_{\bm{q}_{11'}}
\delta^K_{\bm{q}_{232'3'}}
\Delta_{\bm{q}_{123}}(l_i)\Delta_{\bm{q}_{123}'}(l_i')&&\nonumber\\
&&\hspace{-22em}\simeq \delta^K_{l_1l_1'}P(l_1)
T(\bm{l}_2,\bm{l}_3,\bm{l}_2',\bm{l}_3')
\frac{1}{N_{\rm trip}}
\frac{1}{N_{\rm trip}^\prime}\sum_{\bm{q}_i}\Delta_{\bm{q}_{123}}(l_1,l_2,l_3)
\left\{
\sum_{\bm{q}_2',\bm{q}_3'}
\Delta_{-\bm{q}_1+\bm{q}_2'+\bm{q}_3'}\right\}
\nonumber\\
&&\hspace{-22em}\simeq \delta^K_{l_1l_1'}
P(l_1)
T(\bm{l}_2,\bm{l}_3,\bm{l}_2',\bm{l}_3')
\frac{1}{N_{\rm trip}}
\frac{1}{N_{\rm trip}^\prime}\frac{2\times(l_2'\Delta l_2\Delta 
\varphi_{12'})}{(2\upi/\Theta_{\rm s})^2}
\sum_{\bm{l}_i}\Delta_{\bm{l}_{123}}
\nonumber\\
&& \hspace{-22em} \simeq
  \delta^K_{l_1'l_1}
\frac{2\upi}{\Omega_{\rm s}l_1\Delta l_1}
P(l_1)T(\bm{l}_2,\bm{l}_3,\bm{l}_2',\bm{l}_3').
\end{eqnarray}
In the first line of the r.h.s. of the above equation, we use the fact
that the term is non-vanishing only if the two triangle configurations
have the same length on their one side, $l_1=l_1'$ in this case (see
Fig.~\ref{fig:bcov}). In this step, therefore, we drop one summation
over $\sum_{\bm{q}_1'}$ by using the Kronecker delta
$\delta^K_{\bm{q}_1+\bm{q}_1'}$ and introduce the Kronecker delta
$\delta_{l_1l_1'}^K$. We take out the trispectrum from the
summation (assuming it does not largely change within the bin widths),
because the four-point configuration is uniquely specified by the two
triangle configurations; the 4 outer circumference side lengths are
given by $l_2,l_3$, $l_2'$ and $l_3'$, and the diagonal
length is given by $l_1$ \footnote{The two-dimensional trispectrum is
uniquely specified by 5 parameters to characterize the four-point
configuration; e.g. the outer-circumference 4 side lengths plus the
diagonal length.}.  In the second line, we compute the number of the
modes given by the summation
$\sum_{\bm{q}_2',\bm{q}_3'}\Delta_{-\bm{q}_1+\bm{q}_2'+\bm{q}_3'}(l_1',l_2',l_3')$
as we did in the calculation of equation~(\ref{eq:bb1}). The angle
$\varphi_{12'}$ is the angle between the two side lengths $l_1$ and
$l_2'$, and $\Delta \varphi_{12'}$ is the variation due to the bin
widths.  In the third line, we use the fact
$\sum_{\bm{q}_i}\Delta_{\bm{q}_{123}}=N_{\rm trip}(l_i)$ and further
simplify the coefficient $2\times (l_2'\Delta
l_2'\Delta\varphi_{12'})/N_{\rm trip}(l_i')$ as in equation~(\ref{eq:bb1}).

Hence, the terms of $O(PT)$ in the bispectrum covariance are computed
for the limits of $l_i,l_i'\gg l_f$ as
\begin{eqnarray}
{\rm Cov}\left[B(l_i),B(l_i')\right]_{{\rm NG}, O(PT)}=
\delta^K_{l_1l_1'}\frac{2\upi}{\Omega_{\rm s}l_1\Delta
l_1}P(l_1)T(\bm{l}_2,\bm{l}_3,\bm{l}_2',\bm{l}_3') + 
\delta^K_{l_1l_2'}\frac{2\upi}{\Omega_{\rm s}l_1\Delta
l_1}P(l_1)T(\bm{l}_2,\bm{l}_3,\bm{l}_1',\bm{l}_3') 
+ \mbox{7 perms.}
\label{eq:bcovpt}
\end{eqnarray}
Again note that all the trispectra in the terms are uniquely specified by
the two triangle configurations; the 4 outer-circumference side lengths
and the diagonal length (see Fig.~\ref{fig:bcov}). The terms of $O(PT)$
scale with survey area as ${\rm Cov}_{{\rm NG}, O(PT)}\propto
1/\Omega_{\rm s}$. The terms contribute to the diagonal terms of the
bispectrum covariance matrix as well as some off-diagonal terms where
the two triangles have one same length.

\paragraph{Non-Gaussian term of $O(P_6)$}
Finally we
consider the contribution arising from the connected six-point correlation
function as
\begin{eqnarray}
{\rm Cov}[B(l_i),B(l_i')]_{{\rm NG}, O(P_6)}&=&
\frac{1}{\Omega_{\rm s}N_{\rm trip}}
\frac{1}{N_{\rm trip}^\prime}\sum_{\bm{q}_i}\sum_{\bm{q}_i'} 
P_6(\bm{q}_1,..,\bm{q}_3')
\delta^K_{\bm{q}_{1231'2'3'}}
\Delta_{\bm{q}_{123}}(l_i)\Delta_{\bm{q}_{123}'}(l_i')\nonumber\\
&& \hspace{-5em}\simeq
\frac{1}{\Omega_{\rm s}N_{\rm trip}}
\frac{1}{N_{\rm trip}^\prime}\frac{(2\upi)^{12}}{\Omega_{\rm s}^6}
\int\!\!\prod_{i=1}^{3}d^2\bm{q}_i\int\!\!\prod_{i=1}^3d^2\bm{q}_i' 
P_6(\bm{q}_1,\bm{q}_2,\bm{q}_3,\bm{q}_1',\bm{q}_2',\bm{q}_3')
\Delta_{\bm{l}_{123}}(l_1,l_2,l_3)
\Delta_{\bm{l}_{123}'}(l_1',l_2',l_3')\nonumber\\
&&\hspace{-5em}
\simeq \frac{1}{\Omega_{\rm s}}
\int\!\!\frac{d\psi}{2\upi} ~ P_6(\bm{l}_1,\bm{l}_2,\bm{l}_3,\bm{l}_1',\bm{l}_2',\bm{l}_3'; \psi).
\label{eq:bcovp6}
\end{eqnarray}
As shown in Fig.~\ref{fig:bcov}, the six-point configuration in the
Fourier space we consider here is constituted by two triangles.
Indeed, the selection functions $\Delta_{\bm{q}_{123}}$ and
$\Delta_{\bm{q}_{123}^\prime}$ automatically satisfy the condition of
the six-point configuration in the Fourier space:
$\bm{q}_{123}+\bm{q}_{123}^\prime=\bm{0}$. The
unspecified configuration parameter is only the angle between the two
triangles, $\psi$. Hence, the summations
$\sum_{\bm{q}_i}$ and $\sum_{\bm{q}_i^\prime}$ can be replaced with
one-dimensional integration over the angle $\psi$.  The
covariance terms of $O(P_6)$ scale with survey area as ${\rm Cov}_{{\rm
NG}, O(P_6)}\propto 1/\Omega_{\rm s}$ and contribute to both the
diagonal and off-diagonal parts of the bispectrum covariance.

\subsubsection{Flat-sky formula for the bispectrum covariance} 

Summing up all the terms of equations~(\ref{eq:bcov_gauss}),
(\ref{eq:bcovbb}), (\ref{eq:bcovpt}) and (\ref{eq:bcovp6}), 
we obtain the covariance matrix shown in equation~(\ref{eq:bcov}) without
the HSV term.
We should again note that these terms are derived using the
discrete Fourier modes confined within the survey region.
Hence, the non-Gaussian error
contributions in the above equation account only for the mode coupling
between such Fourier eigenmodes. As we have shown in the main part of
this paper, we also need to include the HSV
contribution to the non-Gaussian errors, which arises from a coupling of
the modes within the survey region with the mass density
fluctuations of scales comparable with or larger than the survey
region.

\section{Cross-covariance between lensing power spectrum and bispectrum}
\label{app:pbcov}

Based on the discrete Fourier decomposition formulation, the power
spectrum estimator can be defined in \cite{TakadaBridle:07} as 
\begin{equation}
\hat{P}(l)\equiv \frac{1}{\Omega_{\rm s}N_{\rm pairs}(l)}\sum_{\bm{q};
 q\in l}
\tkappa_{\bm{q}}\tkappa_{-\bm{q}},
\end{equation}
where the summation is over Fourier modes which have the length of $l$
to within the bin width $\Delta l$. 

The cross-covariance between the power spectrum $P(l)$ and the
bispectrum $B(l_1,l_2,l_3)$ is defined as
\begin{eqnarray}
{\rm Cov}\left[P(l),B(l_1,l_2,l_3)\right]
&=& 
\frac{1}{\Omega_{\rm s}N_l}\frac{1}{\Omega_{\rm s}
N_{\rm trip}}\sum_{\bm{q}; q\in l} \sum_{\bm{q}_i}
\skaco{
\tilde{\kappa}_{\bm{q}}
\tilde{\kappa}_{-\bm{q}}
\tilde{\kappa}_{\bm{q}_1}
\tilde{\kappa}_{\bm{q}_2}
\tilde{\kappa}_{\bm{q}_3}
}\Delta_{\bm{q}_{123}}(l_i) - P(l)B(l_1,l_2,l_3).
\label{eq:pbcovdef}
\end{eqnarray}
Thus the cross-covariance depends on the five-point correlation function. 
The five-point correlation function in the above equation can be further
computed as 
\begin{eqnarray}
\skaco{
\tilde{\kappa}_{\bm{q}}
\tilde{\kappa}_{-\bm{q}}
\tilde{\kappa}_{\bm{q}_1}
\tilde{\kappa}_{\bm{q}_2}
\tilde{\kappa}_{\bm{q}_3}
}&=&
\Omega_{\rm s}^2
P(q)B(q_1,q_2,q_3)\delta^K_{\bm{q}_{123}}+
\Omega_{\rm s}^2
P(q)B(q,q_2,q_3)\delta^K_{\bm{q}+\bm{q}_1}
\delta^K_{-\bm{q}+\bm{q}_2+\bm{q}_3}
+\mbox{8 perms.}
\nonumber\\
&&+\Omega_{\rm s}P_5(\bm{q},-\bm{q},\bm{q}_1,\bm{q}_2,\bm{q}_3)
\delta^K_{\bm{q}_{123}}
\end{eqnarray}
Inserting this equation into Eq.(\ref{eq:pbcovdef}) and using the
similar calculation we have used, we can find the flat-sky formula for
the cross-covariance as in equation~(\ref{eq:pbcov}) but without the HSV term.

\section{Dependence of HSV on simulation box size}
\label{app:volume}
As we have studied, the large-scale mass fluctuations
of scales comparable with or larger than a survey area cause a
significant contribution to the non-Gaussian errors of the lensing
field.  However, the ray-tracing simulations we have used are generated
from a finite volume of N-body simulations, and do not 
include contributions from the modes of scales {\em beyond} the N-body
simulation box. Hence the simulation results we showed in the main text
might underestimate the genuine effect of the HSV, which we estimate 
in this appendix. 

\begin{figure}
\begin{center}
 \includegraphics[width=0.48\textwidth]{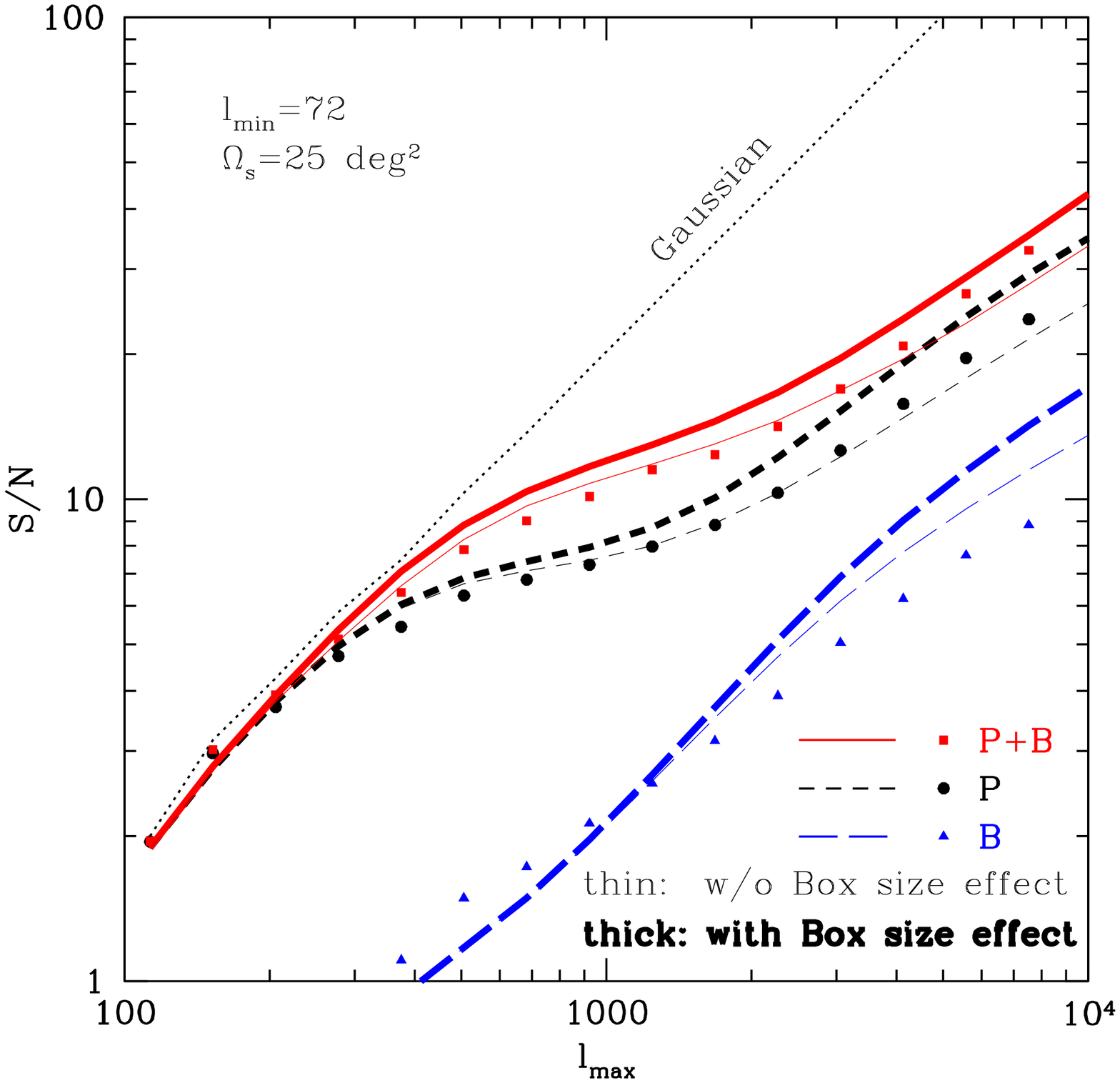}
\caption{The thick curves show the halo model predictions where we used 
the HSV effect taking into account 
of the effect of finite size of N-body simulations used in 
 the ray-tracing simulations (see text for details). 
The thin curves and the simulation results are the same as in the
 left-hand panel of Fig.~\ref{fig:sn}.
 }
\label{fig:boxeffect}
\end{center}
\end{figure}

As described in detail in \cite{Satoetal:09}, the
simulated convergence maps are generated from N-body simulations with
box size of $240 h^{-1}$Mpc up to the redshift of $z_s=1$.  Thus the
simulation maps do not include Fourier modes of $k<k_{\rm box}\equiv
2\upi/240\simeq 0.026$ $h$Mpc$^{-1}$.  Using our halo model approach, we
can estimate the HSV effect seen in the ray-tracing simulations by
inserting $k_{\rm box}$ into the lower bound of the integral of the HSV
terms in equations~(\ref{eq:pscov_hsv}), (\ref{eq:bcov_hsv}) and
(\ref{eq:pbcov_hsv}) instead of zero.  The thick curves in
Fig.~\ref{fig:boxeffect} show the resulting S/N, which differs from
the simulation results by 10-20 per cent.  Comparing the thick and thin curves
manifests that the box size effect is larger at higher multipoles,
because the HSV contribution at higher multipoles arise preferentially
from lensing structures at higher redshifts
(Fig.~\ref{fig:bcov_hsv_dlnM}) and therefore the finite box effect is
more important at higher redshift (as the ratio of the simulation box to
the ray-tracing simulation area becomes smaller).  A possible reason of
the discrepancy between the halo model and the ray-tracing simulation
results is that the halo model calculation ignores some non-Gaussian
contributions; for instance, we included only the one-halo terms to the
four-, five- and six-point correlation functions in the covariance
calculation. Hence, the halo model tends to underestimate the
non-Gaussian error amplitudes, and in turn to overestimate S/N.
However, we would also like to note that the ray-tracing simulations may
not be accurate enough at higher multipoles beyond $l\simeq 6000$
\citep[see][]{Satoetal:09}. Most importantly, in this paper, we were
able to develop the model to describe the non-Gaussian covariance to a
10--20 per cent accuracy, and therefore we leave an issue on the 10-20 per cent
discrepancy for future work.

\end{document}